# Exciton–polaron Umklapp scattering in Wigner crystals

Erfu Liu[1,2†], Matthew Wilson[1†], Jenny Hu[3,4], Alexandra Zimmerman[3,4], Amal Mathew[3,4], Tianyi Ouyang[1], Ao Shi[1], Takashi Taniguchi[5], Kenji Watanabe[6], Tony F. Heinz[3,4], Yia–Chung Chang[7,8*] Chun Hung Lui[1*]

[1]Department of Physics and Astronomy, University of California, Riverside, CA 92521, USA.
[2]School of Physics, National Laboratory of Solid State Microstructures, and Collaborative Innovation Center for Advanced Microstructures, Nanjing University, Nanjing, China.
[3]Department of Applied Physics, Stanford University, Stanford, CA 94305, USA
[4]SLAC National Accelerator Laboratory, Menlo Park, CA 94025, USA
[5]Research Center for Materials Nanoarchitectonics, National Institute for Materials Science, 1–1 Namiki, Tsukuba 305–0044, Japan
[6]Research Center for Electronic and Optical Materials, National Institute for Materials Science, 1–1 Namiki, Tsukuba 305–0044, Japan
[7]Research Center for Applied Sciences, Academia Sinica, Taipei 11529, Taiwan
[8]Department of Physics, National Cheng Kung University, Tainan, 701 Taiwan
[†] These authors contributed equally to the work.
*Corresponding author. Email: yiachang@gate.sinica.edu.tw; joshua.lui@ucr.edu

**Abstract:**

Strong Coulomb interactions in two–dimensional (2D) semiconductors give rise to tightly bound excitons, exciton polarons, and correlated electronic phases such as Wigner crystals (WCs), yet their mutual interplay remains poorly understood. Here we report the observation of multi–branch excitonic Umklapp scattering in both electron and hole WCs realized in ultraclean monolayer WSe₂, exhibiting exceptionally high melting temperatures ($T_c \approx 20$–$30$ K). Robust Wigner crystallization activates multiple finite–momentum optical resonances, including quasilinearly dispersing, light–like excitons and exciton polarons, extending far beyond the single excitonic Umklapp feature reported previously. Helicity–resolved magneto–optical measurements reveal a pronounced valley dependence of the scattering processes. Combined experiment and theory identify a polaron–induced brightening mechanism in which exciton polarons transfer oscillator strength from bright zero–momentum states to otherwise dark finite–momentum states, explaining the emergence of multiple Umklapp branches where conventional exciton–WC scattering is ineffective. These results establish WC polarons as a new quasiparticle paradigm and introduce polaron–induced Umklapp scattering as a general route to accessing finite–momentum many–body excitations in 2D quantum materials.

Polarons—quasiparticles formed when an excitation is dressed by collective many–body degrees of freedom—are a cornerstone of condensed–matter physics[1-4]. In atomically thin semiconductors, exciton polarons have emerged as a powerful framework for interpreting optical spectra in the presence of itinerant charge carriers, where excitons are dressed by a polarized Fermi sea[5-10]. Despite this progress, all exciton polarons explored experimentally to date share a defining characteristic: the dressing medium is a liquid. How polaron physics evolves when the electronic environment instead crystallizes into a strongly correlated Wigner crystal (WC) remains an open and largely unexplored question.

Wigner crystals represent the extreme limit of Coulomb–dominated electronic correlations, in which electrons or holes self–organize into an ordered lattice (Fig. 1A). This crystalline order breaks continuous translational symmetry and introduces a discrete set of lattice vectors in reciprocal space[11-20]. Recent optical studies in atomically thin semiconductors have shown that such charge lattices can weakly scatter excitons[13, 17], enabling Umklapp processes that activate otherwise dark finite–momentum states (Fig. 1B). However, these observations were interpreted within a predominantly excitonic framework and were limited to a single Umklapp feature associated with quadratically dispersing excitons on the electron–doped side[13, 17]. Whether Wigner crystallization qualitatively reshapes polaron physics—and whether it can give rise to fundamentally new optical activation mechanisms beyond conventional exciton–WC scattering—has remained unknown.

In this Article, we demonstrate that the WC regime in a two–dimensional (2D) semiconductor gives rise to a new form of Umklapp scattering governed by many–body polaron dressing. Using ultraclean monolayer $WSe_2$ devices, we realize WCs of both electrons and holes, with unusually high melting temperatures ($T_c \approx 20$–$30$ K), enabling direct comparison of Umklapp processes on the two sides of charge neutrality. In the WC regime, we observe multiple Umklapp branches, far exceeding the single exciton Umklapp line reported previously. These include not only quadratically dispersing excitons but also quasilinearly dispersing, light–like excitons and exciton polarons, revealing a much richer finite–momentum excitation spectrum than previously accessible (Fig. 1B–C). Helicity–resolved magneto–optical measurements further uncover a pronounced valley dependence of Umklapp scattering, reflecting the role of exchange interactions between excitons and the WC. Umklapp processes are strongly enhanced on the hole–doped side when excitons and WCs occupy the same valley and electronic band, while otherwise remaining weak.

Crucially, the observed multiplicity of Umklapp branches — particularly on the electron–doped side — cannot be explained by excitonic Umklapp scattering alone. Instead, it originates from a polaron–induced brightening mechanism in which exciton polarons transfer oscillator strength from bright zero–momentum states to otherwise dark finite–momentum Umklapp states. These results establish WC polarons as a new quasiparticle paradigm and show that electronic crystallization can effectively relax optical selection rules via Umklapp processes in 2D quantum materials.

## Umklapp scattering of quasilinearly dispersing excitons and exciton polarons

Our experiments employ single–gate monolayer $WSe_2$ devices encapsulated in hexagonal boron nitride (BN), with thin graphite serving as contact and gate electrodes (Fig. 2A). A gate voltage $V_g$ is applied to inject electrons or holes into the $WSe_2$ monolayer. We measure reflection spectra $R_s$ from $WSe_2$ and reference spectra $R_r$ from a nearby area without $WSe_2$ to obtain the reflectance contrast, defined as $\Delta R/R = (R_s - R_r)/R_r$. Fig. 2B displays a gate–dependent $\Delta R/R$ map of Device 1, which exhibits the $A^0$ exciton and its exciton polarons on the electron ($A_1^-$, $A_2^-$) and hole ($A^+$) sides [8, 21, 22]. A gate–dependent $\Delta R/R$ trace near 1.72 eV (black arrows and overlying spectrum in Fig. 2B) reveals two features ($A_{eu2}^0$, $A_{hu2}^0$), indicating additional states. To enhance the visibility of these weak features, we calculate the second derivative of the spectra with respect to photon energy. The resulting second–derivative $\Delta R/R$ map (Fig. 2C) then reveals five weak lines, namely, the $A_{hu1}^0$ and $A_{hu2}^0$ lines on the hole side, the $A_{eu2}^0$ line on the electron side, and the $A_{1u}^-$ and $A_{2u}^-$ lines above the exciton polarons on the electron side. We evaluate the energy spacings ($\Delta E$) between the $A_{eu2}^0$, $A_{hu2}^0$, $A_{hu1}^0$ lines and the primary $A^0$ line. The $\Delta E$ values increase nearly linearly with the density of the injected carriers (Fig. 2D). Similar results have been observed in two additional devices (Figs. S3, S4).

Prior studies on monolayer $MoSe_2$ reported a similar weak Umklapp line for excitons on the electron side, attributed to the Umklapp scattering of quadratically dispersing excitons due to the presence of a WC[13]. Following a similar approach, we compare our results with the calculated density–dependent energy of the Umklapp state using an effective exciton mass of $0.8m_e$ for monolayer $WSe_2$ (black lines in Fig. 2D). We find good agreement between the quadratic dispersion and the $A_{hu1}^0$ line, confirming the presence of WC in our sample.

In contrast, the $A_{hu2}^0$ and $A_{eu2}^0$ lines lie well above the energies of the quadratically dispersing exciton, indicating the presence of an additional exciton branch. Theory has predicted that excitons in the K and K′ valleys can couple via electron–hole (e–h) exchange interactions, giving rise to two branches with quadratic and quasilinear dispersions[23-25], consistent with recent experiments[26, 27]. Figure 1B shows these double exciton branches in monolayer $WSe_2$ from our calculations (see Supplementary Section II.6.2 for details). The linear component of the quasilinear branch arises from intra– and intervalley e–h exchange interactions characterized by a common strength $J$. Figure 2D shows the calculated energy positions of the Umklapp lines associated with this branch. Using $J = 160$ and $180$ meV, we obtain good agreement with the observed $A_{hu2}^0$ and $A_{eu2}^0$ lines, respectively, confirming their origin from the quasilinear exciton branch.

In addition, we observe previously unreported Umklapp lines ($A_{1u}^-$, $A_{2u}^-$) associated with the $A_1^-$ and $A_2^-$ exciton polaron. Unlike the $A^0$ exciton with two Umklapp lines, each exciton polaron shows only a single Umklapp line, indicating a single exciton–polaron branch. This behavior arises because exciton polarons in opposite valleys are dressed by distinct Fermi seas. Since their many-body dressing clouds are different, the corresponding exciton–polaron wavefunctions have zero overlap, preventing any coupling between polarons from opposite valleys. Consequently, exciton polarons experience only intravalley e–h exchange interactions. This leads to a single quasilinear dispersion branch with a linear component approximately half that of the $A^0$ exciton, as illustrated in Fig. 1C.

## Robustness of Wigner crystals at elevated density and temperature

Wigner crystals are known as delicate correlated states, which stabilize only at low carrier densities and temperatures[28-37]. Prior theory established that an ideal 2D electron system can only form a WC at low density with a Wigner–Seitz radius $r_s > 31$[18, 38]. Yet the Umklapp scattering signals in our data (Fig. 2C–D) persist up to a hole density of $1 \times 10^{12}$ cm$^{-2}$, corresponding to $r_s$ ~10, well below the theoretical threshold (see Supplementary Section I.6 for details). This apparent breakdown of the theoretical limit hints that lattice defects might help stabilize the WC formation in monolayer WSe$_2$[15-17].

To evaluate the WC robustness against thermal fluctuations, Fig. 2E displays the integrated intensity of the $A^0_{eu2}$, $A^0_{hu2}$ lines as a function of temperature. Both Umklapp lines diminish with increasing temperature and disappear at critical temperatures of approximately $T_c = 27$ K and 21 K on the electron and hole sides, respectively. Since our setup measures only the cold–finger temperature, the sample temperature should be several degrees higher, implying a WC melting temperature near 30 K. This value well exceeds the $T_c$ ~ 11 K reported for monolayer MoSe$_2$[13]. Both the low Wigner–Seitz radius $r_s$ and high melting temperature $T_c$ indicate monolayer WSe$_2$ as an excellent WC platform.

## Umklapp scattering due to exciton–WC scattering

Next, we investigate the physical mechanisms of the observed Umklapp scattering processes with WCs. To this end, we conduct helicity–resolved optical measurements under a vertical magnetic field $B = 17$ T that lifts the valley degeneracy (Fig. 3A–B). Optical helicity enables selective probing of the $K$ or $K'$ valley, while the injection of electrons or holes enables selective population of these valleys[39-42]. The interplay of these factors reveals four distinct exciton–WC scenarios, as illustrated in the right column of Fig. 3 at vertical positions corresponding to the map sections in Fig. 3A–B. In Case 1, both the exciton and WC reside in the $K'$ valley but occupy different bands. In Case 2, the exciton resides in the $K'$ valley, while a hole WC occupies the $K$ valley. In Case 3, the exciton resides in the $K$ valley, while an electron WC resides in the $K'$ valley. In Case 4, both the exciton and hole WC reside in the $K$ valley. A crucial distinction arises in Case 4, where the hole in the exciton resides in the same band as the hole WC, unlike in Cases 1–3 where excitons and WCs are associated with different bands.

Fig. 3A–B reveal a striking feature—the Umklapp line in Case 4 is much brighter than those in Cases 1–3. To understand this distinction, we need to know that an exciton, being charge–neutral,

has near–zero net direct Coulomb interaction with the WC, because its electron and hole components exhibit opposite scattering amplitudes that largely cancel each other out. Therefore, the exciton–WC scattering depends crucially on the exchange interactions, where Cases 1–3 and Case 4 exhibit distinct behavior. In Cases 1–3, the exciton and WC occupy different bands, leading to negligible exciton–WC exchange interaction; as a result, the overall exciton–WC scattering is weak (Fig. 3F). In Case 4, by contrast, the hole of the exciton occupies the same band as the hole WC, leading to a strong long–range exchange interaction between them due to Pauli exclusion, whereas the electron of the exciton occupies a different band with negligible exchange interaction with the hole WC. The uncompensated hole–WC and electron–WC exchange interaction leads to a strong net exciton–WC scattering effect.

To corroborate the picture above, we have calculated the exciton–WC scattering strength in all four cases, incorporating both long–range and short–range, direct and exchange Coulomb interactions. As shown in the left panel of Fig. 3E, the calculated scattering strength in Case 4 is one to two orders of magnitude greater than in Cases 1–3, consistent with our experimental observation in Fig. 3A–B. Using this exciton–WC scattering model, we further simulate an optical conductivity map at zero magnetic field for excitons in monolayer $WSe_2$ (Fig. 4A). Our simulation includes both the quadratic and quasilinear exciton branches, producing two Umklapp lines that match the observed $A_{hu1}^0$ and $A_{hu2}^0$ lines.

While our simulation in Fig. 4A accounts well for the exciton Umklapp lines on the hole side, it produces negligible Umklapp lines on the electron side because of the weak exciton–WC scattering in Cases 1 and 2 (Fig. 3E). Similarly weak optical response is expected for polaron–WC scattering, as it primarily involves the excitonic component of the polaron hopping across different WC sites. Yet our experiment observes not only a pronounced exciton Umklapp line ($A_{eu2}^0$) but also two exciton–polaron Umklapp lines ($A_{1u}^-, A_{2u}^-$) on the electron side. These deviations point to an alternative mechanism—likely involving polaron effects—that governs Umklapp scattering on the electron side.

## Umklapp scattering due to exciton–polaron effect

We construct a polaron model to elucidate the role of exciton–polaron coupling in Umklapp scattering (see Supplementary Section II.7 for details). For clarity, we illustrate our model on the hole–doped regime with only a single exciton polaron. The model includes five basis states: the primary exciton $|X_{s=0}^K\rangle_R$, the primary tetron $|\mathfrak{I}_{s=0}\rangle_R$ and its Umklapp counterpart $|\mathfrak{I}_{s=1}\rangle_R$, as well as two Umklapp excitons $|X_{s=1}^K\rangle_R$ and $|X_{s=1}^{K'}\rangle_R$. Here, a tetron refers to an exciton bound to a WC charge, forming a localized trion with an accompanying WC vacancy (Fig. 1A); unlike conventional tetrons defined by Fermi–sea dressing of an exciton, these tetrons arise from distortions of the WC[6, 43, 44]. All states are associated with the $K$ valley except $|X_{s=1}^{K'}\rangle_R$, which belongs to the $K'$ valley, and the subscript R denotes right–handed optical helicity. The index s =

0 labels the zeroth star in the group theory, corresponding to zero–momentum excitations, whereas $s = 1$ labels the first stars, corresponding to a linear combination of six Umklapp states at the first set of WC reciprocal lattice vectors. The presence of the WC enables coupling between the $s = 0$ and $s = 1$ states, facilitating polaron–mediated interactions.

Using these basis states, we construct the Hamiltonian:

$$\begin{pmatrix} E_{\mathbf{0}}^X & U_{X\mathfrak{Z}}^{00} & U_{X\mathfrak{Z}}^{01} & 0 & 0 \\ U_{X\mathfrak{Z}}^{00} & E_0^{\mathfrak{Z}} & 0 & U_{X\mathfrak{Z}}^{10} & 0 \\ U_{X\mathfrak{Z}}^{01} & 0 & E_1^{\mathfrak{Z}} + \frac{Jg_1}{K} & U_{X\mathfrak{Z}}^{11} & 0 \\ 0 & U_{X\mathfrak{Z}}^{10} & U_{X\mathfrak{Z}}^{11} & E_1^X + \frac{Jg_1}{K} & \frac{Jg_1}{K} \\ 0 & 0 & 0 & \frac{Jg_1}{K} & E_1^X + \frac{Jg_1}{K} \end{pmatrix} \begin{pmatrix} C_{0R}^{\mathrm{K}} \\ C_{0R}^{\mathfrak{Z}} \\ C_{1R}^{\mathfrak{Z}} \\ C_{1R}^{\mathrm{K}} \\ C_{1R}^{\mathrm{K}'} \end{pmatrix} = E_P \begin{pmatrix} C_{0R}^{\mathrm{K}} \\ C_{0R}^{\mathfrak{Z}} \\ C_{1R}^{\mathfrak{Z}} \\ C_{1R}^{\mathrm{K}} \\ C_{1R}^{\mathrm{K}'} \end{pmatrix} \tag{1}$$

Here, the $C$'s are the expansion coefficients and $E_P$ denotes the eigenenergies. The parameters $J$, $g_1$, and $K$ represent the e–h exchange interaction strength, the magnitude of the WC reciprocal lattice vector, and the momentum magnitude at the K point, respectively. The $Jg_1/K$ terms modulate and couple the two Umklapp excitons $|X_{s=1}^{\mathrm{K}}\rangle_R$ and $|X_{s=1}^{\mathrm{K}'}\rangle_R$, producing the two excitonic branches shown in Fig. 1B.

A key feature of Eq. (1) is the presence of off–diagonal terms $U_{X\mathfrak{Z}}$, which hybridize the exciton and tetron states. Among the basic states, only the primary exciton and tetron are optically bright, while the Umklapp states are all initially dark. However, solving Eq. (1) reveals new eigenstates formed as linear combinations of the basis functions, where the Umklapp states become optically active due to mixing with the primary states. This mixing is mediated by the polaron effect, driven by the $U_{X\mathfrak{Z}}$ coupling terms. An analogous Hamiltonian with seven basis states can be constructed for the electron–doped regime, where two tetrons are involved.

Fig. 4B displays the optical conductivity map calculated using this model, which includes the polaron effect but not the exciton–WC scattering effect. The theoretical map exhibits two Umklapp lines for each exciton and one Umklapp line for each polaron. A comparison of Fig. 4A, which includes exciton–WC scattering but no polaron effect, and Fig. 4B, which includes the polaron effect but no exciton–WC scattering, reveals complementary behaviors. While the $A_{hu1}^0$ and $A_{hu2}^0$ lines are strong in Fig. 4A, they are weak in Fig. 4B. This is because Case 4 exhibits strong exciton–WC scattering but no polaron effect, as the hole occupies the same band as the WC, whereas Cases 1–3 exhibit weak exciton–WC scattering but strong polaron effect because the exciton occupies different bands from the WC.

## Simulations with both exciton–WC scattering and exciton–polaron effects

After treating the exciton–WC scattering and exciton–polaron effects separately, we combine their contributions incoherently—assuming that lattice defects and valley disorder of the WC sites destroy any phase coherence between the two—to obtain the total optical conductivity map shown in Fig. 4C (see Supplementary Section II.8 for details). Afterward, we calculate the reflectance contrast spectra by solving the optical interference problem in our device geometry. We also broaden the spectral lines, add a background to account for states outside our energy range, and perform a second–order energy derivative to highlight the weak features. The resulting simulated $d^2(\Delta R/R)/dE^2$ map, as shown in Fig. 2F, shows good agreement with our experimental map in Fig. 2C.

We further calculated the Umklapp states under a magnetic field, which generates distinct Zeeman shifts on different Umklapp states by lifting the valley degeneracy (see Supplementary Section II.9 for details). Fig. 3C–D displays the simulated $d^2(\Delta R/R)/dE^2$ maps at B = 17 T, which agree reasonably with experiment. The simulation–experiment comparison allows us to identify the observed weaker lines in Fig. 3A–B as the $A^0_{eu2}$ and $A^0_{hu2}$ lines and the strong line as the $A^0_{hu1}$ line.

In summary, we uncover a previously inaccessible regime of Umklapp physics in 2D Wigner crystals (WCs), in which polaron effects play a central role. The emergence of multiple Umklapp features—encompassing both excitonic and polaronic excitations with unconventional dispersion—reveals a brightening mechanism whereby many–body interactions redistribute optical strength rather than simply providing momentum. Such behavior lies beyond a single–particle exciton description and underscores the importance of polaron formation in a crystalline electronic environment. Our results identify WC polarons as a distinct class of correlated quasiparticles. More broadly, these findings establish Umklapp processes as a powerful tool for probing and controlling finite–momentum excitations, with implications for interaction–driven phases, valley–selective effects, and nonequilibrium phenomena in 2D quantum materials.

During the preparation of our manuscript, we became aware of related works from other groups[45-48].

**Acknowledgment:** We thank S. A. McGill for assistance in the magneto–optical experiments and H. W. K. Tom for equipment support. C.H.L. acknowledges support from the National Science Foundation (NSF) Division of Materials Research CAREER Award No.1945660 and the American Chemical Society Petroleum Research Fund No. 61640–ND6. Y.C.C. is supported by the National Science and Technology Council (Taiwan) under grant Nos. NSTC 112–2112–M–001–054–MY2 and 114–2112–M–006–030. E.L. acknowledges support from the National Key R&D Program of China (grant no. 2024YFA1410500), the Fundamental Research Funds for the Central Universities, the National Natural Science Foundation of China (grant no. 12374456), the Program for Innovative Talents and Entrepreneur in Jiangsu, and the Jiangsu Provincial Department of Science and Technology (Grant No. BK20253040). Optical spectroscopy at SLAC/Stanford was supported  by the Department of Energy, Office of Basic Energy Sciences, Division of Materials Sciences and Engineering under contract DE–AC02–76SF00515 (T.F.H., J. H., A. M. and A. Z.), with additional support for T.F.H. from the Gordon and Betty Moore Foundation EPiQS Initiative through grant no. GBMF9462. J.H. acknowledges support from an NTT Research Fellowship, and A. Z. acknowledges support from the National Science Foundation Graduate Research Fellowship Program under grant no. DGE–2146755. K.W. and T.T. acknowledge support from the JSPS KAKENHI (Grant Numbers 21H05233 and 23H02052), the CREST (JPMJCR24A5), JST and World Premier International Research Center Initiative (WPI), MEXT, Japan. A portion of this work was performed at the National High Magnetic Field Laboratory, which is supported by the National Science Foundation Cooperative Agreement No. DMR–1644779 and the State of Florida.

**Author contributions:** E.L. fabricated the devices. E.L., M.W., J.H., A.Z., A.M., T.O. and A.S. performed the experiments. M.W. and E.L. analyzed the data. T.T. and K.W. provided boron nitride crystals for device fabrication. T.F.H. supervised the research of J.H., A.Z., and A.M.. Y.C.C. performed the theoretical calculations. C.H.L. supervised the research and coordinated the work. C.H.L., Y.C.C. and M.W. wrote the manuscript.

**Competing interests:**  The authors declare no competing interests.

**Additional information**

**Supplementary information** is available for this paper.

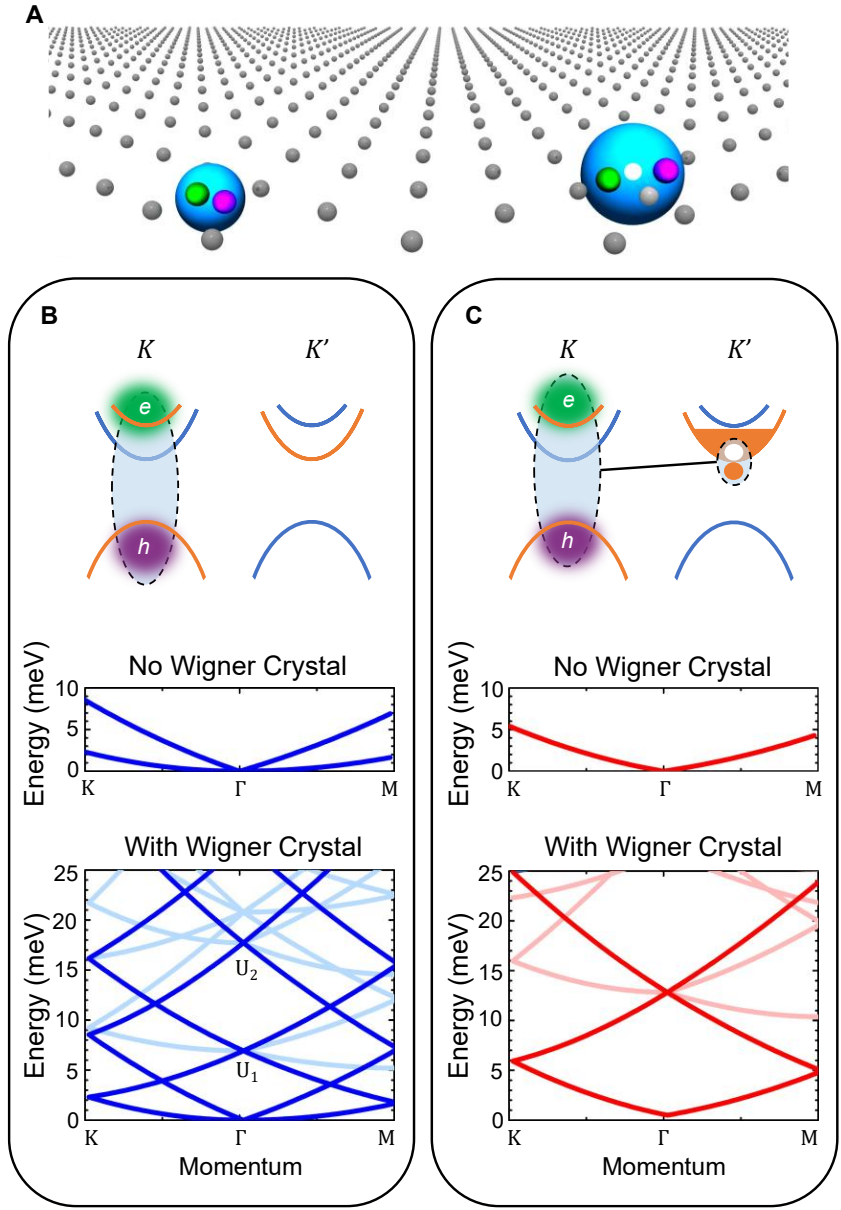

**Figure 1. Umklapp brightening of finite–momentum excitons and exciton polarons by a Wigner crystal.** (A) Schematic illustration of an exciton and an exciton polaron in a Wigner crystal (WC). The polaron consists of an exciton bound to a WC charge, accompanied by a corresponding WC vacancy. (B) Schematic illustration and dispersion of excitons in monolayer WSe₂. Excitonic states from opposite valleys hybridize to form two branches with quadratic and quasilinear dispersions as functions of the center–of–mass (CM) wave vector. The WC induces a zone–folding effect. (C) Schematic illustration and dispersion of the exciton polaron. Because exciton polarons in opposite valleys do not couple owing to their distinct Fermi seas, they exhibit a single branch with quasilinear dispersion.

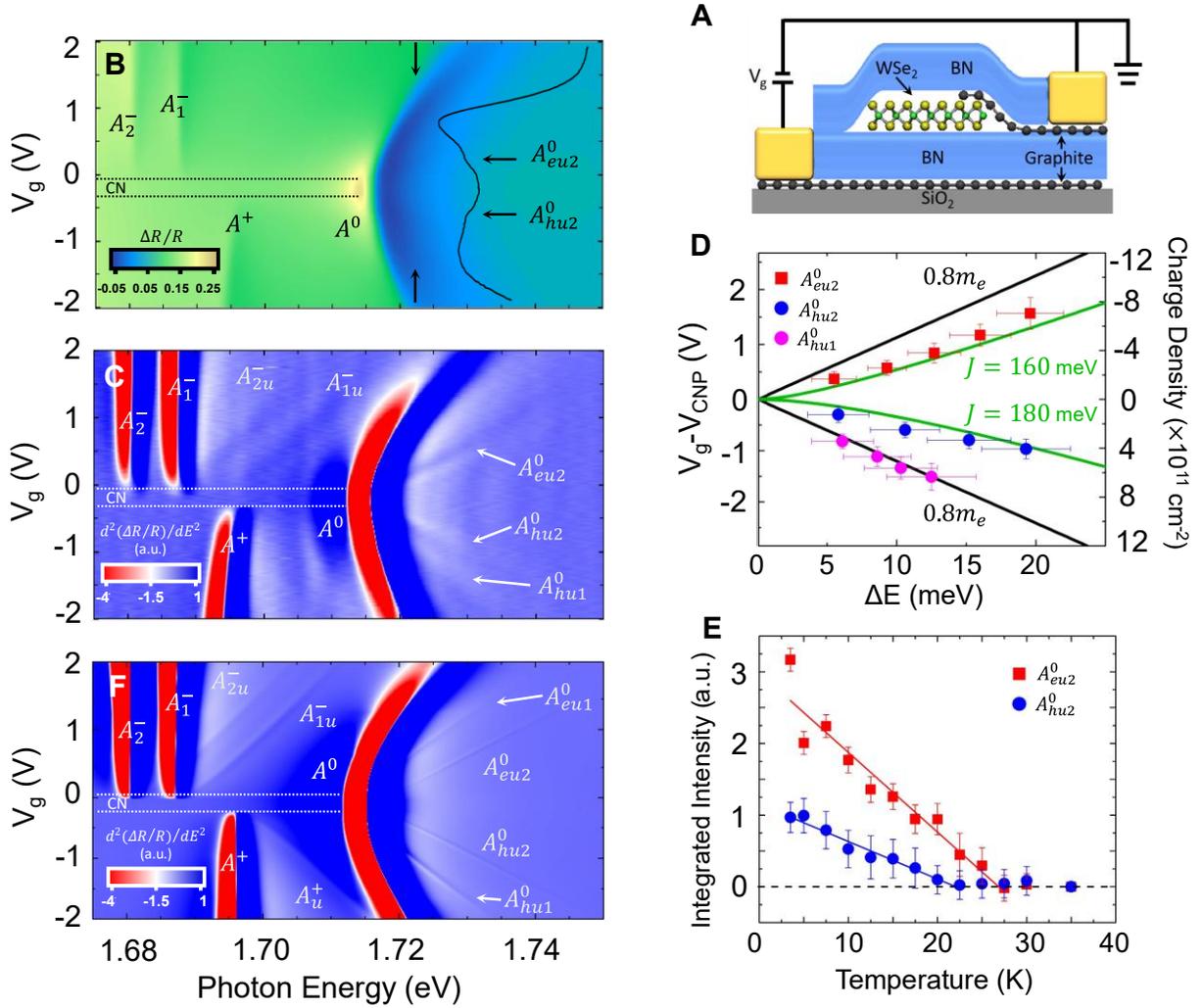

**Figure 2. Excitonic Umklapp scattering in electron and hole Wigner crystals in monolayer WSe₂.** (A) Schematic of a monolayer WSe₂ device. (B) Gate–dependent reflectance contrast ($\Delta R/R$) color map measured at $T$ = 3.5 K, showing the $A^0$ exciton and $A^+$, $A_1^-$, $A_2^-$ exciton polarons. The black trace denotes a gate–dependent profile at 1.722 eV (indicated by arrows), where two weak minima ($A_{eu2}^0$, $A_{hu2}^0$) mark the formation of electron and hole Wigner crystals. (C) Second–order energy derivative of the reflectance contrast map in panel B, $d^2(\Delta R/R)/dE^2$, highlighting multiple Umklapp spectral features. (D) Energy separation $\Delta$E between the $A^0$ resonance and the $A_{eu2}^0$, $A_{hu1}^0$, and $A_{hu2}^0$ Umklapp features. The black (green) lines are the theoretical Umklapp positions based on the quadratic (quasilinear) exciton dispersion. (E) Temperature dependence of the Umklapp scattering signal intensity. (F) Simulated second–derivative reflectance contrast map.

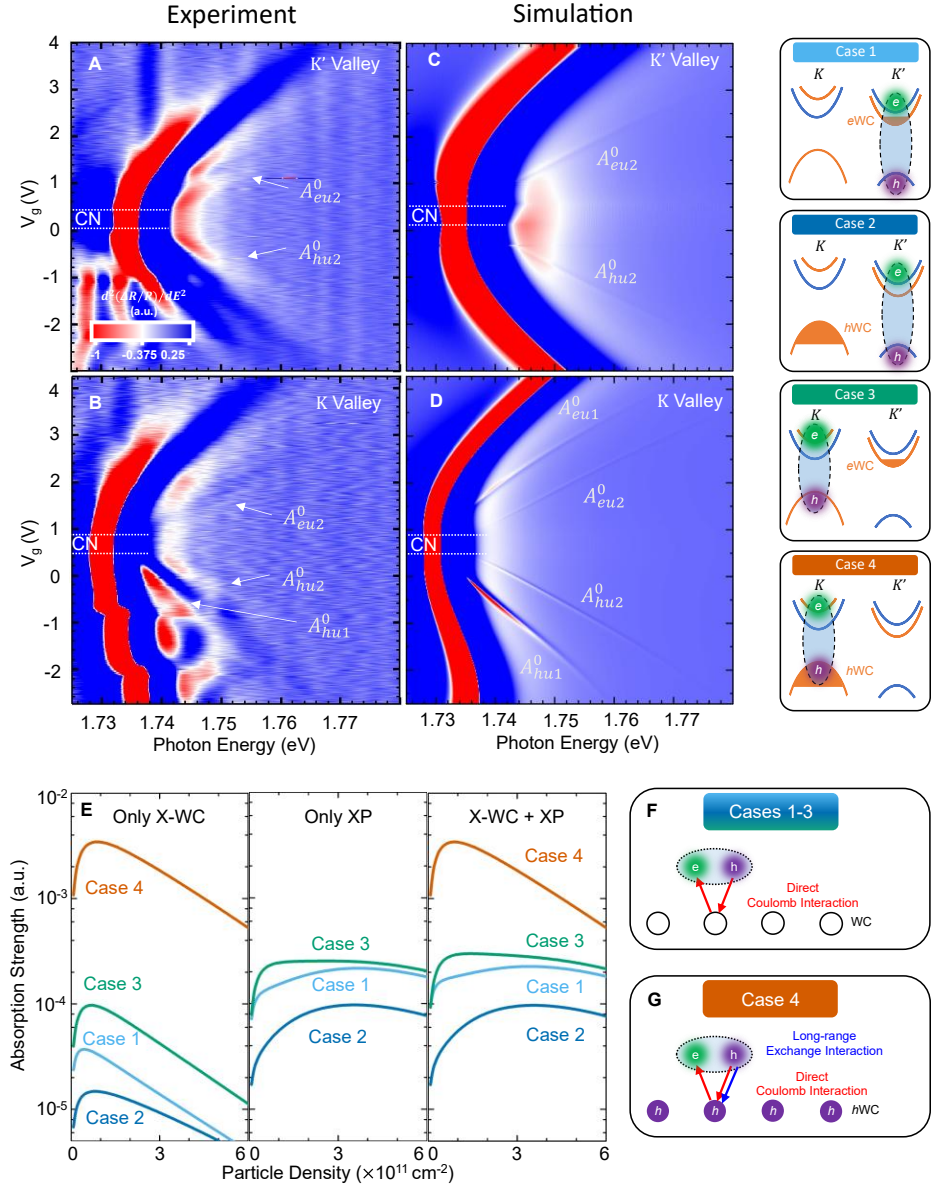

**Figure 3. Valley dependence of Umklapp scattering. (A)** Second–derivative reflectance contrast map, $d^2(\Delta R/R)/dE^2$, measured with left–handed circularly polarized light at a magnetic field of $B = 17$ T, probing optical transitions in the K′ valley. **(B)** Same as panel A, but measured with right–handed circularly polarized light, probing the K valley. **(C, D)** Corresponding simulated spectra for panels A and B, respectively. The simulation neglects the Landau quantization effect. The schematics on the right show the valley configurations of the exciton and the Wigner crystal, defining Cases 1–4 considered in the model. They are arranged vertically to correspond to the map sections in panels A and B. **(E)** Calculated oscillator strength of the Umklapp scattering signal for Cases 1–4 with only exciton–WC scattering (left), only the exciton–polaron effect (middle), and both effects included (right). **(F)** Schematic of exciton–WC scattering processes for Cases 1–3, in which exchange interactions are weak and direct Coulomb interactions cancel. **(G)** Schematic of exciton–WC scattering for Case 4, illustrating that an imbalanced exchange interaction leads to strong net scattering.

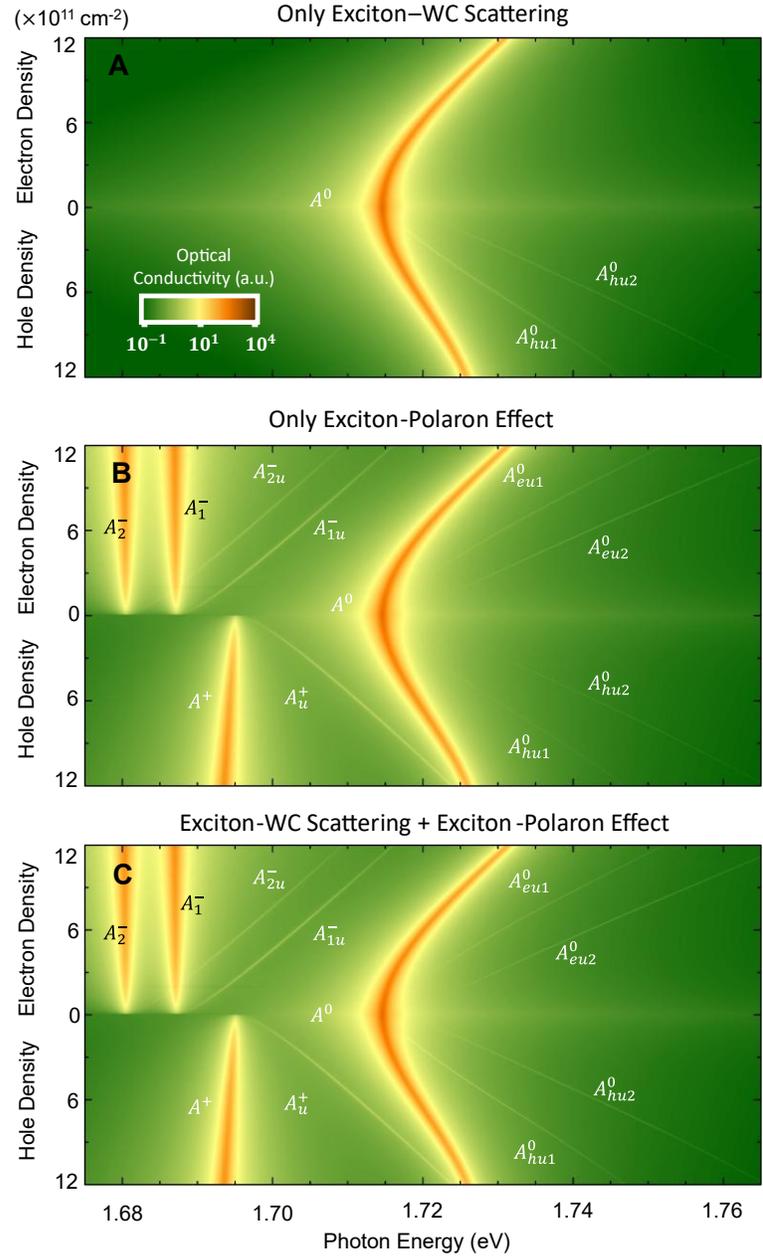

**Figure 4. Simulations of excitonic optical conductivity in monolayer WSe₂ Wigner crystals.**
**(A)** Simulation including only exciton–WC scattering, showing Umklapp features arising from direct exciton–WC coupling. **(B)** Simulation including only exciton–polaron effects, showing the Umklapp optical response due to many–body dressing in the absence of exciton–WC scattering. **(C)** Simulation incorporating both exciton–WC scattering and exciton–polaron effects, reproducing the emergence of multiple Umklapp branches.

**Supplementary Information for**
**"Exciton-Polaron Umklapp Scattering in Wigner Crystals"**

Erfu Liu[1,2†], Matthew Wilson[1†], Jenny Hu[3,4], Alexandra Zimmerman[3,4], Amal Mathew[3,4], Tianyi Ouyang[1], Ao Shi[1], Takashi Taniguchi[5], Kenji Watanabe[6], Tony F. Heinz[3,4], Yia-Chung Chang[7,8*] Chun Hung Lui[1*]

[1]Department of Physics and Astronomy, University of California, Riverside, CA 92521, USA.

[2]School of Physics, National Laboratory of Solid State Microstructures, and Collaborative Innovation Center for Advanced Microstructures, Nanjing University, Nanjing, China.

[3]Department of Applied Physics, Stanford University, Stanford, CA 94305, USA

[4]SLAC National Accelerator Laboratory, Menlo Park, CA 94025, USA

[5]Research Center for Materials Nanoarchitectonics, National Institute for Materials Science, 1-1 Namiki, Tsukuba 305-0044, Japan

[6]Research Center for Electronic and Optical Materials, National Institute for Materials Science, 1-1 Namiki, Tsukuba 305-0044, Japan

[7]Research Center for Applied Sciences, Academia Sinica, Taipei 11529, Taiwan

[8]Department of Physics, National Cheng Kung University, Tainan, 701 Taiwan

[†] These authors contributed equally to the work.

*Corresponding author. Email: yiachang@gate.sinica.edu.tw; joshua.lui@ucr.edu

# Table of Contents

# I. EXPERIMENTAL INFORMATION

## 1. Device fabrication

We fabricate monolayer $WSe_2$ devices with hexagonal boron nitride (BN) encapsulation by the standard mechanical co-lamination of 2D crystals. We use $WSe_2$ bulk crystals from HQ Graphene. We first exfoliate monolayer $WSe_2$, multilayer graphene, and thin BN flakes from their bulk crystals onto $Si/SiO_2$ substrates. Afterward, we apply a polycarbonate-based dry-transfer technique to stack these different 2D crystals together. We use Polydymethylsiloxate (PDMS) from Gel-Pak and Polycarbonate (PC) from Sigma Aldrich to fabricate the polymer stamp. We use the PDMS/PC stamp to first pick up a BN flake, and sequentially pick up multilayer graphene (as the electrodes), monolayer $WSe_2$, thin BN (as the bottom gate dielectric), and multilayer graphene (as the back-gate electrode). This method ensures that the $WSe_2$ monolayer does not come into contact with the polymer during the fabrication process, reducing contamination and bubbles at the interface. We subsequently apply electron-beam lithography techniques to pattern and deposit the gold contacts (100-nm thickness). Finally, we anneal the devices at temperature $T = 300\ °C$ for five hours in an argon environment to improve the interface quality. The BN thickness is measured by atomic force microscopy. The thicknesses of the bottom BN dielectric layer are all $42 \pm 1$ nm for Devices 1 – 3.

## 2. Reflectance contrast experiments

The reflectance contrast measurements for Fig. 2 were conducted with a closed-cycle cryostat (Montana Instruments) and a spectrometer (HRS-500-MS Princeton Instruments) at UC Riverside. The reported temperature in Fig. 2E is the cold-finger temperature, whereas the actual sample temperature should be a few kelvin higher. The experiment for Fig. 3 was conducted with a magneto-optical system with a B =17 T DC superconducting magnet (SCM-3) at the National High Magnetic Field Laboratory in Tallahassee, Florida, USA. The sample temperature is $T \approx 5$ K; the spectrometer is an IsoPlane 320 from Princeton Instruments. In all measurements, we focus broadband white light onto the sample with a spot diameter of about 2 μm using an objective lens (numerical aperture 0.6). The reflected light is collected by the same objective and analyzed by a spectrometer with a charge-coupled-device (CCD) camera. We measure a reflection spectrum on the sample ($R_s$) and a reference reflection spectrum on a nearby area without $WSe_2$ ($R_r$), and obtain the reflectance contrast as $\Delta R/R = (R_s - R_r)/R_r$. We further perform second-order differentiation $\Delta R/R$ with respect to the photon energy to reveal the weak Umklapp features. We simply relate gating voltage to charge density using landau level degeneracy ($\Delta V = n = eB/h$) with data taken from Figs. 4A-B, Figs. S3A and C, and Figs. S3B and D for Devices 1-3 respectively.

## 3. Identification of optical oscillator energies and extraction of background signal

Our measured reflectance contrast ($\Delta R/R$) spectra combine the real and imaginary parts of the conductivity of monolayer WSe$_2$, influenced by optical interference in the BN/WSe$_2$/BN/graphite/SiO$_2$/Si heterostructure. For the strong signals of optical oscillator states ($A^0$, $A_{1,2}^-$, $A^+$) we extract the real and imaginary part of the conductivity by solving the optical problem of the stacked system using the transfer matrix method in linear optics. Using an array of 20 complex Lorentzian functions initially evenly spaced along energy, we fit smoothed line cuts of $\Delta R/R$ along gate voltage using variational analysis. This approach is similar to the Kramers–Kronig constrained variational analysis developed by A.B. Kuzmenko *et al* (*S1*) and adapted by J. van Barren *et al* (*S2*). From this analysis, we simultaneously extract Energy and conductivity details about the strong oscillator states, as well as fitting an experimental charge-density dependent complex conductivity background. We add complex conductivities for theoretically calculated oscillator states onto the experimental background and convert to $\Delta R/R$ for the theory map in Fig. 2f to visually match the experimentally collected $\Delta R/R$ map in (Fig. 2c).

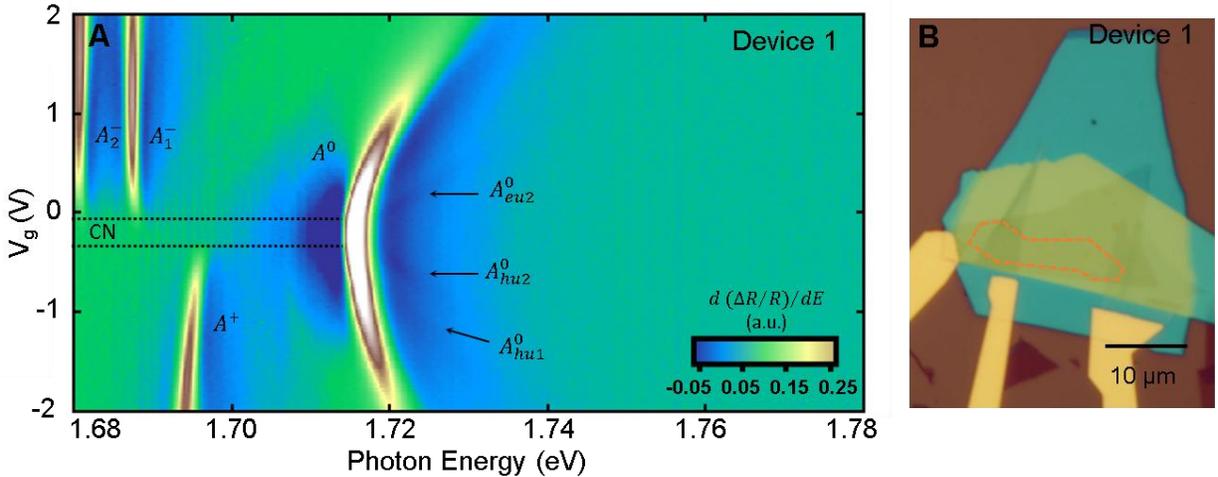

**Figure S1**. (A) First-order energy derivative of reflectance contrast data ($d$ ($\Delta R/R$)$/dE$) of Fig. 2B for Device 1. Neutral exciton Umklapp line signals ($A_{hu1}^0$, $A_{hu2}^0$, $A_{eu2}^0$) become clear, but polaron Umklapp signals ($A_{1u}^-$, $A_{2u}^-$) cannot be resolved. Black dotted lines indicate the voltages that constitute the charge neutrality (CN) gap. (B) Optical image of Device 1. Orange dotted line outlines the monolayer WSe$_2$ layer.

## 4. Comparison of data for three different devices

### 4.1. Results of Device 1

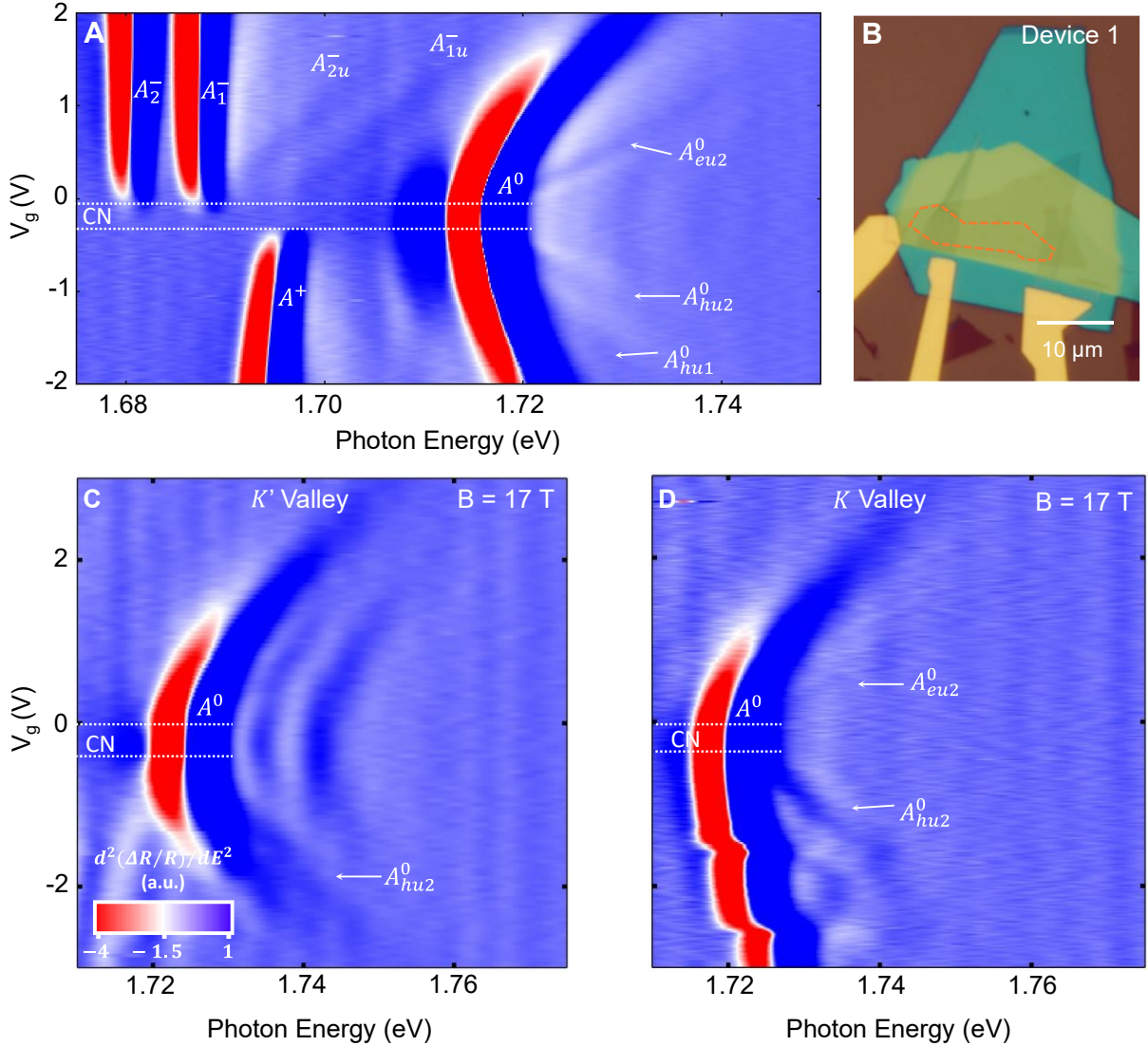

**Figure S2.** (A) Gate-dependent second-derivative reflectance-contrast map, $d^2(\Delta R/R)/dE^2$, of Device 1 (same data as Fig. 2C). Exciton Umklapp lines ($A_{hu1}^0$, $A_{hu2}^0$, $A_{eu2}^0$) and exciton-polaron Umklapp lines ($A_{1u}^-$, $A_{2u}^-$) are indicated. The region between the two white dotted lines marks the charge-neutrality (CN) regime. $T = 3.5$ K. (B) Optical image of Device 1; the orange dotted line outlines the WSe$_2$ monolayer. Scale bar, 10 μm. (C, D) Helicity-resolved $d^2(\Delta R/R)/dE^2$ maps of Device 1 measured under a magnetic field of $B = 17.5$ T, showing the optical response of the K′ and K valleys, respectively. $T = 4$ K. Panels A, C, and D share the same color scale shown in panel C.

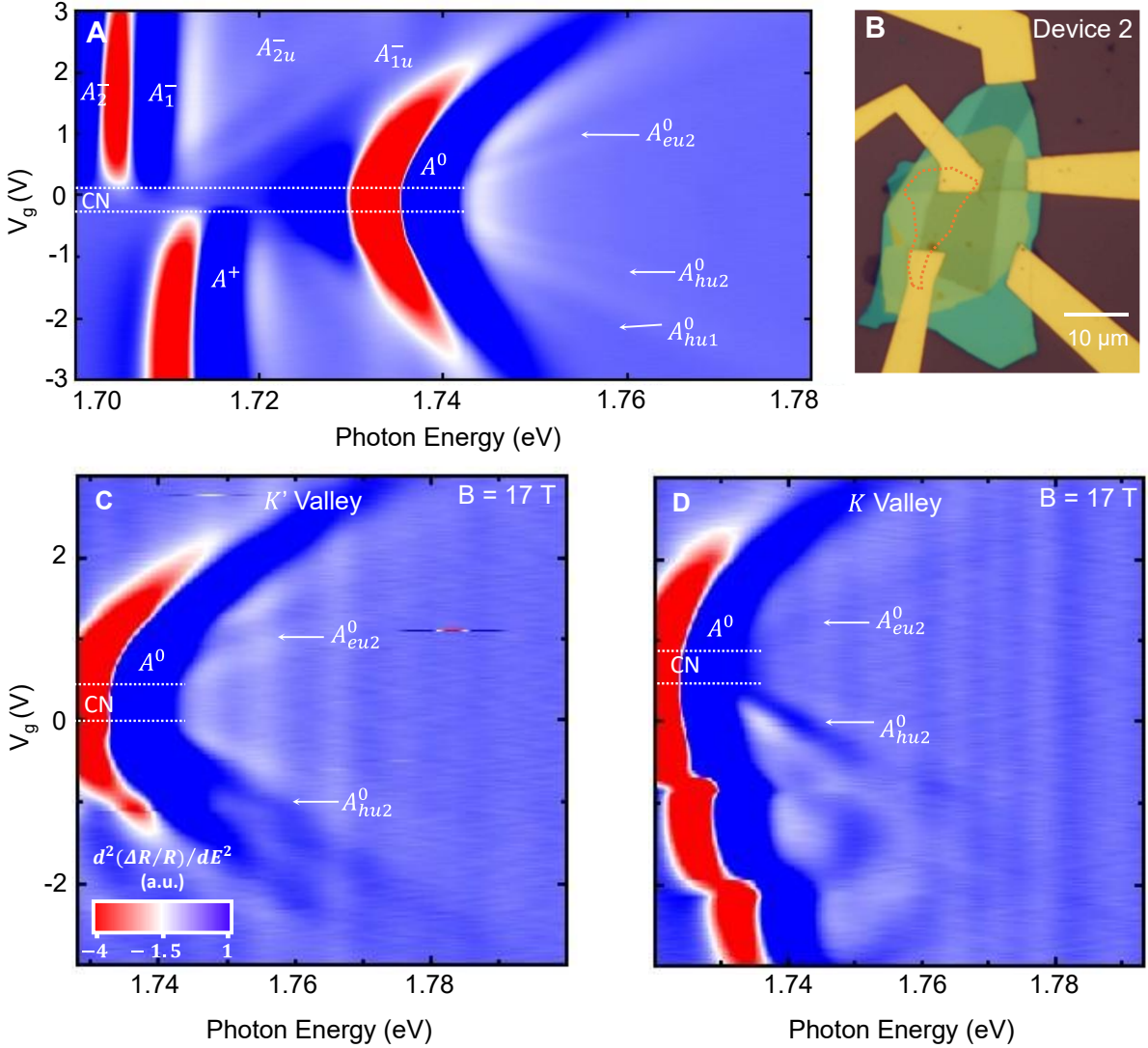

**Figure S3.** (A) Gate-dependent second-derivative reflectance-contrast map, $d^2(\Delta R/R)/dE^2$, of Device 2. Exciton Umklapp lines ($A^0_{hu1}$, $A^0_{hu2}$, $A^0_{eu2}$) and exciton-polaron Umklapp lines ($A^-_{1u}$, $A^-_{2u}$) are indicated. The region between the two white dotted lines marks the charge-neutrality (CN) regime. $T = 3.5$ K. (B) Optical image of Device 2; the orange dotted line outlines the WSe$_2$ monolayer. Scale bar, 10 μm. (C, D) Helicity-resolved $d^2(\Delta R/R)/dE^2$ maps of Device 2 measured under a magnetic field of $B = 17.5$ T, showing the optical response of the K′ and K valleys, respectively. $T = 4$ K. Panels A, C, and D share the same color scale shown in panel C.

### 4.3. Results of Device 3

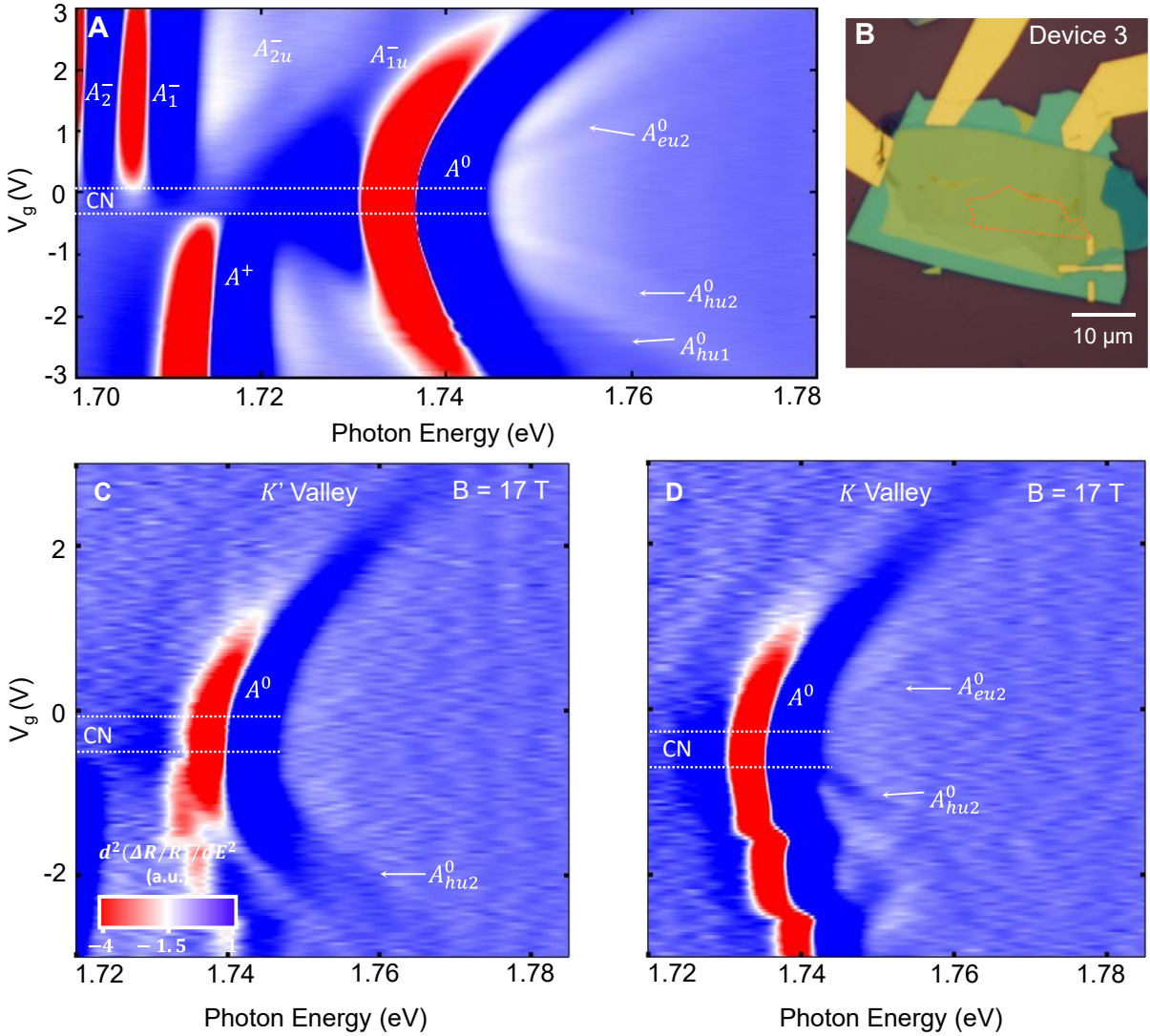

**Figure S4.** (A) Gate-dependent $d^2(\Delta R/R)/dE^2$ map of Device 3. $T = 3.5$ K. (B) Optical image of Device 3; the orange dotted line outlines the WSe$_2$ monolayer. Scale bar, 10 μm. (C, D) Helicity-resolved $d^2(\Delta R/R)/dE^2$ maps of Device 3 measured under a magnetic field of $B = 17.5$ T, showing the optical response of the K′ and K valleys, respectively. $T = 4$ K. Panels A, C, and D share the same color scale shown in panel C.

### 5. Temperature-Dependent Data Maps

Figure 2E in the main text shows the temperature dependence of the intensity of two Umklapp lines. Here Figure S5 presents the $d^2(\Delta R/R)/dE^2$ maps of Device 1 at selected temperatures, illustrating the melting of the Wigner crystals.

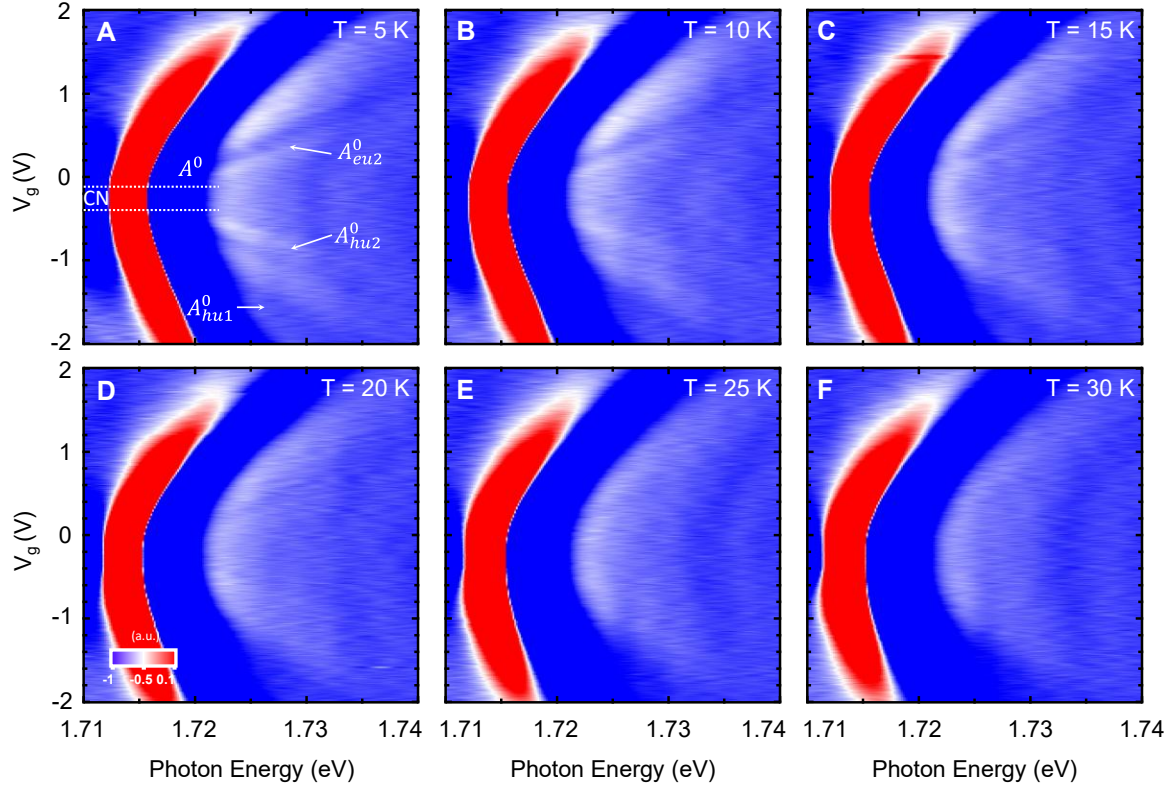

**Figure S5**. (A-F) Gate-dependent $d^2(\Delta R/R)/dE^2$ maps of Device 1 at $T = 5$, 10, 15, 20, 25, and 30 K.

## 6. Estimation of the Wigner–Seitz radius

The Wigner–Seitz radius $r_s$ measures the average interparticle separation in units of the effective Bohr radius and characterizes the ratio of Coulomb interaction energy to kinetic energy. It is given by $r_s = m^* e^2 / (4\pi\varepsilon\varepsilon_0\hbar^2\sqrt{\pi N})$, where $N$ is the carrier density, $m^*$ the effective mass, $\varepsilon$ the background dielectric constant, $\varepsilon_0$ the vacuum permittivity, $e$ the elementary charge, and $\hbar$ the reduced Planck constant (*S3-5*). In our estimation, we take $m^* \approx 0.4m_0$ ($m_0$ is the free electron mass) for both electrons and holes in monolayer WSe₂. In our device geometry, the surrounding dielectric medium BN yields an effective dielectric constant $\varepsilon = \sqrt{\varepsilon_\parallel \varepsilon_\perp}$, with in-plane and out-of-plane dielectric constants $\varepsilon_\perp = 3.04$ and $\varepsilon_\parallel = 6.93$, respectively (*S6*). Experimentally, the Umklapp signals persist up to a hole density of $N = 1 \times 10^{12}$ cm⁻²; this corresponds to a Wigner–Seitz radius of $r_s \approx 10$. This value is well below the theoretical crystallization threshold $r_s > 31$ in two dimensions (*S3, 4*). The apparent breakdown of this theoretical limit suggests that lattice defects may assist in stabilizing Wigner-crystal formation in monolayer WSe₂.

## II. THEORETICAL SIMULATION

### 1. Theory of excitons in monolayer WSe₂

The exciton states in monolayer WSe₂ can be reasonably described by an effective-mass model (*S6*). For a WSe₂ monolayer encapsulated by boron nitride (BN), the effective electron-electron interaction potential can be well approximated by the Keldysh potential with the following expression in the reciprocal space (*S7*):

$$\tilde{v}_d(q) = \frac{e^2}{2A\kappa\varepsilon_0 q(1+q\rho_0)} \tag{1}$$

In this paper, we consistently use the "~" hat to indicate that the function is in the reciprocal space. In the long-wavelength limit ($q \to 0$), the Keldysh potential has the following expression in the real space (*S7*):

$$v_d(r) = \frac{e^2}{2\kappa\varepsilon_0\rho_0}\left[H_0\left(\frac{r}{\rho_0}\right) - N_0\left(\frac{r}{\rho_0}\right)\right] \tag{2}$$

Here $A$ is the sample area; $\kappa$ is the average static dielectric constant of the BN-encapsulated sample; $H_0$ and $N_0$ denote the Struve and Neuman functions of zeroth order, respectively. According to prior studies, the average dielectric constant is $\kappa \sim 4$ in the long-wavelength limit for the Coulomb interaction in monolayer WSe₂ encapsulated by hexagonal boron nitride (BN) (*S6*).

The bright exciton in monolayer WSe₂ consists of a hole in the spin-up (spin-down) valence band and an electron in the spin-up (spin-down) conduction band in the K (K′) valley. The K-valley bright exciton ground state, with zero center-of-mass (CM) momentum, can be expressed as:

$$\left|X_0^K\right\rangle = \sum_{\mathbf{k}} \tilde{\varphi}_0(\mathbf{k})\left|u_{c\uparrow,\mathbf{K+k}}; u_{v\uparrow,\mathbf{K+k}}\right\rangle \tag{3}$$

Here $u_{c\uparrow,\mathbf{K+k}}$ ($u_{v\uparrow,\mathbf{K+k}}$) is the Bloch state with wave vector $\mathbf{k}$ in the spin-up ($\uparrow$) conduction ($c$) (valence, $v$) band in the K valley. The $\mathbf{k}$ vector is measured from the K point. Similarly, the K′-valley exciton ground state is expressed as:

$$\left|X_0^{K'}\right\rangle = \sum_{\mathbf{k}} \tilde{\varphi}_0(\mathbf{k})\left|u_{c\downarrow,\mathbf{K'+k}}; u_{v\downarrow,\mathbf{K'+k}}\right\rangle \tag{4}$$

The effective-mass Hamiltonian of either the KK or K′K′ exciton reads:

$$H_X(\mathbf{r}, \mathbf{R}) = H_X^{CM}(\mathbf{R}) + H_X^{in}(\mathbf{r}), \tag{5}$$

Here

$$H_X^{CM}(\mathbf{R}) = -\frac{\hbar^2 \nabla_{\mathbf{R}}^2}{2m_X} \qquad (6)$$

denotes the Hamiltonian for the exciton center-of-mass (CM) motion with the CM coordinate $\mathbf{R} = (m_e \mathbf{r}_e + m_h \mathbf{r}_h)/m_X \equiv (\gamma_e \mathbf{r}_e + \gamma_h \mathbf{r}_h)$, where $m_X = m_h + m_e$ is the total exciton mass and $\gamma_e = \frac{m_e}{m_X}$ and $\gamma_h = \frac{m_h}{m_X}$. The envelope function for the CM motion with momentum $\mathbf{Q}$ is given by the plane wave

$$\frac{1}{\sqrt{A}} e^{i\mathbf{Q}\cdot\mathbf{R}} = \frac{1}{\sqrt{A}} e^{i\mathbf{Q}\cdot(\gamma_e \mathbf{r}_e + \gamma_h \mathbf{r}_h)} \qquad (7)$$

Here $A$ is the sample area. The total exciton states, including both the internal and CM motion, become:

$$\begin{cases} \left| X_{\mathbf{Q}}^{K} \right\rangle = \sum_{\mathbf{k}} \tilde{\varphi}_0(\mathbf{k}) \left| u_{c\uparrow, \mathbf{K} + \gamma_e \mathbf{Q} + \mathbf{k}}; u_{v\uparrow, \mathbf{K} - \gamma_h \mathbf{Q} + \mathbf{k}} \right\rangle \\ \left| X_{\mathbf{Q}}^{K'} \right\rangle = \sum_{\mathbf{k}} \tilde{\varphi}_0(\mathbf{k}) \left| u_{c\uparrow, \mathbf{K}' + \gamma_e \mathbf{Q} + \mathbf{k}}; u_{v\uparrow, \mathbf{K}' - \gamma_h \mathbf{Q} + \mathbf{k}} \right\rangle \end{cases} \qquad (8)$$

On the other hand,

$$H_X^{in}(\mathbf{r}) = -\frac{\hbar^2 \nabla_{\mathbf{r}}^2}{2\mu} - v_d(\mathbf{r}) + \Delta_X \delta(\mathbf{r}) \qquad (9)$$

is the Hamiltonian for the internal (*in*) motion of the exciton ($X$). Here $\mu$ is the reduced mass of the exciton. $\mathbf{r} = \mathbf{r}_e - \mathbf{r}_h$ is the relative coordinate between the electron and the hole. $-v_d(\mathbf{r})$ denotes the direct electron-hole interaction in monolayer WSe$_2$ as given in Eq. (2); The term $\Delta_X \delta(\mathbf{r})$ is caused by the electron-hole exchange (EHX) interaction (*S8, 9*); $\Delta_X$ is a constant evaluated at the valley extrema (*S6*). The electron and hole effective masses in the bright KK or K'K' exciton of monolayer WSe$_2$ are taken to be $m_e = 0.37 m_0$ and $m_h = 0.44\, m_0$ (*S10*), respectively, where $m_0$ is the free electron mass. The corresponding exciton reduced mass is $\mu = 0.2 m_0$ that is consistent with the value adopted in Ref. (*S6*). While the bright excitons involve the upper conduction band, the electron Wigner crystals are associated with the lower conduction band, whose effective mass is close to the hole effective mass. Thus, for both electron and hole Wigner crystals, the effective masses are taken to be the same, *i.e.* $m^* = 0.44\, m_0$.

The exciton eigenstates with zero angular momentum can be calculated via the Rayleigh-Ritz variational method (*S11*) by choosing a set of Gaussian functions as the basis for expansion:

$$f_n(\mathbf{r}) = e^{-\alpha_n r^2} \qquad (10)$$

With this basis, the exciton ground state in the K or K' valley can be written as

$$\varphi_0(\mathbf{r}) = \sum_n C_n \, e^{-\alpha_n \, r^2} \tag{11}$$

which is the same for the KK and K'K' excitons. $C_n$ are the expansion coefficients determined by variational principle. The number of Gaussian functions with different exponents $\alpha_n$ is chosen to be $N = 20$. Higher $N$ is not needed as it will only improve the ground state energy by less than 0.1%.

## 2. Charge distribution of Wigner crystal in monolayer WSe₂

When carriers are injected into monolayer WSe₂, they can form a Wigner crystal (WC) in the low-density regime. The Wigner crystal is a solid phase of electrons, in which the charges form a 2D lattice. Let's consider a Wigner crystal with $N_{wc}$ sites (hereafter we use the subscript "$wc$" to denote association with a Wigner crystal). Each site is labeled by an index $j = 1, 2 \dots N_{wc}$ and the corresponding lattice vectors are denoted by $\mathbf{R}_j$.

We model the local electron density distribution of the Wigner crystal at the $j$-th site by a normalized Gaussian function

$$\eta(\mathbf{r} - \mathbf{R}_j) = \frac{\beta}{\pi} e^{-\beta|\mathbf{r} - \mathbf{R}_j|^2} \tag{12}$$

Here $\beta$ is a variational parameter, which is related to the standard deviation (s) of the distribution as $\beta = \frac{1}{2s^2}$. The corresponding trial wave function of the localized electron is

$$\psi_{wc}(\mathbf{r} - \mathbf{R}_j) = \sqrt{\beta/\pi} \, e^{-\beta|\mathbf{r} - \mathbf{R}_j|^2/2} \tag{13}$$

We represent the Wigner crystal by a 2D array of such localized charge distribution. The total density can be expanded in terms of the normalized plane-wave basis functions $\frac{1}{\sqrt{A}} e^{i\mathbf{g} \cdot \mathbf{r}}$ as

$$\eta_{tot}(\mathbf{r}) = \sum_j \eta(\mathbf{r} - \mathbf{R}_j) = \frac{1}{\sqrt{A}} \sum_{\mathbf{g}} \tilde{\eta}(\mathbf{g}) \sum_j e^{i\mathbf{g} \cdot (\mathbf{r} - \mathbf{R}_j)} = \frac{N_{wc}}{\sqrt{A}} \sum_{\mathbf{g}} \tilde{\eta}(\mathbf{g}) \, e^{i\mathbf{g} \cdot \mathbf{r}} \tag{14}$$

Here $\mathbf{g}$ denotes the reciprocal lattice vectors of the Wigner crystal and

$$\tilde{\eta}(\mathbf{g}) = \frac{1}{\sqrt{A}} \frac{\beta}{\pi} \int e^{-i\mathbf{g} \cdot \mathbf{r}} \, e^{-\beta r^2} = \frac{1}{\sqrt{A}} e^{-\mathbf{g}^2/(4\beta)} \tag{15}$$

The Wigner crystal produces a Coulomb potential energy on an electron at $\mathbf{r}_e$ as:

$$V_{wc}(\mathbf{r}_e) = \sum_j \int d\,\mathbf{r}' v_d \,(\mathbf{r}_e - \mathbf{r}')\eta(\mathbf{r}' - \mathbf{R}_j) \tag{16}$$

Here $v_d(\mathbf{r})$, given by Eq. (2), denotes the effective electron-electron interaction potential energy; The Fourier transform of $V_{wc}(\mathbf{r}_e)$ gives

$$\tilde{V}_{wc}(\mathbf{q}) = \frac{1}{\sqrt{A}}\sum_j e^{i\mathbf{q}\cdot\mathbf{R}_j}\,\tilde{v}_d(\mathbf{q})\tilde{\eta}(\mathbf{q}),\tag{17}$$

where $\tilde{v}_d(\mathbf{q})$ is given by Eq. (1) and $\tilde{\eta}(\mathbf{q})$ is the Fourier transform of $\eta(\mathbf{r})$.

We can determine the variational parameter $\beta$ by minimizing the total energy of the Wigner crystal. If we neglect the inter-site overlapping integrals, the kinetic energy per site (within the effective-mass approximation and in atomic units) is given by

$$K(\beta) = -\frac{\hbar^2}{2m^*}\int d\,\mathbf{r}\,\psi_{wc}(\mathbf{r})\nabla^2\,\psi_{wc}(\mathbf{r}) = \frac{\hbar^2}{2m^*}\beta,\tag{18}$$

where $m^*$ is the effective mass of the Wigner-crystal particle. After considering the Coulomb interaction, the total energy of a Wigner crystal with $N_{wc}$ particles is given by

$$E_{tot}(\beta) = N_{wc}\left\{\frac{\hbar^2}{2m^*}\beta + \frac{1}{2}\int d\,\mathbf{r}\left[\eta(\mathbf{r}-\mathbf{R}_1)-\eta_0\right]\int d\,\mathbf{r}'v_d\,(\mathbf{r}-\mathbf{r}')\sum_{j\neq1}[\eta(\mathbf{r}'-\mathbf{R}_j)-\eta_0]\right\}$$

$$= N_{wc}\left\{\frac{\hbar^2}{2m^*}\beta + \frac{1}{2}\sum_{\mathbf{q}}\tilde{\eta}(\mathbf{q})\tilde{\eta}(\mathbf{q})\tilde{v}_d(\mathbf{q})\sum_{j\neq1}e^{i\mathbf{q}\cdot\mathbf{R}_j}\right\}$$

$$= N_{wc}\left\{\frac{\hbar^2}{2m^*}\beta + \frac{1}{2}\left[N_{wc}\sum_{\mathbf{g}}\tilde{\eta}(\mathbf{g})\tilde{\eta}(\mathbf{g})\tilde{v}_d(\mathbf{g}) - \frac{A}{4\pi^2}\int d\mathbf{q}\,\tilde{\eta}(\mathbf{q})\tilde{\eta}(\mathbf{q})\tilde{v}_d(\mathbf{q})\right]\right\}\tag{19}$$

Here we use the electron at the $j=1$ site as the reference for the integration; $\eta_0$ denotes the background charge density to make the sample charge neutral; $A$ is the sample area; $\mathbf{g}$ denotes the reciprocal lattice vectors of the Wigner crystal; $\tilde{\eta}(\mathbf{q}) = e^{-q^2/(4\beta)}$ is the Fourier transform of $\eta(\mathbf{r})$. As there is only one electron per site, we don't need to consider the self-interacting potential within the same site (*i.e.* excluding the $j=1$ term in the summation). By using the variational method, the minimum energy occurs at

$$\frac{\partial E_{total}(\beta)}{N_{wc}\partial\beta} = \frac{\hbar^2}{2m^*} + \frac{1}{4\beta^2}\left[N_{wc}\sum_{\mathbf{g}}\mathbf{g}^2\tilde{\eta}(\mathbf{g})\tilde{\eta}(\mathbf{g})\tilde{v}_d(\mathbf{g}) - \frac{A}{4\pi^2}\int q^2d\,\mathbf{q}\,\tilde{\eta}(\mathbf{q})\tilde{\eta}(\mathbf{q})\tilde{v}_d(\mathbf{q})\right] = 0\tag{20}$$

In our calculation, we consider a Wigner crystal that has a triangular lattice with lattice constant $a_{wc}$, which corresponds to an average inter-electron distance of $0.93a_{wc}$. After solving Eq. (20), we obtain the standard deviation $s = \sqrt{1/(2\beta)}$ of the Gaussian distribution $\eta(\mathbf{r})$. Prior research has derived a Wigner-Seitz radius of $r_s > 31$ to be the criterium to form Wigner crystals in 2D (*S4, 12*); This corresponds to $a_{wc} = 28.4$ nm in our system and the associated $s$ is $0.17\,a_{wc}$ in our calculation. When $a_{wc}$ increases from 28.4 nm up to 100 nm (*i.e.* $r_s$ from 31 to 109), $s$ decreases from 0.17 $a_{wc}$ to 0.119 $a_{wc}$.

$r_s$ and $a_{wc}$ are related by

$$\pi r_s^2 a_B^{*2} = A_{wc} = \sqrt{3} a_{wc}^2 / 2 \tag{21}$$

Here $a_B^* = \frac{4\pi \varepsilon_0 \kappa \hbar^2}{m^* e^2}$ is the Bohr radius of the Wigner-crystal particle and $A_{wc}$ is the area of a primitive cell of the Wigner crystal. For our monolayer WSe$_2$ system with $m^* = 0.44\ m_0$, $a_B^* = 0.481$ nm and $a_{wc} \approx 1.905 r_s a_B^* \approx 0.916\ r_s$ nm.

Our above calculation is done with the Keldysh potential given by Eq. (1), which will be applied in our whole theory in this paper. But similar $s$ values are obtained when we use the ideal 2D Coulomb potential. In this ideal case which considers zero layer thickness of monolayer WSe$_2$, our calculation shows that $s$ only decreases smaller than ~0.4% for the range of $a_{wc}$ from 28.4 nm to 100 nm (*i.e.* $r_s$ from 31 to 109). This suggests that the long-range electron-electron interaction plays a dominant role in the formation of Wigner crystals in the low-density regime.

To check if it is reasonable to neglect the overlap integrals between neighboring sites, we calculate the overlap integral of two Gaussian functions separated by a distance of $a_{wc}$ (the nearest-neighbor distance in the triangular lattice) and obtain

$$\langle \psi_{wc}(\mathbf{r}) | \psi_{wc}(\mathbf{r} - \mathbf{R}_1) \rangle = e^{-a_{wc}^2 / 8s^2} \tag{22}$$

The overlap integrals range from $e^{-8.83} \approx 1.46 \times 10^{-4}$ to $e^{-4.325} \approx 0.013$ for $a_{wc}$ between 100 and 28.4 nm (*i.e.* $r_s$ from 100 to 31). Therefore, our approximation to neglect the nearest-neighbor wavefunction overlap is reasonable for $a_{wc} \geq 28.4$ nm.

We note that the simple model adopted here cannot be used to find the critical density for the formation of the Wigner crystal. Although the inter-site overlap is still insignificant (~1 %) at the critical density, it increases exponentially when the charge density increases beyond the critical density. Consequently, we must include the inter-site exchange terms in the overlap integral and in the inter-particle Coulomb interactions to reliably compute the total energy. Despite such limitation, this simple model allows us to determine the optimized trial wavefunction at charge densities below the critical density for both the Keldysh and ideal 2D Coulomb interaction potentials.

We note that a previous model calculation based on local density approximation (LDA) within the density functional theory (DFT) predicts an optimized Gaussian wavefunction with $s \approx 0.3a_{wc}$ at the critical density with $r_s = 31$ (*S4*); their $s$ value is 1.765 times larger than our result $s \approx 0.17a_{wc}$ at $r_s = 31$. In the LDA-DFT approach, the nearest-neighbor overlap is significant to reduce the total energy of the Wigner crystal. We are uncertain which approach is more accurate over the entire density range of interest in our experiment. For the sake of comparison, we will calculate the exciton Umklapp scattering based on two sets of Wigner crystal parameters. The first set is based on our own $s$ values, which are obtained through a semiclassical approach (Model 1). The second set employs the extrapolated DFT-LDA outcomes (Model 2), which involve our $s$ values scaled by a factor of 1.765 (Since the DFT-LDA result is available only for $r_s = 31$, we just use the scaling factor to extrapolate the $s$ values at other densities).

Thus far, we only consider the conventional criterion $r_s > 31$ in the calculation. Such a criterion is derived for ideal 2D systems with zero sample thickness. The criterion could change in a quasi-2D material with finite sample thickness and Keldysh potential. Indeed, our experiment shows signatures of Wigner crystals even at $r_s \sim 15$, well below the conventional criterion. Such an enhancement of WC formation might be caused by the presence of defects and the Keldysh potential (*S7*). To extend our simple model to the regime below $r_s = 31$, we need to consider the behavior of the WC-site charge distribution beyond the $r_s = 31$ condition. As $r_s$ approaches 31 from large value (*i.e.* increasing density), the optimized $s$ value in Model 1 decreases due to increasing inter-site repulsion. But once $r_s$ goes below 31, we expect $s$ to increase exponentially and the charges to become a Fermi liquid due to the inter-site exchange and correlation effect, which is neglected in our model. Here, we emulate such a behavior phenomenologically by multiplying $s$ with a factor $exp(\zeta/a_{wc}^2)$, where $\zeta$ is an empirical parameter. With $\zeta = 100 \ nm^2$ we get a good fit to experimental results. After we get the s value in Model 1, we simply scale it by a factor of 1.765 and use it as the s value for Model 2 in the corresponding regime of $r_s < 31$.

## 3. Exciton Umklapp scattering with different Wigner crystals

The Umklapp scattering of excitons with a Wigner crystal can induce the zone-folding effect, giving rise to a weak feature above the exciton resonance in the absorption spectrum. Here we consider the Umklapping scattering between excitons and Wigner crystals with different band configurations, which can be accessed through electrostatic gating and the aplication of magnetic field.

As shown in Fig. 3 in the main paper, a vertical magnetic field can lift the degeneracy of the K and K' valleys. The injected electrons will fill the lower spin-up conduction band in the K' valley, and the injected holes will fill the spin-up valence band in the K valley. Since an exciton (an photoexcited electron-hole pair) is charge-neutral, the long-range part of the direct Coulomb interaction between the exciton and the Wigner crystal is zero because the electron-electron and hole-electron scattering has opposite amplitudes and hence cancel each other. However, their long-range part of the exchange Coulomb interaction may or may not cancel each other, depending on their band configuration.

Our experiment invovles four different cases, as illustrated in Fig. 3 of the main paper. In Case 1-2, the photoexcited electron (hole) and Fermi-sea electron (hole) have different spins and hence no exchange interaction. In Case 3, the photoexcited electron and the Fermi-sea electron have the same spin, but since they are at different valleys, their exchange interaction is one order of magnitude smaller than when they are in the same valley. In contrast, in Case 4, the photoexcited hole and the Fermi-sea hole have the same spin and also in the same valley, their exchange interaction will be sizable.

The total Hamiltonian ($H_{wc}^X$) of a photoexcited exciton ($X$) interacting with all the particles in a Wigner crystal ($wc$) is given as

$$H_{wc}^X\big(\mathbf{r}_e, \mathbf{r}_h, \{\mathbf{r}_j\}\big) = H_{wc} + H_X^{in}(\mathbf{r}_e - \mathbf{r}_h) - \frac{\hbar^2 \nabla_\mathbf{R}^2}{2m_X} + v_{wc}^X \qquad (23)$$

Here $\mathbf{r}_e$ and $\mathbf{r}_h$ denote the coordinates of the electron and hole in the exciton; $\{\mathbf{r}_j\}$ denote the coordinates of all $N_{wc}$ particles with index $j = 1, 2 \dots N_{wc}$ in the Wigner crystal; $H_{wc}$ is the Hamiltonian of the Wigner crystal that includes the kinetic and potential energy of the particles within; $H_X^{in}$ is the Hamiltonian for the internal motion of the exciton as given in Eq. (9); $-\frac{\hbar^2 \nabla_\mathbf{R}^2}{2m_X}$ denotes the kinetic energy of the CM motion of the exciton.

$$v_{wc}^X = \pm \sum_j \big[v(\mathbf{r}_h - \mathbf{r}_j) - v(\mathbf{r}_e - \mathbf{r}_j)\big] \qquad (24)$$

denotes the potential energy of the screened Coulomb interaction between the photoexcited electron and hole in the exciton and all of the particles in the Wigner crystal. Here $v$ denotes the potential energy of Coulomb interaction between two particles of the same charge (screened by background charges in the solid); "±" denotes the sign of the charges in the Wigner crystal ("−" for an electron Wigner crystal and "+" for a hole Wigner crystal).

We define the interacting state between an exciton (*e.g.* at the K valley) and a Wigner crystal as a linear combination of the following many-body basis states

$$|WC; X_{\mathbf{k}_1+\mathbf{g}}^K\rangle = \sum_{\mathbf{k}} \tilde{\varphi}_0(\mathbf{k}) a_{c\uparrow,\mathbf{K}+\mathbf{k}+\gamma_e(\mathbf{k}_1+\mathbf{g})}^+ a_{v\uparrow,\mathbf{K}+\mathbf{k}-\gamma_h(\mathbf{k}_1+\mathbf{g})}^- |WC\rangle \tag{25}$$

Here $X_{\mathbf{k}_1+\mathbf{g}}$ denotes an exciton with CM wave vector $\mathbf{k}_1 + \mathbf{g}$, where $\mathbf{k}_1$ is a wavevector within the first Brillouin zone of the Wigner crystal and $\mathbf{g}$ is a reciprocal lattice vector of the Wigner crystal; $\tilde{\varphi}_0(\mathbf{k})$ is the envelope function for the internal motion of the exciton ground state. The operator $a_{c\uparrow,\mathbf{K}+\mathbf{k}+\gamma_e(\mathbf{k}_1+\mathbf{g})}^+$ creates an electron with a wavevector $\mathbf{K}+\mathbf{k}+\gamma_e(\mathbf{k}_1+\mathbf{g})$ in the spin-up ($\uparrow$) conduction band ($c$) in the K valley and the other operator $a_{v\uparrow,\mathbf{K}+\mathbf{k}-\gamma_h(\mathbf{k}_1+\mathbf{g})}^-$ annihilates an electron with a wavevector $\mathbf{K}+\mathbf{k}-\gamma_h(\mathbf{k}_1+\mathbf{g})$ in the spin-up ($\uparrow$) valence band ($v$) in the K valley. The corresponding electron Bloch states are denoted as $|u_{c\uparrow,\mathbf{K}+\mathbf{k}+\gamma_e(\mathbf{k}_1+\mathbf{g})}\rangle$ and $|u_{v\uparrow,\mathbf{K}+\mathbf{k}-\gamma_h(\mathbf{k}_1+\mathbf{g})}\rangle$, respectively. For the K'-valley exciton $|X_{\mathbf{k}_1+\mathbf{g}}^{K'}\rangle$, we simply replace the $\mathbf{K}$ vector in Eq. (25) by $\mathbf{K'} = -\mathbf{K}$ and flip the spins.

$|WC\rangle$ denotes the quantum state of the Wigner crystal, including all the particles within. It is expressed as

$$|WC\rangle = \Pi_j \sum_{\mathbf{q}} \tilde{\psi}_{wc}(\mathbf{q}) \frac{1}{\sqrt{A}} e^{i\mathbf{q}\cdot\mathbf{R}_j} a_{\uparrow,\mp\mathbf{K}+\mathbf{q}}^{\pm} |0\rangle. \tag{26}$$

Here $|0\rangle$ denotes the ground state of monolayer WSe$_2$ without any injected carriers.

$$\tilde{\psi}_{wc}(\mathbf{q}) = \sqrt{4\pi/A\beta}\, e^{-q^2/2\beta} \tag{27}$$

is the Fourier transform of the particle wavefunction localized at a WC site as given in Eq. (13). Eq. (24) denotes either the electron or hole Wigner crystal. The electron Wigner crystal is associated with the operator $a_{\uparrow,-\mathbf{K}+\mathbf{q}}^+$ that creates a spin-up ($\uparrow$) electron with wave vector $-\mathbf{K}+\mathbf{q}$ in the K'-valley conduction band (Cases 1 and 3 in Fig. 3); The hole Wigner crystal is associated with the operator $a_{\uparrow,\mathbf{K}+\mathbf{q}}^-$ that annihilates a spin-up ($\uparrow$) electron with wave vector $\mathbf{K}+\mathbf{q}$ in the K-valley valence band (Cases 2 and 4). These electron and hole Wigner crystals correspond to the band-filling configurations of our sample under positive magnetic field as illustrated in Case 1- 4 in Fig. 3.

By using the basis defined in Eq. (25), the matrix elements of the Hamiltonian in Eq. (23) are expressed as:

$$\langle WC; X_{\mathbf{k}_1+\mathbf{g}}|H_{wc}^X|WC; X_{\mathbf{k}_1+\mathbf{g'}}\rangle = \left[E_{wc} + \left(\frac{\hbar^2|\mathbf{k}_1+\mathbf{g}|^2}{2m_X} + E_0^X\right)\right]\delta_{\mathbf{g},\mathbf{g'}} + \langle WC; X_{\mathbf{k}_1+\mathbf{g}}|v_{wc}^X|WC; X_{\mathbf{k}_1+\mathbf{g'}}\rangle. \tag{28}$$

Here $E_{wc}$ is the energy of the Wigner crystal; $E_0^X$ is the ground-state energy of the exciton internal motion at zero CM momentum; $\frac{\hbar^2 |\mathbf{k}_1 + \mathbf{g}|^2}{2m_X}$ is the energy of the exciton CM motion in the effective-mass approximation. $v_{wc}^X$, given in Eq. (24), describes the screened Coulomb interaction between the photoexcited electron and hole in the exciton and all the particles in the Wigner crystal.

Our theory above can be used to describe excitons and Wigner crystals with different valley configurations. Below we will consider four specific cases, as illustrated in Fig. S6 for zero magnetic field (and also in Fig. 3 for finite magnetic field). Our calculations below consider no magnetic field (e.g. no Zeeman shift or Landau quantization), but all WC particles are considered to reside within a single valley to distinguish Cases 1-4.

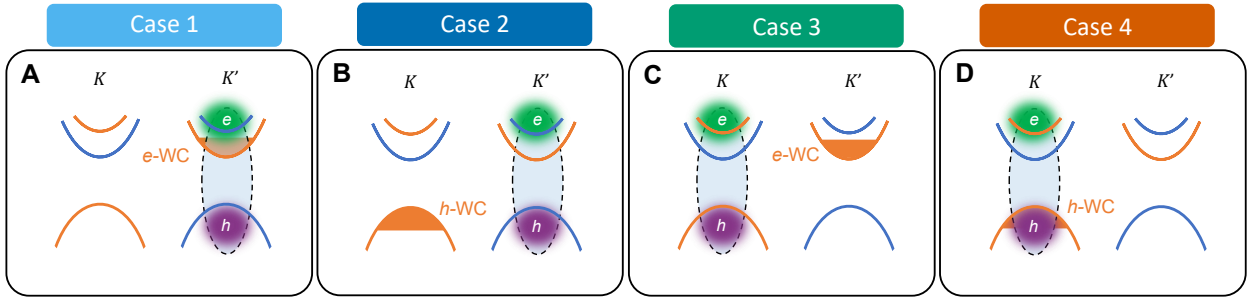

**Figure S6**. Schematic of Cases 1 – 4 at zero magnetic field. The orange (blue) lines denote bands with spin-up (spin-down) configurations.

## Case 1: Umklapp scattering between a $\mathrm{K}'$-valley exciton and a $\mathrm{K}'$-valley electron Wigner crystal

Case 1 in Fig. 3 and Fig. S6A shows the configuration of a $\mathrm{K}'$-valley exciton and an electron Wigner crystal in the lower conduction band (spin up) of the $\mathrm{K}'$ valley. The state of the particle at the $j$-th site of the Wigner crystal is:

$$|j\rangle = \frac{1}{\sqrt{A}} \sum_{\mathbf{q}} \widetilde{\psi}_{wc}(\mathbf{q}) e^{-i\mathbf{q}\cdot\mathbf{R}_j} a_{\uparrow,\mathbf{K}'+\mathbf{q}}^+ |0\rangle \qquad (29)$$

This state can be expressed as a wavefunction in the real-space coordinate as

$$\langle \mathbf{r}|j\rangle = \frac{1}{\sqrt{A}} \sum_{\mathbf{q}} \widetilde{\psi}_{wc}(\mathbf{q}) e^{-i\mathbf{q}\cdot\mathbf{R}_j} u_{c\uparrow,\mathbf{K}'+\mathbf{q}}(\mathbf{r}). \qquad (30)$$

By using Eqs. (8,24,25,30), we can express the scattering matrix elements between the K′-valley exciton and the $j$-th site of the Wigner crystal as:

$$\left\langle j; X_{\mathbf{k}_1+\mathbf{g}}^{K'} \middle| v_{wc}^X \middle| j; X_{\mathbf{k}_1+\mathbf{g}'}^{K'} \right\rangle = \sum_{\mathbf{k},\mathbf{k}',\mathbf{q},\mathbf{q}'} \tilde{\varphi}_0(\mathbf{k}) \tilde{\varphi}_0(\mathbf{k}') \tilde{\psi}_{wc}(\mathbf{q}) \tilde{\psi}_{wc}(\mathbf{q}') e^{i(\mathbf{q}-\mathbf{q}')\cdot\mathbf{R}_j} \times$$

$$\left\{ \left\langle u_{c\downarrow,\mathbf{K}'+\mathbf{k}+\gamma_e(\mathbf{k}_1+\mathbf{g})} u_{v\downarrow,\mathbf{K}'+\mathbf{k}-\gamma_h(\mathbf{k}_1+\mathbf{g})} u_{c\uparrow,\mathbf{K}'+\mathbf{q}} \middle| v(\mathbf{r}_e - \mathbf{r}) \middle| u_{c\downarrow,\mathbf{K}'+\mathbf{k}'+\gamma_e(\mathbf{k}_1+\mathbf{g}')} u_{v\downarrow,\mathbf{K}'+\mathbf{k}'-\gamma_h(\mathbf{k}_1+\mathbf{g}')} u_{c\uparrow,\mathbf{K}'+\mathbf{q}'} \right\rangle - \right.$$

$$\left. \left\langle u_{c\downarrow,\mathbf{K}'+\mathbf{k}+\gamma_e(\mathbf{k}_1+\mathbf{g})} u_{v\downarrow,\mathbf{K}'+\mathbf{k}-\gamma_h(\mathbf{k}_1+\mathbf{g})} u_{c\uparrow,\mathbf{K}'+\mathbf{q}} \middle| v(\mathbf{r}_h - \mathbf{r}) \middle| u_{c\downarrow,\mathbf{K}'+\mathbf{k}'+\gamma_e(\mathbf{k}_1+\mathbf{g}')} u_{v\downarrow,\mathbf{K}'+\mathbf{k}'-\gamma_h(\mathbf{k}_1+\mathbf{g}')} u_{c\uparrow,\mathbf{K}'+\mathbf{q}'} \right\rangle \right\} \quad (31)$$

In the curly bracket of Eq. (31), The first (second) term describes the scattering between the photoexcited electron (hole) and a particle in the Wigner crystal. The first and second Bloch states are associated, respectively, with the photoexcited electron and hole in the spin-down bands. The third Bloch state $u_{c\uparrow,\mathbf{K}'+\mathbf{q}}$ is associated with the Wigner-crystal electron (spin up).

Since $v_{wc}^X$ only involves two-particle scattering, the third particle not involved in the scattering can be taken out of the bracket. We can simplify Eq. (31) as:

$$\left\langle j; X_{\mathbf{k}_1+\mathbf{g}}^{K'} \middle| v_{wc}^X \middle| j; X_{\mathbf{k}_1+\mathbf{g}'}^{K'} \right\rangle = \sum_{\mathbf{k},\mathbf{k}',\mathbf{q},\mathbf{q}'} \tilde{\varphi}_0(\mathbf{k}) \tilde{\varphi}_0(\mathbf{k}') \tilde{\psi}_{wc}(\mathbf{q}) \tilde{\psi}_{wc}(\mathbf{q}') e^{i(\mathbf{q}-\mathbf{q}')\cdot\mathbf{R}_j} \times$$

$$\left\{ \left\langle u_{c\downarrow,\mathbf{K}'+\mathbf{k}+\gamma_e(\mathbf{k}_1+\mathbf{g})} u_{c\uparrow,\mathbf{K}'+\mathbf{q}} \middle| v(\mathbf{r}_e - \mathbf{r}) \middle| u_{c\downarrow,\mathbf{K}'+\mathbf{k}'+\gamma_e(\mathbf{k}_1+\mathbf{g}')} u_{c\uparrow,\mathbf{K}'+\mathbf{q}'} \right\rangle \delta_{\mathbf{k}-\gamma_h\mathbf{g},\ \mathbf{k}'-\gamma_h\mathbf{g}'} - \right.$$

$$\left. \left\langle u_{v\downarrow,\mathbf{K}'+\mathbf{k}-\gamma_h(\mathbf{k}_1+\mathbf{g})} u_{c\uparrow,\mathbf{K}'+\mathbf{q}} \middle| v(\mathbf{r}_h - \mathbf{r}) \middle| u_{v\downarrow,\mathbf{K}'+\mathbf{k}'-\gamma_h(\mathbf{k}_1+\mathbf{g}')} u_{c\uparrow,\mathbf{K}'+\mathbf{q}'} \right\rangle \delta_{\mathbf{k}+\gamma_e\mathbf{g},\ \mathbf{k}'+\gamma_e\mathbf{g}'} \right\} \quad (32)$$

We note that in Eq. (32) there is no exchange interaction between the two scattering particles since they have different spin in the Bloch states involved.

By summing over the scattering with every site in the Wigner crystal, the scattering matrix elements between the K′K′ exciton and the whole Wigner crystal are then

$$\left\langle WC; X_{\mathbf{k}_1+\mathbf{g}}^{K'} \middle| v_{wc}^X \middle| WC; X_{\mathbf{k}_1+\mathbf{g}'}^{K'} \right\rangle = \sum_j \left\langle j; X_{\mathbf{k}_1+\mathbf{g}}^{K'} \middle| v_{wc}^X \middle| j; X_{\mathbf{k}_1+\mathbf{g}'}^{K'} \right\rangle \quad (33)$$

To calculate the above scattering matrix elements, which in this case only include the direct Coulomb interaction (no exchange terms), we apply a supercell method in the density functional theory (DFT). This DFT method considers a stack of infinite number of WSe$_2$ layers with large space in between, thus giving rise to a three-dimensional (3D) crystal with 3D unit cells (called supercells).

In this artificial crystal, the scattering matrix elements (between a photoexcited electron and a Wigner-crystal electron) have the expression (*S13*):

$$\left\langle u_{c\downarrow,\mathbf{K}'+\mathbf{k}+\gamma_e(\mathbf{k}_1+\mathbf{g})}\, u_{c\uparrow,\mathbf{K}'+\mathbf{q}} \middle| W \middle| u_{c\downarrow,\mathbf{K}'+\mathbf{k}'+\gamma_e(\mathbf{k}_1+\mathbf{g}')}\, u_{c\uparrow,\mathbf{K}'+\mathbf{q}'} \right\rangle$$

$$= \frac{1}{\Omega} \sum_{\mathbf{G},\mathbf{G}'} W_{\mathbf{G},\mathbf{G}'}(\Delta\mathbf{q}) \left\langle u_{c\downarrow,\mathbf{K}'+\mathbf{k}+\gamma_e(\mathbf{k}_1+\mathbf{g})}\, \middle|\, e^{i(\mathbf{G}-\Delta\mathbf{q})\cdot\mathbf{r}}\, \middle|\, u_{c\downarrow,\mathbf{K}'+\mathbf{k}'+\gamma_e(\mathbf{k}_1+\mathbf{g}')} \right\rangle \left\langle u_{c\uparrow,\mathbf{K}'+\mathbf{q}}\, \middle|\, e^{-i(\mathbf{G}'+\Delta\mathbf{q})\cdot\mathbf{r}}\, \middle|\, u_{c\uparrow,\mathbf{K}'+\mathbf{q}'} \right\rangle \delta_{\Delta\mathbf{q},-\Delta\mathbf{g}}$$

$$(34)$$

Here $\Omega$ is the volume of the sample in the supercell method; $\mathbf{G}$ and $\mathbf{G}'$ denote the reciprocal lattice vectors of the superlattice ($\mathbf{G} = \mathbf{G}_\parallel + \mathbf{G}_\perp$), which consist of the in-plane reciprocal lattice vectors ($\mathbf{G}_\parallel$) of monolayer WSe$_2$ and the out-of-plane reciprocal lattice vectors ($\mathbf{G}_\perp$) for the supercell; $\Delta\mathbf{q} = \mathbf{k} - \mathbf{k}' + \gamma_e(\mathbf{g} - \mathbf{g}') = \mathbf{q}' - \mathbf{q}$ and $\Delta\mathbf{g} = \mathbf{g}' - \mathbf{g}$ are the difference between the wave vectors of the Bloch states (*i.e.* the momentum transfer of the scattering processes); We use $\delta_{\Delta\mathbf{q},-\Delta\mathbf{g}}$ to impose momentum conservation during the scattering; $W_{\mathbf{G},\mathbf{G}'}(\Delta\mathbf{q})$ is the scattering matrix element of $W$ between two 3D plane waves with wave vectors $\mathbf{K}' + \mathbf{k} + \gamma_e(\mathbf{k}_1+\mathbf{g}) + \mathbf{G}$ and $\mathbf{K}' + \mathbf{k}' + \gamma_e(\mathbf{k}_1+\mathbf{g}') + \mathbf{G}'$, which has the expression (*S14*):

$$W_{\mathbf{GG}'}(\Delta\mathbf{q}) = \frac{e^2 \varepsilon_{\mathbf{GG}'}^{-1}(\Delta\mathbf{q})}{\varepsilon_0 |\Delta\mathbf{q}+\mathbf{G}| |\Delta\mathbf{q}+\mathbf{G}'|} \qquad (35)$$

Here $\varepsilon_{\mathbf{GG}'}^{-1}(\Delta\mathbf{q})$ is an inverse dielectric matrix defined in Eq. (12) in Ref. (*S14*). We further apply the LASTO code via a supercell method to calculate the Bloch states with 3D augmented plane waves as the underlying basis functions (*S15, 16*).

The primary contribution to the $W$ matrix elements in Eq. (34) comes from the long-range term (*i.e.* the $\mathbf{G}_\parallel = \mathbf{G}'_\parallel = 0$ term associated with zero in-plane reciprocal lattice vector of monolayer WSe$_2$). The long-range ($L$) component of the $W$ matrix elements (between a photoexcited electron and a Wigner-crystal electron) in Eq. (34) can be approximated as

$$W_{ee}^L(\Delta\mathbf{g})$$

$$= \frac{1}{\Omega} \sum_{\mathbf{G}_\perp} W_{\mathbf{G},\mathbf{G}}(\Delta\mathbf{g}) \left\langle u_{c\downarrow,\mathbf{K}'+\mathbf{k}+\gamma_e(\mathbf{k}_1+\mathbf{g})}\, \middle|\, e^{i(\mathbf{G}_\perp-\Delta\mathbf{g})\cdot\mathbf{r}}\, \middle|\, u_{c\downarrow,\mathbf{K}'+\mathbf{k}'+\gamma_e(\mathbf{k}_1+\mathbf{g}')} \right\rangle \left\langle u_{c\uparrow,\mathbf{K}'+\mathbf{q}}\, \middle|\, e^{-i(\mathbf{G}_\perp-\Delta\mathbf{g})\cdot\mathbf{r}}\, \middle|\, u_{c\uparrow,\mathbf{K}'+\mathbf{q}'} \right\rangle$$

$$= \frac{e^2}{2\Lambda\kappa(\Delta\mathbf{g})\varepsilon_0|\Delta\mathbf{g}|} \int dz_e \int dz\, e^{-|\Delta\mathbf{g}||z_e-z|} \widetilde{\eta}_{c\downarrow,\mathbf{K}';\,c\downarrow,\mathbf{K}'}(\Delta\mathbf{g},z_e)\, \widetilde{\eta}_{c\uparrow,\mathbf{K}';\,c\uparrow,\mathbf{K}'}(-\Delta\mathbf{g},z)$$

$$(36)$$

with

$$\widetilde{\eta}_{c\downarrow,\mathbf{K}';\,c\downarrow,\mathbf{K}'}(\Delta\mathbf{g},z_e) = \int d\boldsymbol{\rho}_e \left| u_{c\downarrow,\mathbf{K}'}(\boldsymbol{\rho}_e,z_e) \right|^2 e^{i\Delta\mathbf{g}\cdot\boldsymbol{\rho}_e}$$

$$\widetilde{\eta}_{c\uparrow,K';\,c\uparrow,K'}\,(-\Delta\mathbf{g},z) = \int d\boldsymbol{\rho}\,\left|u_{c\uparrow,K'}(\boldsymbol{\rho},z)\right|^2 e^{-i\Delta\mathbf{g}\cdot\boldsymbol{\rho}} \tag{37}$$

Here $L_c$ is the length of the supercell along the $z$ axis; $\mathbf{r}_e = (\boldsymbol{\rho}_e, z_e)$ denotes the coordinate of the photoexcited electron; $\mathbf{r} = (\boldsymbol{\rho}, z)$ denotes the coordinate of the WC electron. We make two approximations in Eqs. (36-37). First, we neglect the weak dependence of the integrals on the small wave vectors $\mathbf{k}$, $\mathbf{k}'$, and $\mathbf{k}_1$. Second, we replace the dielectric matrix $\varepsilon_{\mathbf{GG}'}(\Delta\mathbf{g})$ by a local dielectric function $\kappa(\Delta\mathbf{g})\delta_{\mathbf{GG}'}$; this approximation is reasonable for dealing with Wannier excitons (*S17*). We note that the DFT calculation is carried out in a 3D supercell with a large spacer placed between adjacent WSe$_2$ Monolayers. Thus, the charge density $\widetilde{\eta}$ can be written in terms of plane waves $e^{i\mathbf{G}_\perp \cdot z}$ along the $z$ axis. The numerical results for the in-plane averaged charge density $\widetilde{\eta}_{b,K';\,b,K'}(\mathbf{0},z)$ for the three bands of interest ( $b = c\uparrow$, $c\downarrow$, and $v\downarrow$ ) for monolayer WSe$_2$ are shown in Fig. S7.

After we calculate the long-range term ($\mathbf{G}_\parallel = 0$), we move on to calculate the short-range terms ($\mathbf{G}_\parallel \neq 0$). The short-range ($S$) components of the direct Coulomb scattering matrix elements can be expressed as:

$$W_{ee}^S = \frac{1}{L_c}\sum_{\mathbf{G}_\parallel \neq 0,\mathbf{G}_\perp}\frac{e^2}{\varepsilon_0|\mathbf{G}|^2}\langle u_{c\downarrow,K'}|e^{i\mathbf{G}\cdot\mathbf{r}}|\,u_{c\downarrow,K'}\rangle\langle u_{c\uparrow,K'}|e^{-i\mathbf{G}\cdot\mathbf{r}}|\,u_{c\uparrow,K'}\rangle \tag{38}$$

Here we have also neglected the weak dependence of the Bloch states on the small wave vectors $\mathbf{k}$, $\mathbf{g}$, and $\mathbf{k}_1$. We have also neglected the dielectric screening, which is insignificant for the short-range terms.

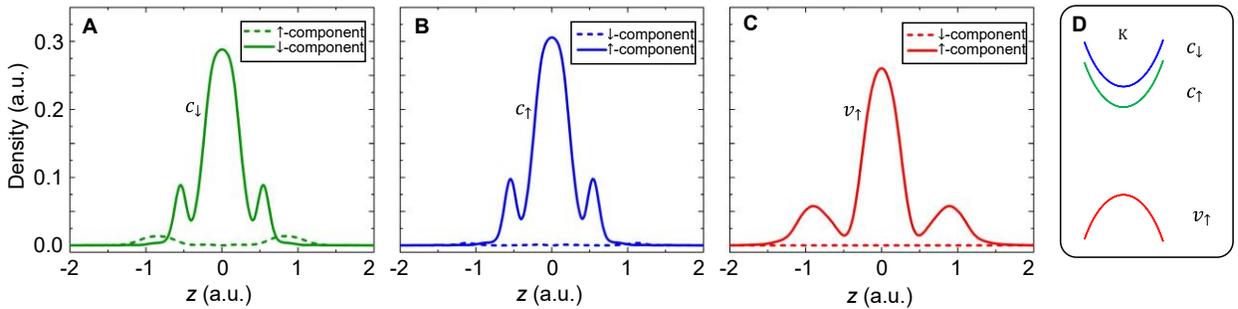

**Figure S7**. (A-C) In-plane-averaged electron density along the vertical direction (z) for the states at the K point in monolayer WSe$_2$, including the states in (A) the lower conduction band ($c_\downarrow$), (B) the higher conduction band ($c_\uparrow$), and (C) the valence band ($v_\uparrow$). These states have both the spin-up and spin-down components, but one of the spins dominates. The dominant spin is denoted by the subscripts of the state symbols. (D) The schematic band configuration in the K valley of monolayer WSe$_2$.

After we consider the electron-WC scattering, the hole-WC scattering can be calculated in a similar way. The long-range term of the direct scattering matrix elements between the photoexcited hole and a Wigner crystal electron is given as:

$$W_{he}^L(\Delta \mathbf{g}) = \frac{e^2}{A\kappa(\Delta \mathbf{g})\varepsilon_0|\Delta \mathbf{g}|} \int dz_h \int dz \, e^{-|\Delta \mathbf{g}||z_h - z|} \widetilde{\eta}_{v\downarrow K',v\downarrow K'}(\Delta \mathbf{g}, z_h) \widetilde{\eta}_{c\uparrow K',c\uparrow K'}(-\Delta \mathbf{g}, z) \quad (39)$$

Here $\widetilde{\eta}_{v\downarrow,K'\,;\,v\downarrow,K'}$ is defined similarly as in Eq. (37) with the band index $c$ replaced by $v$ to represent the valence band.

The short-range terms of the direct Coulomb scattering matrix elements between the photoexcited hole and a Wigner-crystal electron are given as:

$$W_{he}^S = \frac{1}{L_c} \sum_{\mathbf{G}_\parallel \neq 0, \mathbf{G}_\perp} \frac{e^2}{\varepsilon_0|\mathbf{G}|^2} \langle u_{v\downarrow,K'}|e^{i\mathbf{G}\cdot\mathbf{r}}|u_{v\downarrow,K'}\rangle \langle u_{c\uparrow,K'}|e^{-i\mathbf{G}\cdot\mathbf{r}}|u_{c\uparrow,K'}\rangle \quad (40)$$

After summing over all the long-range and short-range terms of electron-WC and hole-WC scattering, the scattering matrix elements of exciton-WC scattering can be expressed as:

$$\left\langle WC; X_{\mathbf{k}_1+\mathbf{g}}^{K'} \middle| v_{wc}^X \middle| WC; X_{\mathbf{k}_1+\mathbf{g}'}^{K'} \right\rangle = \frac{1}{A} \sum_{\mathbf{k},\mathbf{k}',\mathbf{q},\mathbf{q}',j} \widetilde{\varphi}_0(\mathbf{k}) \widetilde{\varphi}_0(\mathbf{k}') \widetilde{\psi}_{wc}(\mathbf{q}) \widetilde{\psi}_{wc}(\mathbf{q}') e^{i(\mathbf{q}-\mathbf{q}')\cdot\mathbf{R}_j} \times$$

$$\left\{ [W_{ee}^L(\Delta \mathbf{g}) + W_{ee}^S] \delta_{\mathbf{k}-\gamma_h\mathbf{g},\mathbf{k}'-\gamma_h\mathbf{g}'} - [W_{he}^L(\Delta \mathbf{g}) + W_{he}^S] \delta_{\mathbf{k}+\gamma_e\mathbf{g},\mathbf{k}'+\gamma_e\mathbf{g}'} \right\} \delta_{\Delta\mathbf{q},-\Delta\mathbf{g}} \quad (41)$$

We note that the long-range terms $W_{ee}^L(\Delta \mathbf{g})$ and $W_{he}^L(\Delta \mathbf{g})$ have similar Keldysh form; their difference is small and will only give rise to weak exciton-WC scattering that is comparable to the scattering due to the short-range interaction terms $W_{ee}^S$ and $W_{he}^S$. Therefore, the net exciton-WC scattering is weak in Case 1.

## Case 2: Umklapp scattering between a K′-valley exciton and the K-valley hole Wigner crystal

After we consider Case 1 above, we will consider Case 2 as illustrated in Fig. 3 and Fig. S6B. In this case, the Wigner crystal consists of holes on the spin-up valence band in the K valley, whereas the exciton is still the K′K′ exciton as in Case 1. The particles participating in the scattering process have opposite spins, and therefore there is no exchange interaction in this case. The calculation of the scattering matrix elements is essentially the same as in Case 1; we only need to replace the conduction Bloch state $u_{c\uparrow,K'}$ of the Wigner crystal in Eqs. (29-40) by the valence Bloch state $u_{v\uparrow,K}$ and switch the sign of charge of the Wigner crystal.

The scattering matrix elements of exciton-WC scattering can be expressed as:

$$\left\langle \mathrm{WC}; X^{K'}_{\mathbf{k}_1+\mathbf{g}} \middle| v^X_{wc} \middle| \mathrm{WC}; X^{K'}_{\mathbf{k}_1+\mathbf{g}'} \right\rangle = \frac{1}{A} \sum_{\mathbf{k},\mathbf{k}',\mathbf{q},\mathbf{q}',j} \tilde{\varphi}_0(\mathbf{k}) \tilde{\varphi}_0(\mathbf{k}') \tilde{\psi}_{wc}(\mathbf{q}) \tilde{\psi}_{wc}(\mathbf{q}') e^{i(\mathbf{q}-\mathbf{q}')\cdot\mathbf{R}_j} \times$$

$$\{-[W^L_{eh}(\Delta\mathbf{g}) + W^S_{eh}]\delta_{\mathbf{k}-\gamma_h\mathbf{g},\,\mathbf{k}'-\gamma_h\mathbf{g}'} + [W^L_{hh}(\Delta\mathbf{g}) + W^S_{hh}]\delta_{\mathbf{k}+\gamma_e\mathbf{g},\,\mathbf{k}'+\gamma_e\mathbf{g}'}\}\delta_{\Delta\mathbf{q},-\Delta\mathbf{g}} \qquad (42)$$

Here $W^L_{eh}$ $(W^L_{hh})$ is the long-range term of the scattering matrix elements between the photoexcited electron (hole) and a Wigner-crystal hole; they have similar expressions as $W^L_{ee}$ and $W^L_{he}$ given in Eqs. (36) and (39) with the replacement of $u_{c\uparrow,\mathbf{K}'}$ by $u_{v\uparrow,\mathbf{K}}$ for the WC particle. $W^S_{eh}$ $(W^S_{hh})$ are the short-range terms of the scattering matrix elements between the photoexcited electron (hole) and a WC hole. They can be obtained from Eqs. (38) and (40), respectively, by replacing $u_{c\uparrow,\mathbf{K}'}$ with $u_{v\uparrow,\mathbf{K}}$. Similar to Case 1, the net exciton-WC scattering is weak in Case 2.

## Case 3: Umklapp scattering between a K-valley exciton and the K′-valley electron Wigner crystal

We next consider Case 3 in Fig. 3 and Fig. S6C. In this case, we consider an exciton in the K valley, which consists of an electron in the spin-up conduction band $u_{c\uparrow,\mathbf{K}}$ and a hole in the spin-up valence band $u_{v\uparrow,\mathbf{K}}$, and the electron Wigner crystal in the spin-up conduction band $u_{c\uparrow,\mathbf{K}'}$ of the K′ valley. In contrast to Cases 1 and 2, which have no exchange interaction, here we need to consider the exchange interaction in Case 3 because the particles have the same spin.

The matrix elements of exciton-WC scattering are expressed as:

$$\left\langle \mathrm{WC}; X^{K}_{\mathbf{k}_1+\mathbf{g}} \middle| v^X_{wc} \middle| \mathrm{WC}; X^{K}_{\mathbf{k}_1+\mathbf{g}'} \right\rangle = \frac{1}{A} \sum_{\mathbf{k},\mathbf{k}',\mathbf{q},\mathbf{q}',j} \tilde{\varphi}_0(\mathbf{k}) \tilde{\varphi}_0(\mathbf{k}') \tilde{\psi}_{wc}(\mathbf{q}) \tilde{\psi}_{wc}(\mathbf{q}') e^{i(\mathbf{q}-\mathbf{q}')\cdot\mathbf{R}_j} \times$$

$$\Big\{ \left\langle u_{c\uparrow,\mathbf{K}+\mathbf{k}+\gamma_e(\mathbf{k}_1+\mathbf{g})} u_{c\uparrow,\mathbf{K}'+\mathbf{q}} \middle| v(\mathbf{r}_e-\mathbf{r}) \middle| u_{c\uparrow,\mathbf{K}+\mathbf{k}'+\gamma_e(\mathbf{k}_1+\mathbf{g}')} u_{c\uparrow,\mathbf{K}'+\mathbf{q}'} \right\rangle_{\mathcal{A}} \delta_{\mathbf{k}-\gamma_h\mathbf{g},\;\mathbf{k}'-\gamma_h\mathbf{g}'} -$$

$$\left\langle u_{v\uparrow,\mathbf{K}+\mathbf{k}-\gamma_h(\mathbf{k}_1+\mathbf{g})} u_{c\uparrow,\mathbf{K}'+\mathbf{q}} \middle| v(\mathbf{r}_h-\mathbf{r}) \middle| u_{v\uparrow,\mathbf{K}+\mathbf{k}'-\gamma_h(\mathbf{k}_1+\mathbf{g}')} u_{c\uparrow,\mathbf{K}'+\mathbf{q}'} \right\rangle_{\mathcal{A}} \delta_{\mathbf{k}+\gamma_e\mathbf{g},\;\mathbf{k}'+\gamma_e\mathbf{g}'} \Big\} \qquad (43)$$

Here the subscript $\mathcal{A}$ denotes that we use the antisymmetric wavefunctions of the bra and ket states, because we need to consider the exchange effect of identical particles here. Each term above consists of the components for direct and exchange Coulomb interactions.

For the scattering between the photoexcited electron and a WC electron, the matrix elements are expressed as:

$$\left\langle u_{c\uparrow,\mathbf{K}+\mathbf{k}+\gamma_e(\mathbf{k}_1+\mathbf{g})} u_{c\uparrow,\mathbf{K}'+\mathbf{q}} \middle| v(\mathbf{r}_e-\mathbf{r}) \middle| u_{c\uparrow,\mathbf{K}+\mathbf{k}'+\gamma_e(\mathbf{k}_1+\mathbf{g}')} u_{c\uparrow,\mathbf{K}'+\mathbf{q}'} \right\rangle_{\mathcal{A}}$$

$$= \left\langle u_{c\uparrow,\mathbf{K}+\mathbf{k}+\gamma_e(\mathbf{k}_1+\mathbf{g})} u_{c\uparrow,\mathbf{K}'+\mathbf{q}} \middle| W \middle| u_{c\uparrow,\mathbf{K}+\mathbf{k}'+\gamma_e(\mathbf{k}_1+\mathbf{g}')} u_{c\uparrow,\mathbf{K}'+\mathbf{q}'} \right\rangle$$

$$- \left\langle u_{c\uparrow,\mathbf{K}+\mathbf{k}+\gamma_e(\mathbf{k}_1+\mathbf{g})} u_{c\uparrow,\mathbf{K}'+\mathbf{q}} \middle| V_{ex} \middle| u_{c\uparrow,\mathbf{K}+\mathbf{k}'+\gamma_e(\mathbf{k}_1+\mathbf{g}')} u_{c\uparrow,\mathbf{K}'+\mathbf{q}'} \right\rangle \tag{44}$$

Here W denotes the screened direct Coulomb interaction and $V_{ex}$ denotes the exchange interaction. The W term consists of the long-range and short-range terms:

$$\left\langle u_{c\uparrow,\mathbf{K}+\mathbf{k}+\gamma_e(\mathbf{k}_1+\mathbf{g})} u_{c\uparrow,\mathbf{K}'+\mathbf{q}} \middle| W \middle| u_{c\uparrow,\mathbf{K}+\mathbf{k}'+\gamma_e(\mathbf{k}_1+\mathbf{g}')} u_{c\uparrow,\mathbf{K}'+\mathbf{q}'} \right\rangle = [\overline{W}_{ee}^L(\Delta\mathbf{g}) + \overline{W}_{ee}^S]\delta_{\Delta\mathbf{q},-\Delta\mathbf{g}} \tag{45}$$

Here the long-range term $\overline{W}_{ee}^L$ and the short-range term $\overline{W}_{ee}^S$ can be obtained from Eq. (36) and (38), respectively, by replacing $u_{c\downarrow,\mathbf{K}'}$ there with $u_{c\uparrow,\mathbf{K}}$ because we are considering the K-valley exciton in this case.

The exchange $V_{ex}$ term here is a short-range term, because it involves large momentum transfer between a K-valley electron and a K′-valley electron. We can therefore neglect the dependence on the small wave vectors (e.g. $\mathbf{k}, \mathbf{k}', \mathbf{g}, \mathbf{g}'$) and simply approximate the $V_{ex}$ matrix elements as a constant $\lambda_{ee}$:

$$\left\langle u_{c\uparrow,\mathbf{K}+\mathbf{k}+\gamma_e(\mathbf{k}_1+\mathbf{g})} u_{c\uparrow,\mathbf{K}'+\mathbf{q}} \middle| V_{ex} \middle| u_{c\uparrow,\mathbf{K}+\mathbf{k}'+\gamma_e(\mathbf{k}_1+\mathbf{g}')} u_{c\uparrow,\mathbf{K}'+\mathbf{q}'} \right\rangle$$

$$\approx \frac{1}{\Omega}\sum_{\mathbf{G}}\frac{e^2}{\varepsilon_0|\mathbf{G}+\Delta\mathbf{K}|^2}\left|\left\langle u_{c\uparrow,\mathbf{K}} \middle| e^{i(\mathbf{G}+\Delta\mathbf{K})\cdot\mathbf{r}} \middle| u_{c\uparrow,\mathbf{K}'} \right\rangle\right|^2 \equiv \lambda_{ee} \tag{46}$$

Here $\Delta\mathbf{K} = \mathbf{K}' - \mathbf{K}$ is the momentum transfer for the scattering process.

After we consider the electron-WC scattering, we can calculate the hole-WC scattering by simply replacing $u_{c\uparrow,\mathbf{K}}$ in Eqs. (45-46) with $u_{v\uparrow,\mathbf{K}}$ and obtain the corresponding matrix elements $\overline{W}_{he}^L(\Delta\mathbf{g})$, $\overline{W}_{he}^S$, and $\lambda_{he}$.

In summary, the matrix elements of the exciton-WC scattering are:

$$\left\langle \text{WC}; X_{\mathbf{k}_1+\mathbf{g}}^K \middle| v_{wc}^X \middle| \text{WC}; X_{\mathbf{k}_1+\mathbf{g}'}^K \right\rangle = \frac{1}{A}\sum_{\mathbf{k},\mathbf{k}',\mathbf{q},\mathbf{q}',j} \tilde{\varphi}_0(\mathbf{k})\tilde{\varphi}_0(\mathbf{k}')\tilde{\psi}_{wc}(\mathbf{q})\tilde{\psi}_{wc}(\mathbf{q}')e^{i(\mathbf{q}-\mathbf{q}')\cdot\mathbf{R}_j} \times$$

$$\{[\overline{W}_{ee}^L(\Delta\mathbf{g}) + \overline{W}_{ee}^S - \lambda_{ee}]\delta_{\mathbf{k}-\gamma_h\mathbf{g},\ \mathbf{k}'-\gamma_h\mathbf{g}'} - [\overline{W}_{he}^L(\Delta\mathbf{g}) + \overline{W}_{he}^S - \lambda_{he}]\delta_{\mathbf{k}+\gamma_e\mathbf{g},\ \mathbf{k}'+\gamma_e\mathbf{g}'}\}\delta_{\Delta\mathbf{q},-\Delta\mathbf{g}} \tag{47}$$

Similar to Cases 1-2, the long-range terms $\overline{W}_{ee}^L(\Delta\mathbf{g})$ and $\overline{W}_{he}^L(\Delta\mathbf{g})$ are similar. The short-ranged terms $\overline{W}_{ee}^S$, $\overline{W}_{he}^S$, $\lambda_{ee}$, and $\lambda_{he}$ are one or two orders of magnitude smaller than the long-range terms. So, the net exciton-WC scattering is also weak in Case 3.

**Case 4: Umklapp scattering between a K-valley exciton and the K-valley hole Wigner crystal**

Finally, we consider the case with both the exciton and Wigner crystal in the K valley, as illustrated by Case 4 in Fig. 3 and Fig. S6D. The most distinguished feature of this case is that the photoexcited hole and the injected holes that form the Wigner crystal occupy the same band; the large overlap between them is expected to give rise to strong exchange interaction. In addition, the photoexcited electron has the same spin as the valence band that hosts the Wigner crystal; they therefore have the electron-hole exchange effect.

The matrix elements of the exciton-WC scattering can be obtained from Eq. (43) in Case 3 by replacing $u_{c\uparrow,\mathbf{K}'}$ with $u_{v\uparrow,\mathbf{K}}$ and flipping the signs of Coulomb potentials (because here we consider the hole Wigner crystal):

$$\left\langle \mathrm{WC}; X^K_{\mathbf{k}_1+\mathbf{g}} \middle| v^X_{wc} \middle| \mathrm{WC}; X^K_{\mathbf{k}_1+\mathbf{g}'} \right\rangle = \frac{1}{A}\sum_{\mathbf{k},\mathbf{k}',\mathbf{q},\mathbf{q}',j} \tilde{\varphi}_0(\mathbf{k})\tilde{\varphi}_0(\mathbf{k}')\tilde{\psi}_{wc}(\mathbf{q})\tilde{\psi}_{wc}(\mathbf{q}')e^{i(\mathbf{q}-\mathbf{q}')\cdot\mathbf{R}_j} \times$$

$$\left\{ -\left\langle u_{c\uparrow,\mathbf{K}+\mathbf{k}+\gamma_e(\mathbf{k}_1+\mathbf{g})}u_{v\uparrow,\mathbf{K}+\mathbf{q}} \middle| v(\mathbf{r}_e - \mathbf{r}) \middle| u_{c\uparrow,\mathbf{K}+\mathbf{k}'+\gamma_e(\mathbf{k}_1+\mathbf{g}')}u_{v\uparrow,\mathbf{K}+\mathbf{q}'} \right\rangle_{\mathcal{A}} \delta_{\mathbf{k}-\gamma_h\mathbf{g},\ \mathbf{k}'-\gamma_h\mathbf{g}'} + \right.$$

$$\left. \left\langle u_{v\uparrow,\mathbf{K}+\mathbf{k}\,-\gamma_h(\mathbf{k}_1+\mathbf{g})}u_{v\uparrow,\mathbf{K}+\mathbf{q}} \middle| v(\mathbf{r}_h - \mathbf{r}) \middle| u_{v\uparrow,\mathbf{K}+\mathbf{k}'-\gamma_h(\mathbf{k}_1+\mathbf{g}')}u_{v\uparrow,\mathbf{K}+\mathbf{q}'} \right\rangle_{\mathcal{A}} \delta_{\mathbf{k}+\gamma_e\mathbf{g},\ \mathbf{k}'+\gamma_e\mathbf{g}'} \right\} \qquad (48)$$

For the scattering between the K-valley photoexcited electron and a K-valley WC hole, the matrix elements are:

$$\left\langle u_{c\uparrow,\mathbf{K}+\mathbf{k}+\gamma_e(\mathbf{k}_1+\mathbf{g})}u_{v\uparrow,\mathbf{K}+\mathbf{q}} \middle| v(\mathbf{r}_e - \mathbf{r}) \middle| u_{c\uparrow,\mathbf{K}+\mathbf{k}'+\gamma_e(\mathbf{k}_1+\mathbf{g}')}u_{v\uparrow,\mathbf{K}+\mathbf{q}'} \right\rangle_{\mathcal{A}}$$

$$= \left\langle u_{c\uparrow,\mathbf{K}+\mathbf{k}+\gamma_e(\mathbf{k}_1+\mathbf{g})}u_{v\uparrow,\mathbf{K}+\mathbf{q}} \middle| W \middle| u_{c\uparrow,\mathbf{K}+\mathbf{k}'+\gamma_e(\mathbf{k}_1+\mathbf{g}')}u_{v\uparrow,\mathbf{K}+\mathbf{q}'} \right\rangle$$

$$- \left\langle u_{c\uparrow,\mathbf{K}+\mathbf{k}+\gamma_e(\mathbf{k}_1+\mathbf{g})}u_{v\uparrow,\mathbf{K}+\mathbf{q}} \middle| V_{ex} \middle| u_{c\uparrow,\mathbf{K}+\mathbf{k}'+\gamma_e(\mathbf{k}_1+\mathbf{g}')}u_{v\uparrow,\mathbf{K}+\mathbf{q}'} \right\rangle \qquad (49)$$

Here W and $V_{ex}$ denote the direct and exchange interaction. Similar to Case 3, the direct Coulomb scattering term is obtained as

$$\left\langle u_{c\uparrow,\mathbf{K}+\mathbf{k}+\gamma_e(\mathbf{k}_1+\mathbf{g})}u_{v\uparrow,\mathbf{K}+\mathbf{q}} \middle| W \middle| u_{c\uparrow,\mathbf{K}+\mathbf{k}'+\gamma_e(\mathbf{k}_1+\mathbf{g}')}u_{v\uparrow,\mathbf{K}+\mathbf{q}'} \right\rangle = \frac{1}{A}[\overline{W}^L_{eh}(\Delta\mathbf{g}) + \overline{W}^S_{eh}]\delta_{\Delta\mathbf{q},-\Delta\mathbf{g}} \quad (50)$$

Here the long-range term $\overline{W}^L_{eh}$ and the short-range term $\overline{W}^S_{eh}$ can be obtained from Eq. (36) and (38), respectively, by replacing $u_{c\uparrow,\mathbf{K}'}$ there with $u_{v\uparrow,\mathbf{K}}$ (for the Wigner crystal) and replacing $u_{c\downarrow,\mathbf{K}'}$ there with $u_{c\uparrow,\mathbf{K}}$ (for the photoexcited electron).

The $V_{ex}$ term in Eq. (49) is:

$$\left\langle u_{c\uparrow,\mathbf{K}+\mathbf{k}+\gamma_e(\mathbf{k}_1+\mathbf{g})}u_{\nu\uparrow,\mathbf{K}+\mathbf{q}}\left|V_{ex}\right|u_{c\uparrow,\mathbf{K}+\mathbf{k}'+\gamma_e(\mathbf{k}_1+\mathbf{g}')}u_{\nu\uparrow,\mathbf{K}+\mathbf{q}'}\right\rangle$$

$$=\frac{1}{\Omega}\sum_{\mathbf{G}}\frac{e^2}{\kappa(\mathbf{G}-\mathbf{\Delta}_e)\varepsilon_0|\mathbf{G}-\mathbf{\Delta}_e|^2}\left\langle u_{c\uparrow,\mathbf{K}+\mathbf{k}+\gamma_e(\mathbf{k}_1+\mathbf{g})}\left|e^{i(\mathbf{G}-\mathbf{\Delta}_e)\cdot\mathbf{r}}\right|u_{\nu\uparrow,\mathbf{K}+\mathbf{q}'}\right\rangle\left\langle u_{\nu\uparrow,\mathbf{K}+\mathbf{q}}\left|e^{-i(\mathbf{G}-\mathbf{\Delta}_e)\cdot\mathbf{r}}\right|u_{c\uparrow,\mathbf{K}+\mathbf{k}'+\gamma_e(\mathbf{k}_1+\mathbf{g}')}\right\rangle$$

$$=\frac{1}{\Omega}\sum_{\mathbf{G}}\frac{e^2}{\kappa(\mathbf{G}-\mathbf{\Delta}_e)\varepsilon_0|\mathbf{G}-\mathbf{\Delta}_e|^2}\left\langle u_{c\uparrow,\mathbf{K}+\mathbf{q}'-\mathbf{\Delta}_e}\left|e^{i(\mathbf{G}-\mathbf{\Delta}_e)\cdot\mathbf{r}}\right|u_{\nu\uparrow,\mathbf{K}+\mathbf{q}'}\right\rangle\left\langle u_{\nu\uparrow,\mathbf{K}+\mathbf{q}}\left|e^{-i(\mathbf{G}-\mathbf{\Delta}_e)\cdot\mathbf{r}}\right|u_{c\uparrow,\mathbf{K}+\mathbf{q}-\mathbf{\Delta}_e}\right\rangle \quad (51)$$

Here $\mathbf{\Delta}_e = \mathbf{q}' - \mathbf{k} - \gamma_e(\mathbf{k}_1 + \mathbf{g}) = \mathbf{q} - \mathbf{k}' - \gamma_e(\mathbf{k}_1 + \mathbf{g}')$ is the momentum difference between the photo-generated electron and the WC hole in the scattering processes.

In the summation over different $\mathbf{G}$ in Eq. (51), the $\mathbf{G}_\parallel = \mathbf{0}$ term (the long-range contribution) is:

$$\frac{1}{AL_c}\sum_{G_\perp}\frac{e^2}{\kappa(\mathbf{G}_\perp-\mathbf{\Delta}_e)\varepsilon_0|\mathbf{G}_\perp-\mathbf{\Delta}_e|^2}\left\langle u_{c\uparrow,\mathbf{K}+\mathbf{q}'-\mathbf{\Delta}_e}\left|e^{i(\mathbf{G}_\perp-\mathbf{\Delta}_e)\cdot\mathbf{r}}\right|u_{\nu\uparrow,\mathbf{K}+\mathbf{q}'}\right\rangle\left\langle u_{\nu\uparrow,\mathbf{K}+\mathbf{q}}\left|e^{-i(\mathbf{G}_\perp-\mathbf{\Delta}_e)\cdot\mathbf{r}}\right|u_{c\uparrow,\mathbf{K}+\mathbf{q}-\mathbf{\Delta}_e}\right\rangle$$

$$\approx\frac{1}{\Omega}\sum_{G_\perp}\frac{e^2}{\kappa(\mathbf{0})\varepsilon_0|\mathbf{G}_\perp-\mathbf{\Delta}_e|^2}\left|\left\langle u_{c\uparrow,\mathbf{K}-\mathbf{\Delta}_e}\left|e^{i(\mathbf{G}_\perp-\mathbf{\Delta}_e)\cdot\mathbf{r}}\right|u_{\nu\uparrow,\mathbf{K}}\right\rangle\right|^2\equiv\frac{1}{A}\lambda_{eh}^0(\mathbf{\Delta}_e). \quad (52)$$

Here $L_c$ is the length of the supercell along the $z$ axis. The above equation neglects the dependence on $\mathbf{q}$ and $\mathbf{q}'$ because the result changes little ($\sim$1%) with $\mathbf{q}$ and $\mathbf{q}'$ over the reciprocal-space spread of the WC hole wavefunction [$i.e.$ $\sqrt{\beta}$ as defined in Eq. (13), which is $\sim$1% of the WSe$_2$ Brillouin zone]. We have also approximated $\kappa(\mathbf{G}_\perp - \mathbf{\Delta}_e)$ by $\kappa(0)$, since the dominant contributions are provided by small $\mathbf{G}_\perp - \mathbf{\Delta}_e$ here. The calculated results of $\lambda_{eh}^0(\mathbf{\Delta}_e)$ are shown in Fig. S8. Since $\lambda_{eh}^0(\mathbf{\Delta}_e)$ is a smooth function of $|\mathbf{\Delta}_e|$ over the range covered by the exciton wavefunction in the reciprocal space, we can approximate $\lambda_{eh}^0(\mathbf{\Delta}_e)$ by its average weighted by the probability distribution of exciton internal motion as given by $\bar{\lambda}_{eh}^0 = \int d\mathbf{k}|\tilde{\varphi}_0(\mathbf{k})|^2 \lambda_{eh}^0(\mathbf{k})$.

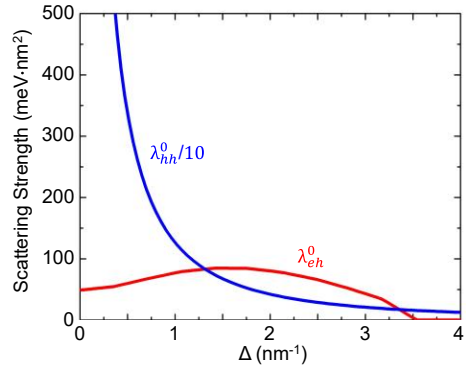

**Figure S8**. Interaction strength $\lambda_{eh}^0(\mathbf{\Delta})$ for the electron-hole exchange interaction and $\lambda_{hh}^0(\mathbf{\Delta})$ for the hole-hole exchange interaction in Case 4. We divide $\lambda_{hh}^0$ by a factor of 10 for clarity.

On the other hand, the $\mathbf{G}_\parallel \neq \mathbf{0}$ terms (the short-range contribution) in Eq. (51), are:

$$\frac{1}{\Omega}\sum_{\mathbf{G}_\parallel \neq 0, G_\perp} \frac{e^2}{\kappa(\mathbf{G}-\Delta_e)\varepsilon_0|\mathbf{G}-\Delta_e|^2} \left\langle u_{c\uparrow,\mathbf{K}+\mathbf{q}'-\Delta_e}\left|e^{i(\mathbf{G}-\Delta_e)\cdot\mathbf{r}}\right|u_{v\uparrow,\mathbf{K}+\mathbf{q}'}\right\rangle \left\langle u_{v\uparrow,\mathbf{K}+\mathbf{q}}\left|e^{-i(\mathbf{G}-\Delta_e)\cdot\mathbf{r}}\right|u_{c\uparrow,\mathbf{K}+\mathbf{q}-\Delta_e}\right\rangle$$

$$\approx \frac{1}{\Omega}\sum_{\mathbf{G}_\parallel \neq 0, G_\perp} \frac{e^2}{\varepsilon_0|\mathbf{G}|^2} \left|\left\langle u_{c\uparrow,\mathbf{K}}\left|e^{i\mathbf{G}\cdot\mathbf{r}}\right|u_{v\uparrow,\mathbf{K}}\right\rangle\right|^2 \equiv \frac{1}{A}\lambda'_{eh} \tag{53}$$

Here we neglect the dependence on the small wave vectors (e.g. $\mathbf{k}$, $\mathbf{k}'$, $\mathbf{g}$, $\mathbf{g}'$) because the $\mathbf{G}_\parallel \neq \mathbf{0}$ terms describe short-range interaction with large momentum transfer. Moreover, $\kappa(\mathbf{G}-\Delta_e)$ converges to 1 when the wave vector exceeds the Brillouin zone. Consequently, we can approximate the sum as a constant $\lambda'_{eh}$. By combining the $\mathbf{G}_\parallel = \mathbf{0}$ and $\mathbf{G}_\parallel \neq \mathbf{0}$ terms, the net exchange matrix element is $\lambda_{eh} = \lambda^0_{eh}(\Delta_e) + \lambda'_{eh} \approx \bar{\lambda}^0_{eh} + \lambda'_{eh}$.

For the scattering between the K-valley photoexcited hole and a K-valley WC hole, the matrix elements are:

$$\left\langle u_{v\uparrow,\mathbf{K}+\mathbf{k}-\gamma_h(\mathbf{k}_1+\mathbf{g})}u_{v\uparrow,\mathbf{K}+\mathbf{q}}\left| v(\mathbf{r}_h-\mathbf{r})\right|u_{v\uparrow,\mathbf{K}+\mathbf{k}'-\gamma_h(\mathbf{k}_1+\mathbf{g}')}u_{v\uparrow,\mathbf{K}+\mathbf{q}'}\right\rangle_{\mathcal{A}}$$

$$= \left\langle u_{v\uparrow,\mathbf{K}+\mathbf{k}-\gamma_h(\mathbf{k}_1+\mathbf{g})}u_{v\uparrow,\mathbf{K}+\mathbf{q}}\left| W\right|u_{v\uparrow,\mathbf{K}+\mathbf{k}'-\gamma_h(\mathbf{k}_1+\mathbf{g}')}u_{v\uparrow,\mathbf{K}+\mathbf{q}'}\right\rangle$$

$$- \left\langle u_{v\uparrow,\mathbf{K}+\mathbf{k}-\gamma_h(\mathbf{k}_1+\mathbf{g})}u_{v\uparrow,\mathbf{K}+\mathbf{q}}\left| V_{ex}\right|u_{v\uparrow,\mathbf{K}+\mathbf{k}'-\gamma_h(\mathbf{k}_1+\mathbf{g}')}u_{v\uparrow,\mathbf{K}+\mathbf{q}'}\right\rangle \tag{54}$$

Similar to Eq. (50), the direct Coulomb scattering term is obtained as:

$$\left\langle u_{v\uparrow,\mathbf{K}+\mathbf{k}-\gamma_h(\mathbf{k}_1+\mathbf{g})}u_{v\uparrow,\mathbf{K}+\mathbf{q}}\left| W\right|u_{v\uparrow,\mathbf{K}+\mathbf{k}'-\gamma_h(\mathbf{k}_1+\mathbf{g}')}u_{v\uparrow,\mathbf{K}+\mathbf{q}'}\right\rangle = \frac{1}{A}[\bar{W}^L_{hh}(\Delta\mathbf{g}) + \bar{W}^S_{hh}]\delta_{\Delta\mathbf{q},-\Delta\mathbf{g}} \tag{55}$$

Here the long-range term $\bar{W}^L_{hh}$ and the short-range term $\bar{W}^S_{hh}$ can be obtained from Eq. (36) and (38), respectively, by replacing $u_{c\uparrow,\mathbf{K}'}$ there with $u_{v\uparrow,\mathbf{K}}$ (for the Wigner crystal) and replacing $u_{c\downarrow,\mathbf{K}'}$ there with $u_{v\uparrow,\mathbf{K}}$ (for the photoexcited hole).

Similar to Eq. (51), the $V_{ex}$ term (for hole-hole exchange) is:

$$\left\langle u_{v\uparrow,\mathbf{K}+\mathbf{k}-\gamma_h(\mathbf{k}_1+\mathbf{g})}u_{v\uparrow,\mathbf{K}+\mathbf{q}}\left| V_{ex}\right|u_{v\uparrow,\mathbf{K}+\mathbf{k}'-\gamma_h(\mathbf{k}_1+\mathbf{g}')}u_{v\uparrow,\mathbf{K}+\mathbf{q}'}\right\rangle$$

$$= \frac{1}{\Omega}\sum_{\mathbf{G}} \frac{e^2}{\varepsilon_0\kappa(\mathbf{G}-\Delta_h)|\mathbf{G}-\Delta_h|^2} \left\langle u_{v\uparrow,\mathbf{K}+\mathbf{k}-\gamma_h(\mathbf{k}_1+\mathbf{g})}\left|e^{i(\mathbf{G}-\Delta_h)\cdot\mathbf{r}}\right|u_{v\uparrow,\mathbf{K}+\mathbf{q}'}\right\rangle \left\langle u_{v\uparrow,\mathbf{K}+\mathbf{q}}\left|e^{-i(\mathbf{G}-\Delta_h)\cdot\mathbf{r}}\right|u_{v\uparrow,\mathbf{K}+\mathbf{k}'-\gamma_h(\mathbf{k}_1+\mathbf{g}')}\right\rangle \tag{56}$$

Here $\Delta_h = \mathbf{q}' - \mathbf{k} + \gamma_h(\mathbf{k}_1+\mathbf{g}) = \mathbf{q} - \mathbf{k}' + \gamma_h(\mathbf{k}_1+\mathbf{g}')$ is the momentum difference between the photogenerated hole and the WC hole in the scattering processes.

In the summation of Eq. (56), the $\mathbf{G}_\parallel = \mathbf{0}$ term (the long-range contribution) is:

$$\frac{1}{A}\lambda_{hh}^0(\mathbf{\Delta}_h)$$

$$= \frac{1}{\Omega}\sum_{\mathbf{G}_\perp}\frac{e^2}{\varepsilon_0\kappa(\mathbf{G}_\perp - \mathbf{\Delta}_h)|\mathbf{G}_\perp - \mathbf{\Delta}_h|^2}\left\langle u_{v\uparrow,\mathbf{K}+\mathbf{k}-\gamma_h(\mathbf{k}_1+\mathbf{g})}\left|e^{i(\mathbf{G}_\perp-\mathbf{\Delta}_h)\cdot\mathbf{r}}\right|u_{v\uparrow,\mathbf{K}+\mathbf{q}'}\right\rangle\left\langle u_{v\uparrow,\mathbf{K}+\mathbf{q}}\left|e^{-i(\mathbf{G}_\perp-\mathbf{\Delta}_h)\cdot\mathbf{r}}\right|u_{v\uparrow,\mathbf{K}+\mathbf{k}'-\gamma_h(\mathbf{k}_1+\mathbf{g}')}\right\rangle \quad (57)$$

Remarkably, although $\lambda_{hh}^0$ is an exchange term, it has the same form as Eq. (36) for the long-range direct Coulomb scattering, except that here we are dealing with the momentum difference ($\mathbf{\Delta}_h$) between a photo-generated hole and a WC hole, but in Eq. (36) we deal with the momentum difference ($\mathbf{\Delta g}$) between two WC particles (or two photo-generated carriers). We numerically calculate $\lambda_{hh}^0$ with the approximation that $\kappa(\mathbf{G}_\perp - \mathbf{\Delta}_h) \approx \kappa(\mathbf{0})$. The numerical result of $\lambda_{hh}^0$ exhibits a Keldysh form. Since this calculation is just for free-standing WSe$_2$, to more accurately account for the dielectric screening in our BN-encapsulated sample, we set $\lambda_{hh}^0$ as a Keldysh potential with $\kappa$ and $\rho_0$ suitable for our sample. The comparison of $\lambda_{eh}^0(\mathbf{\Delta}_e)$ and $\lambda_{hh}^0(\mathbf{\Delta}_h)$ is shown in Fig. S8.

It is remarkable to note that Eq. (57) for hole-hole exchange produces much stronger scattering than Eq. (52) for electron-hole exchange. For illustration, let's compare the $\mathbf{G}_\perp = \mathbf{0}$ term in both cases. In Eq. (52), the $\mathbf{G}_\perp = \mathbf{0}$ term is:

$$\frac{1}{\Omega}\frac{e^2}{\kappa(\mathbf{0})\varepsilon_0|\mathbf{\Delta}_e|^2}\left|\left\langle u_{c\uparrow,\mathbf{K}-\mathbf{\Delta}_e}\left|e^{-i\mathbf{\Delta}_e\cdot\mathbf{r}}\right|u_{v\uparrow,\mathbf{K}}\right\rangle\right|^2$$

$$\approx \frac{1}{\Omega}\frac{e^2}{\kappa(\mathbf{0})\varepsilon_0|\mathbf{\Delta}_e|^2}\left|\left\langle u_{c\uparrow,\mathbf{K}}\left|1\right|u_{v\uparrow,\mathbf{K}}\right\rangle - i\mathbf{\Delta}_e\cdot\left\langle u_{c\uparrow,\mathbf{K}}\left|\mathbf{r}\right|u_{v\uparrow,\mathbf{K}}\right\rangle\right|^2$$

$$= \frac{1}{\Omega}\frac{e^2}{\kappa(\mathbf{0})\varepsilon_0}\left|\left\langle u_{c\uparrow,\mathbf{K}}\left|\mathbf{r}\right|u_{v\uparrow,\mathbf{K}}\right\rangle\right|^2 = \frac{1}{\Omega}\frac{e^2}{\kappa(\mathbf{0})\varepsilon_0}\left|\frac{\hbar}{m_0}\frac{\left\langle u_{c\uparrow,\mathbf{K}}\left|\mathbf{p}\right|u_{v\uparrow,\mathbf{K}}\right\rangle}{E_{c\uparrow}(\mathbf{K})-E_{v\uparrow}(\mathbf{K})}\right|^2. \quad (58)$$

Here we have applied the Taylor expansion of the inner product with respect to $\mathbf{\Delta}_e$ and used the commutator relation $[\mathbf{r}, H] = \frac{i\hbar}{m_0}\mathbf{p}$. The results show two key features. First, the inner product $\left\langle u_{c\uparrow,\mathbf{K}}\left|1\right|u_{v\uparrow,\mathbf{K}}\right\rangle$ is zero because the two states (in different bands) are orthogonal to each other. The next-order $\mathbf{\Delta}_e^2$ term cancels the $1/|\mathbf{\Delta}_e|^2$ factor. Consequently, the result becomes non-singular with respect to $\mathbf{\Delta}_e$; That means, the Fourier transform into the real space will not give a long-range tail. Therefore, the scattering is not long-range. In contrast, the $\mathbf{G}_\perp = \mathbf{0}$ term in Eq. (57) is:

$$\frac{1}{\Omega}\frac{e^2}{\varepsilon_0\kappa(\mathbf{\Delta}_h)|\mathbf{\Delta}_h|^2}\left\langle u_{v\uparrow,\mathbf{K}+\mathbf{q}'-\mathbf{\Delta}_h}\left|e^{-i\mathbf{\Delta}_h\cdot\mathbf{r}}\right|u_{v\uparrow,\mathbf{K}+\mathbf{q}'}\right\rangle\left\langle u_{v\uparrow,\mathbf{K}+\mathbf{q}}\left|e^{i\mathbf{\Delta}_h\cdot\mathbf{r}}\right|u_{v\uparrow,\mathbf{K}+\mathbf{q}-\mathbf{\Delta}_h}\right\rangle \quad (59)$$

In the limit of $\mathbf{\Delta}_h \to \mathbf{0}$, the inner products above approach one, and the result becomes:

$$\frac{1}{\Omega}\frac{e^2}{\varepsilon_0\kappa(\Delta_h)|\Delta_h|^2}\left\langle u_{\nu\uparrow,\mathbf{K}+\mathbf{q}'}\Big|1\Big|u_{\nu\uparrow,\mathbf{K}+\mathbf{q}'}\right\rangle\left\langle u_{\nu\uparrow,\mathbf{K}+\mathbf{q}}\Big|1\Big|u_{\nu\uparrow,\mathbf{K}+\mathbf{q}}\right\rangle = \frac{1}{\Omega}\frac{e^2}{\varepsilon_0\kappa(\Delta_h)|\Delta_h|^2} \tag{60}$$

Unlike in Eq. (58), the inner products in Eq. (60) are 1 because the two states (in the same band) are identical. Consequently, the $1/|\Delta_h|^2$ factor remains, and the result becomes singular with respect to $\Delta_h$. That means, the Fourier transform into real space will give a long-range tail. Therefore, the scattering is long-range.

From our illustration above, we can perceive that Eq. (57) for hole-hole exchange describes long-range scattering, whereas Eq. (52) for electron-hole exchange does not. The former therefore contribute much more significantly to the scattering strength than the latter.

After discussing the $\mathbf{G}_\parallel = 0$ term (long-range contribution) in Eq. (56), we now consider the $\mathbf{G}_\parallel \neq 0$ terms (*i.e.* the short-range contribution) in Eq. (56), which are:

$$\frac{1}{\Omega}\sum\nolimits_{\mathbf{G}_\parallel\neq0,G_\perp}\frac{e^2}{\varepsilon_0\kappa(\mathbf{G}-\Delta_h)|\mathbf{G}-\Delta_h|^2}\left\langle u_{\nu\uparrow,\mathbf{K}+\mathbf{k}\,-\gamma_h(\mathbf{k}_1+\mathbf{g})}\Big|e^{\mathrm{i}(\mathbf{G}-\Delta_h)\cdot\mathbf{r}}\Big|u_{\nu\uparrow,\mathbf{K}+\mathbf{q}'}\right\rangle\left\langle u_{\nu\uparrow,\mathbf{K}+\mathbf{q}}\Big|e^{-\mathrm{i}(\mathbf{G}-\Delta_h)\cdot\mathbf{r}}\Big|u_{\nu\uparrow,\mathbf{K}+\mathbf{k}'-\gamma_h(\mathbf{k}_1+\mathbf{g}')}\right\rangle$$

$$\approx \frac{1}{\Omega}\sum\nolimits_{\mathbf{G}\neq0,G_\perp}\frac{e^2}{\varepsilon_0|\mathbf{G}|^2}\left|\left\langle u_{\nu\uparrow,\mathbf{K}}|e^{\mathrm{i}\mathbf{G}\cdot\mathbf{r}}|u_{\nu\uparrow,\mathbf{K}}\right\rangle\right|^2 \equiv \frac{1}{A}\lambda'_{hh}. \tag{61}$$

As the $\mathbf{G}_\parallel \neq 0$ (short-range) terms involve large momentum transfer, we can neglect the dependence on the small wave vectors (*e.g.* $\mathbf{k}$, $\mathbf{k}'$, $\mathbf{k}_1$, $\mathbf{g}$, $\mathbf{g}'$) and also approximate $\kappa(\mathbf{G}-\Delta_h)$ as 1 because $\kappa$ converges to 1 when the momentum exceeds the Brillouin zone. Consequently, the short-range contribution in $V_{ex}$ can be approximated by a constant $\lambda'_{hh}$ as shown in Eq. (61).

In summary, the matrix elements between an exciton and the hole Wigner crystal are:

$$\left\langle \mathrm{WC};\mathrm{X}^{\mathrm{K}}_{\mathbf{k}_1+\mathbf{g}}\Big|v^X_{wc}\Big|\mathrm{WC};\mathrm{X}^{\mathrm{K}}_{\mathbf{k}_1+\mathbf{g}'}\right\rangle$$

$$= \frac{1}{A}\sum\nolimits_{\mathbf{k},\mathbf{k}',\mathbf{q},\mathbf{q}',j}\tilde{\varphi}_0(\mathbf{k})\tilde{\varphi}_0(\mathbf{k}')\tilde{\psi}_{wc}(\mathbf{q})\tilde{\psi}_{wc}(\mathbf{q}')e^{\mathrm{i}(\mathbf{q}-\mathbf{q}')\cdot\mathbf{R}_j} \times \big\{-[\bar{W}^L_{eh}(\Delta\mathbf{g})+\bar{W}^S_{eh}-\lambda_{eh}]\,\delta_{\mathbf{k}-\gamma_h\mathbf{g},\;\mathbf{k}'-\gamma_h\mathbf{g}'} + [\bar{W}^L_{hh}(\Delta\mathbf{g})+\bar{W}^S_{hh}-\lambda^0_{hh}(\Delta_h)-\lambda'_{hh}]\delta_{\mathbf{k}+\gamma_e\mathbf{g},\;\mathbf{k}'+\gamma_e\mathbf{g}'}\big\}\delta_{\Delta\mathbf{q},-\Delta\mathbf{g}}. \tag{62}$$

Similar to the Cases 1-3, here the long-range direct Coulomb terms $\bar{W}^L_{eh}(\Delta\mathbf{g})$ and $\bar{W}^L_{hh}(\Delta\mathbf{g})$ are similar, and they only give a small net contribution to the exciton-WC scattering. But unlike Cases 1-3, here there is a long-range exchange term $\lambda^0_{hh}$ that gives more contribution than all the other terms combined. Therefore, the exciton-WC scattering in Case 4 is much stronger than Cases 1-3, consistent with the experimental results in Figs. 3A and B of the main paper.

The essential message here is that when one carrier in the exciton occupies the same band as the Wigner crystal, the Umklapp scattering will be greatly enhanced because of the existence of long-range exchange interaction. In monolayer WSe$_2$, such enhancement occurs only on the hole side and not on the electron side because the bright exciton occupies the upper conduction band. However, the electron-side enhancement is expected to occur in other transition metal dichalcogenides (e.g. monolayer MoSe$_2$) where the bright exciton occupies the lower conduction band.

We note that all of the long-range terms $W_{\mu\nu}^L(\mathbf{q})$ ($\mu\nu = eh, hh, ee, he$) calculated according to Eqs. (36), (39), and (42) can be well approximated by Keldysh potentials with the same effective dielectric function $\kappa(\mathbf{0})$ but different $\rho_0$ parameters, as expressed in Eq. (1). As K and K' valleys are related by the time-reversal symmetry, we have $\overline{W}_{\mu\nu}^L(\mathbf{q}) = W_{\mu\nu}^L(\mathbf{q})$. The effective electron-hole interaction for the bright KK exciton in BN-encapsulated monolayer WSe$_2$ has been well described by a Keldysh potential of the form $v_d(q) = \frac{e^2}{2A\kappa(0)\varepsilon_0 q}\frac{1}{(1+\rho_0 q)}$ with $\rho_0 = 1.1\ nm$ and $\kappa(0) \approx 4$ (*S9*). Here, we assume that $\overline{W}_{eh}^L(\mathbf{q}) = \tilde{v}_d(\mathbf{q})$ for BN-encapsulated WSe$_2$ and calculate the ratios among different long-range interaction terms according to the DFT calculation for monolayer WSe$_2$. It is convenient to define $W_{\mu\nu}^L(\mathbf{q}) = R_{\mu\nu}(q)\overline{W}_{eh}^L(\mathbf{q})$, where $R_{\mu\nu}(q)$ ($\mu\nu = eh, hh, ee, he$) are the correction functions that account for the different charge distributions of different bands as shown in Fig. S7. The calculated results for $R_{ee}(q)$, $R_{hh}(q)$, and $R_{he}(q)$ are shown in Fig. S9.

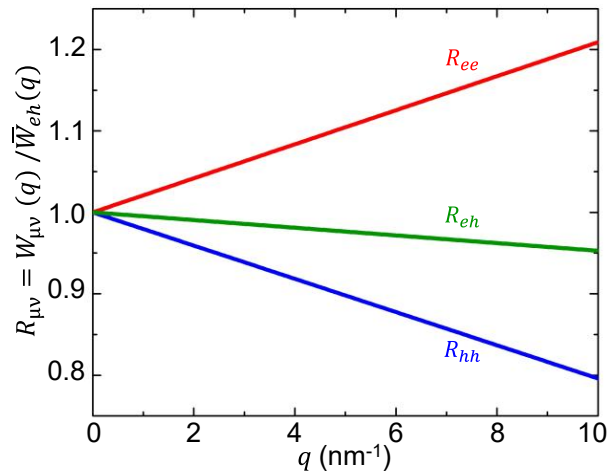

**Figure S9**. The ratios $R_{\mu\nu}(q) = W_{\mu\nu}^L(\mathbf{q})/\overline{W}_{eh}^L(\mathbf{q})$ with $\mu\nu = ee, hh, he$ for the long-range Coulomb interaction. They can be well fitted by straight lines for wave vectors q within the range of interest.

## 4. Numerical Results for Scattering Strengths

By using the exciton wavefunction in Eq. (11) and the localized-electron wavefunction in the Wigner crystal in Eq. (13), we can calculate the matrix element in Eq. (41) for Case 1 as:

$$\left\langle \text{WC}; X_{\mathbf{k}_1+\mathbf{g}}^{K'} \middle| v_{wc}^X \middle| \text{WC}; X_{\mathbf{k}_1+\mathbf{g}'}^{K'} \right\rangle \approx \frac{A}{A_{wc}} \tilde{v}_d\left(\Delta\mathbf{g}\right)\tilde{\eta}(\Delta\mathbf{g})[R_{ee}(\Delta\mathbf{g})\zeta_0(\gamma_h\Delta\mathbf{g}) - R_{he}(\Delta\mathbf{g})\zeta_0(\gamma_e\Delta\mathbf{g})]$$

$$+ \frac{1}{A_{wc}}\tilde{\eta}(\Delta\mathbf{g})[\, W_{ee}^S\zeta_0(\gamma_h\Delta\mathbf{g}) - W_{he}^S\zeta_0(\gamma_e\Delta\mathbf{g})] \tag{63}$$

Here $A$ is the area of the monolayer WSe$_2$ sample; $A_{wc}$ is the area of the WC unit cell; $\tilde{v}_d\left(\Delta\mathbf{g}\right)$ is the Keldysh potential as defined in Eq. (1); $\tilde{\eta}(\Delta\mathbf{g})$ is the Fourier transform of the charge density distribution $\eta(\mathbf{r})$ on one WC site as defined in Eq. (12); $\zeta_0(\mathbf{q})$ is the Fourier transform of the particle density for the exciton internal motion, as defined below

$$\zeta_0(\mathbf{q}) = \int d\mathbf{r}\, \varphi_0^*(\mathbf{r})\varphi_0(\mathbf{r})e^{-i\mathbf{q}\cdot\mathbf{r}} \tag{64}$$

Similarly, we obtain the matrix element in Eq. (42) for Case 2 as

$$\left\langle \text{WC}; X_{\mathbf{k}_1+\mathbf{g}}^{K'} \middle| v_{wc}^X \middle| \text{WC}; X_{\mathbf{k}_1+\mathbf{g}'}^{K'} \right\rangle \approx -\frac{A}{A_{wc}} \tilde{v}_d\left(\Delta\mathbf{g}\right)\tilde{\eta}(\Delta\mathbf{g})[\zeta_0(\gamma_h\Delta\mathbf{g}) - R_{hh}(\Delta\mathbf{g})\zeta_0(\gamma_e\Delta\mathbf{g})]$$

$$-\tilde{\eta}(\Delta\mathbf{g})[\, W_{eh}^S\zeta_0(\gamma_h\Delta\mathbf{g}) - W_{hh}^S\zeta_0(\gamma_e\Delta\mathbf{g})] \tag{65}$$

and we obtain the matrix element in Eq. (47) for Case 3 as:

$$\left\langle \text{WC}; X_{\mathbf{k}_1+\mathbf{g}}^{K} \middle| v_{wc}^X \middle| \text{WC}; X_{\mathbf{k}_1+\mathbf{g}'}^{K} \right\rangle \approx \frac{A}{A_{wc}} \tilde{v}_d\left(\Delta\mathbf{g}\right)\tilde{\eta}(\Delta\mathbf{g})[R_{ee}(\Delta\mathbf{g})\zeta_0(\gamma_h\Delta\mathbf{g}) - R_{he}(\Delta\mathbf{g})\zeta_0(\gamma_e\Delta\mathbf{g})]$$

$$+ \frac{1}{A_{wc}}\tilde{\eta}(\Delta\mathbf{g})[\, (\overline{W}_{ee}^S - \lambda_{ee})\zeta_0(\gamma_h\Delta\mathbf{g}) - (\overline{W}_{he}^S - \lambda_{he})\zeta_0(\gamma_e\Delta\mathbf{g})] \tag{66}$$

Finally, we obtain the matrix element in Eq. (62) for Case 4 as:

$$\left\langle \text{WC}; X_{\mathbf{k}_1+\mathbf{g}}^{K} \middle| v_{wc}^X \middle| \text{WC}; X_{\mathbf{k}_1+\mathbf{g}'}^{K} \right\rangle \approx -\frac{A}{A_{wc}} \tilde{v}_d\left(\Delta\mathbf{g}\right)\tilde{\eta}(\Delta\mathbf{g})[\zeta_0(\gamma_h\Delta\mathbf{g}) - R_{hh}(\Delta\mathbf{g})\zeta_0(\gamma_e\Delta\mathbf{g})]$$

$$+ \frac{1}{A_{wc}}\tilde{\eta}(\Delta\mathbf{g})[(\overline{W}_{eh}^S - \lambda_{eh})\zeta_0(\gamma_h\Delta\mathbf{g}) - (\overline{W}_{hh}^S - \lambda'_{hh})\zeta_0(\gamma_e\Delta\mathbf{g})] - V_{hh}^{ex}(\mathbf{g}, \mathbf{g}') \tag{67}$$

Here

$$V_{hh}^{ex}(\mathbf{g}, \mathbf{g}') = \frac{1}{A_{wc}}\sum_{\mathbf{q}',\mathbf{k}} \tilde{\varphi}_0^*(\mathbf{k})\tilde{\psi}_{wc}(\mathbf{q}')\tilde{\psi}_{wc}^*(\mathbf{q}' - \Delta\mathbf{g})\, \tilde{\varphi}_0(\mathbf{k} - \gamma_e\Delta\mathbf{g})\lambda_{hh}^0(\Delta_h) \tag{68}$$

denotes the long-range exchange interaction term in the matrix element. To simplify $V_{hh}^{ex}$, we approximate that $\lambda_{hh}^0\big(\Delta_h = \mathbf{q}' - \mathbf{k} + \gamma_h(\mathbf{k}_1 + \mathbf{g})\big) \approx \lambda_{hh}^0(\mathbf{q}' - \mathbf{k})$ because the exciton CM momentum $\mathbf{k}_1$ is zero in the absorption process and the range of the WC reciprocal wave vector $\mathbf{g}$ is one order of magnitude smaller than the range of the exciton internal-motion wave vector **k**. With this approximation, $V_{hh}^{ex}$ becomes a function of $\Delta \mathbf{g}$ only.

We have calculated the short-range terms with the LASTO package based on DFT. The results are listed in Table S1.

| $W_{he}^S = \overline{W}_{he}^S$ | $W_{ee}^S = \overline{W}_{ee}^S$ | $W_{hh}^S = \overline{W}_{hh}^S = \lambda'_{hh}$ | $W_{eh}^S = \overline{W}_{eh}^S$ |
|---|---|---|---|
| 1.5 | 205 | 39.1 | 35.8 |
| $\lambda_{he}$ | $\lambda_{ee}$ | $\overline{\lambda}_{eh}^0$ | $\lambda'_{eh}$ |
| 1.2 | 403 | 62 | 58.1 |

**Table S1.** The calculated scattering strengths (in units of meV $\cdot$ nm$^2$) due to short-range contributions from the direct Coulomb terms ($W_{\mu\nu}^S$, $\overline{W}_{\mu\nu}^S$) and exchange Coulomb terms ($\lambda_{\mu\nu}$) for various types of scattering ($\mu\nu = he, ee, hh, eh$).

The matrix elements in all four cases are real-value functions of $\Delta \mathbf{g}$, which is the discrete momentum transfer to the exciton by the Wigner crystal. The scattering potential $V_{wc}(\Delta \mathbf{g})$ due to the Wigner crystal for the four different cases are given by the matrix elements in Eqs. (63, 65-67). They are the Fourier transform of the WC potential for calculating the miniband structure of the exciton CM states. Fig. S10 displays $V_{wc}(\mathbf{q})$ at continuous momentum transfer $\mathbf{q}$ for two different WC lattice constants $a_{wc} = 28.4 \: and \: 56.8 \: nm$, corresponding to $r_s = 31$ and $62$.

When we compare the results of all four cases in Fig. S10, we find that $V_{wc}(\Delta \mathbf{g})$ is the strongest in Case 4 and the weakest in Case 2, while $V_{wc}(\Delta \mathbf{g})$ in Cases 1 and 3 are stronger than Case 2 by a factor $4 \sim 8$ and weaker than Case 4 by a factor $6 \sim 12$. For each lattice constant, the Wigner crystal has a different optimized on-site Gaussian distribution of charges, which can be obtained by using the semiclassical approach (Model 1) or by extrapolating the DFT-LDA result (Model 2). For the former, the Gaussian standard deviation is $s = 0.16 a_{wc}$ for $a_{wc} = 28.4 \: nm$ and $s = 0.13 a_{wc}$ for $a_{wc} = 56.8 \: nm$; For the latter, $s = 0.3 a_{wc}$ for $a_{wc} = 31 \: nm$ and $s = 0.25 a_{wc}$ for $a_{wc} = 62 \: nm$. Generally, using smaller $s$ will lead to $V_{wc}(\mathbf{q})$ that covers a wider **q** range. The ratios of $V_{wc}(\mathbf{q})$ among the four cases remain almost the same in both approaches.

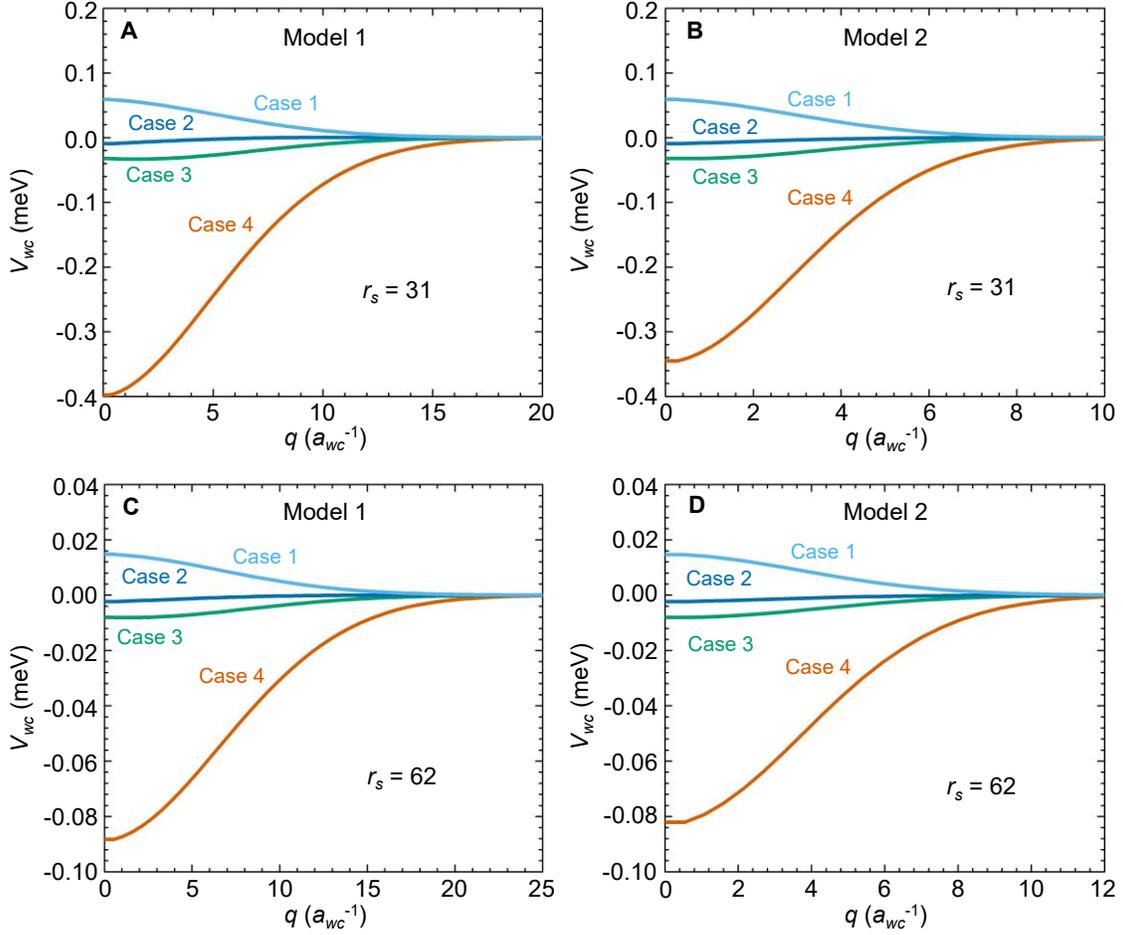

**Figure S10**. Calculated exciton-WC interaction potentials $V_{wc}(q)$ for Wigner crystals of two different lattice constants (A-B) $a_{wc} = 28.4$ nm and (C-D) $a_{wc} = 56.8$ nm, and of different standard deviation of on-site charge distribution, including (A) $s = 0.17\ a_{wc}$ and (C) $s = 0.136\ a_{wc}$ for the semiclassical model (Model 1) as well as (B) $s = 0.30\ a_{wc}$ and (D) $s = 0.255\ a_{wc}$ for the extrapolated DFT-LDA results (Model 2).

As the scattering potential is weak (< 1 meV in Fig. S10), we will use the perturbation theory to calculate the strength of Umklapp scattering. For the exciton CM motion, the basis is defined in Eq. (25). The basis functions for the exciton CM envelope function are plane waves $\Phi_{\mathbf{k_1+g}}^{(0)}(\mathbf{R}_X) = e^{i(\mathbf{k_1+g})\cdot\mathbf{R}_X}$, where $\mathbf{k_1 + g}$ is the CM wave vector and $\mathbf{R}_X$ is the CM coordinate; the corresponding CM kinetic energy is $\frac{\hbar^2|\mathbf{k_1+g}|^2}{2m_X}$. The perturbed wave function to the first order is given by

$$\Phi_{\mathbf{k_1+g}}^{(1)} = \Phi_{\mathbf{k_1+g}}^{(0)} + \frac{2m_X}{\hbar^2} \sum_{\mathbf{g'}} \Phi_{\mathbf{k_1+g'}}^{(0)} \frac{\left\langle WC; X_{\mathbf{k_1+g}}^{\alpha} \middle| v_{wc}^X \middle| WC; X_{\mathbf{k_1+g'}}^{\alpha} \right\rangle}{|\mathbf{k_1+g}|^2 - |\mathbf{k_1+g'}|^2} \quad (\alpha = K, K') \tag{69}$$

We are interested in the case with $\mathbf{k_1} = \mathbf{0}$ and $\mathbf{g'} = \mathbf{0}$ because a visible photon has negligible momentum. In this case, the probability for an exciton CM state at a finite $\mathbf{g}$ to be scattered to the ground state with $\mathbf{k_1} = \mathbf{0}$ and $\mathbf{g'} = \mathbf{0}$ is

$$P(\mathbf{g}) = \left|\frac{2m_X}{\hbar^2}\right|^2 \frac{\left|\left\langle WC; X_{\mathbf{g}}^{\alpha} \middle| v_{wc}^X \middle| WC; X_{\mathbf{g'}}^{\alpha} \right\rangle\right|^2}{|\mathbf{g}|^4} \tag{70}$$

The result shows that $P(\mathbf{g})$ decreases rapidly with $|\mathbf{g}|$.

The Umklapp scattering involves numerous scattering wave vectors. Some of the wave vectors are connected by time-reversal and three-fold rotational symmetry and thus have the same magnitude but different directions (They are called a star in group theory). Generally, each Umklapp peak has contributions from all different stars. But in our case where the scattering potential is much smaller than the separation of the Umklapp peaks, each Umklapp peak is contributed dominantly by just one star. Therefore, for simplicity, we just describe the s-th Umklapp exciton peak with one star, in which we denote an individual vector as $\mathbf{g_s}$ and their collection as $\{\mathbf{g_s}\}$. Different $\mathbf{g_s}$ in the same star give the same probability $P(\mathbf{g_s})$ and the total probablity is $d_s P(\mathbf{g_s})$, where $d_s$ (the degeneracy factor) denotes the number of $\mathbf{g_s}$ in the star.

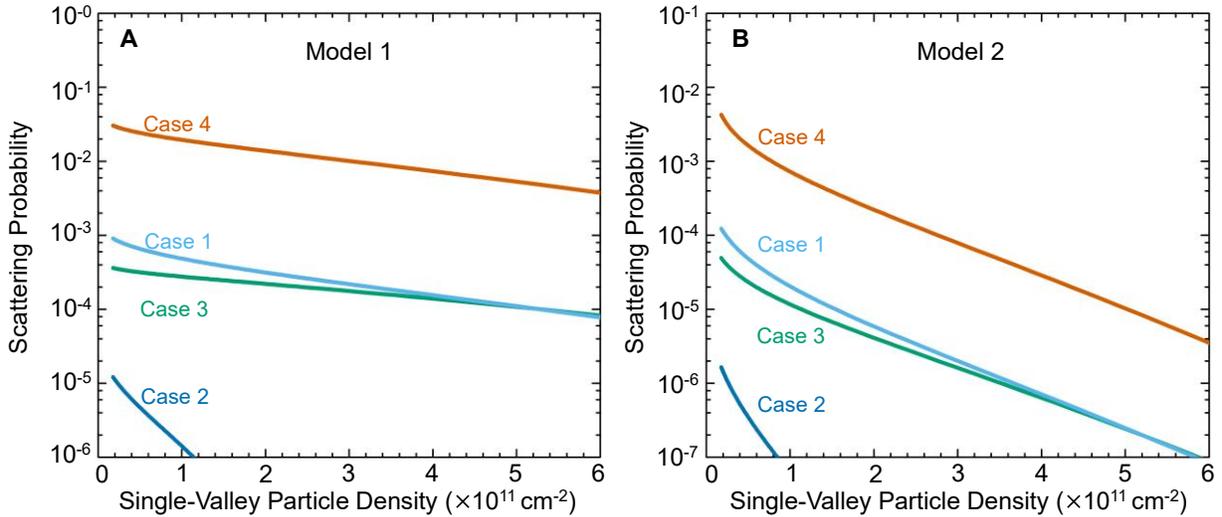

**Figure S11**. Calculated exciton-WC Umklapp scattering probability at varying particle density in monolayer WSe$_2$. (A) Probability obtained by the semiclassical model (Model 1). (B) Probability obtained by using the extrapolated DFT-LDA results (Model 2).

As our experiment only observes the first Umklapp peak, we only need to consider $P(\mathbf{g_1})$ There are total six $\mathbf{g_1}$ in the first star $\{\mathbf{g_1}\}$; So, the degeneracy factor is $d_1 = 6$. Fig. S11 displays our numerical results of $d_1 P(\mathbf{g_1})$ as a function of the particle density $n_{wc} = 1/A_{wc}$ for all four cases, where $A_{wc} = \sqrt{3} a_{wc}^2 / 2$ is the area of the Wigner-crystal primitive cell. The scattering probability obtained by using Model 1 is about 10 times higher than that obtained by Model 2, though both models give similar ratios of probability among the four cases.

## 5. Strength of the excitonic Umklapp peaks

In Eq. (25), $X_{\mathbf{k_1 + g}}^{K}$ represents the exciton eigen states of monolayer WSe$_2$ (with CM momentum $\mathbf{k_1 + g}$) in the absence of the Wigner crystal. When the Wigner crystal is present, the Umklapp scattering will mix excitons with CM momenta differing by a reciprocal vector $\mathbf{g}$ of the Wigner crystal. This will give rise to different exciton minibands with band index $n$ and exciton eigen states $\psi_{n\mathbf{k_1}}^{K}$ as expressed below:

$$|\text{WC}; \psi_{n,\mathbf{k_1}}^{K}\rangle = \sum_{\mathbf{g}} F_{n,\mathbf{k_1}}^{K}(\mathbf{g}) |\text{WC}; X_{\mathbf{k_1 + g}}^{K}\rangle. \tag{71}$$

Here $F_{n\mathbf{k_1}}(\mathbf{g})$ is the expansion coefficient. Acting the Hamiltonian operator $H_{wc}^{X}$ from Eq. (23) on the states given in Eq. (71) yields the following Schrödinger equation:

$$\sum_{\mathbf{g'}} \left\langle \text{WC}; X_{\mathbf{k_1 + g}}^{K} \middle| H_{wc}^{X} \middle| \text{WC}; X_{\mathbf{k_1 + g'}}^{K} \right\rangle F_{n,\mathbf{k_1}}^{K}(\mathbf{g'}) = [E_{wc} + E_0^{X} + E_n^{K}(\mathbf{k_1 + g})] F_{n,\mathbf{k_1}}^{K}(\mathbf{g}) \tag{72}$$

Here the Hamiltonian matrix elements are given in Eq. (28) with details described in four cases afterward; $E_n^{K}(\mathbf{k_1})$ denotes the miniband energy of the KK exciton CM motion in the presence of the Wigner crystal. The expansion coefficient $F_{n,\mathbf{k_1}}^{K}(\mathbf{g'})$ is obtained by solving Eq. (72).

The absorption spectra of the KK exciton are calculated according to Fermi's golden rule:

$$S^{K}(\omega) \propto \sum_{n} \left| \left\langle \text{WC} \middle| \hat{\mathbf{e}} \cdot \mathbf{p} \middle| \text{WC}; \psi_{n,\mathbf{k_1 = 0}}^{K} \right\rangle \right|^2 \delta(\hbar\omega - E_n^{K}(\mathbf{k_1 = 0}))$$

$$= \sum_{n} \left| F_{n,\mathbf{k_1 = 0}}^{K}(\mathbf{g' = 0}) \left\langle \text{WC} \middle| \hat{\mathbf{e}} \cdot \mathbf{p} \middle| \text{WC}; X_{\mathbf{g' = 0}}^{K} \right\rangle \right|^2 \delta(\hbar\omega - E_n^{K}(\mathbf{k_1 = 0})) \tag{73}$$

Here $\hat{\mathbf{e}}$ denotes the polarization of the light; $\mathbf{p}$ is the momentum operator. Only exciton components $X_{\mathbf{k_1 + g'}}^{K}$ with $\mathbf{k_1 + g' = 0}$ contribute to the optical absorption due to the negligible photon momentum of visible light. The formula can be applied to the K′K′ excitons by replacing the superscript K with K′.

The oscillator strength ($f_s$) of the s-th Umklapp exciton peak can be calculated approximately as:

$$f_s \propto d_s P(\mathbf{g_s}) \qquad (74)$$

Here $P(\mathbf{g_s})$ represents the probability of exciton-WC scattering with wave vector $\mathbf{g_s}$; $d_s$ denotes the number of $\mathbf{g_s}$ in the star $\{\mathbf{g_s}\}$. We can calculate it as $P(\mathbf{g_s}) = \left| F_{n,\mathbf{k_1}=\mathbf{0}}^K(\mathbf{g'}=\mathbf{0}) \right|^2$ by solving Eq. (72) for the states (denoted by $n$) corresponding to the star $\{\mathbf{g_s}\}$. Indeed, we have solved Eq. (72) with 140 stars but find that the improvement is insignificant. Therefore, it is sufficient to consider only the first and second stars in our calculation, which correspond to the primary exciton and the lowest Umklapp exciton states, respectively. The result is the same as the perturbation result in Eq. (70). As the oscillator strength is proportional to the scattering probability, they will have the same density-dependent behavior (Fig. S11).

## 6. Mixing of KK and K′K′excitons by the Wigner crystal due to electron-hole exchange

### 6.1. Matrix equation for exciton-WC scattering with intervalley mixing

Generally, an exciton with finite center-of-mass (CM) momentum is the linear combination of all possible electron-hole pairs with the same CM momentum in the Brillouin zone, and we cannot precisely define which valley it belongs to. But when the CM momentum is zero, the exciton can exhibit the rotational symmetry of the system. In this special case, the exciton is formed by the linear combination of electron-hole pairs according to a point-group representation, and it will not be mixed with other excitons with different symmetry representations. In monolayer WSe₂, the optical transitions at the K and K′ points have opposite chiral symmetry. As a result, the states near the K and K′ points can form two excitons with opposite chiral symmetry. As these two excitons are not mixed, we can define them as the KK and K′K′ excitons. We usually only consider such excitons with zero CM momentum in the optical absorption processes.

In the presence of Wigner crystal, however, excitons with finite CM momenta become optically active due to Umklapp scattering processes. Such excitons with finite CM momenta are no longer purely KK or K′K′ excitons because they are the linear combination of electron-hole pairs in both valleys. In another words, the KK and K′K′ electron-hole pairs are mixed in the exciton. Our previous analysis has not considered such mixing effect between the KK and K′K′ electron-hole pairs. Here we will take this effect into account and examine its influence on the absorption spectra.

According to previous studies (*S5, 18*), the coupling between the KK and K′K′ excitons is caused by the long-range electron-hole exchange interaction ($V_{ehx}$) (which is prominent in our Case 4). By using the basis defined in Eq. (25), we can express the $V_{ehx}$ matrix elements as (*S18*):

$$\left\langle \text{WC}; X^{K}_{\mathbf{k}_1+\mathbf{g}} \middle| V_{ehx} \middle| \text{WC}; X^{K'}_{\mathbf{k}_1+\mathbf{g}'} \right\rangle \approx \delta_{\mathbf{g},\mathbf{g}'} \left( \frac{J}{K} \right) \left( \text{g}_x - i\text{g}_y \right)^2 / \text{g} = \delta_{\mathbf{g},\mathbf{g}'} \left( \frac{J}{K} \right) \text{g} e^{-i2\varphi_{\mathbf{g}}} \tag{75}$$

Here $J$ denotes the coupling strength between the KK and K′K′ excitons; $K$ is the magnitude of the wavevector between the $\Gamma$ and K points. From now on we will set $\mathbf{k}_1 = \mathbf{0}$ because we consider the optical absorption (the visible photon has negligible momentum). Therefore, the reciprocal lattice vector $\mathbf{g} = (\text{g}_x, \text{g}_y) = \text{g}(\cos\varphi_{\mathbf{g}}, \sin\varphi_{\mathbf{g}})$ of the Wigner crystal already represents the exciton CM wavevector here. Only excitons with the same CM wavevector $\mathbf{g}$ can be coupled due to the translational symmetry of the Wigner crystal. The long-range electron-hole exchange interaction also introduces a correction term to the energy dispersion of the exciton CM motion in the form (*S5, 18*):

$$\left\langle \text{WC}; X^{K}_{\mathbf{g}} \middle| V_{ehx} \middle| \text{WC}; X^{K}_{\mathbf{g}'} \right\rangle = \left\langle \text{WC}; X^{K'}_{\mathbf{g}} \middle| V_{ehx} \middle| \text{WC}; X^{K'}_{\mathbf{g}'} \right\rangle \approx \delta_{\mathbf{g},\mathbf{g}'} \frac{J}{K} \text{g} \tag{76}$$

By adding the $V_{ehx}$ terms into Eq. (72) and setting $\mathbf{k}_1 = \mathbf{0}$, we revise the Schrödinger equation as:

$$\begin{cases} \sum_{\mathbf{g}'} \left\{ \left\langle \text{WC}; X^{K}_{\mathbf{g}} \middle| H^{X}_{wc} + V_{ehx} \middle| \text{WC}; X^{K}_{\mathbf{g}'} \right\rangle F^{K}_{n,0}(\mathbf{g}') + \left\langle \text{WC}; X^{K}_{\mathbf{g}} \middle| V_{ehx} \middle| \text{WC}; X^{K'}_{\mathbf{g}'} \right\rangle F^{K'}_{n,0}(\mathbf{g}') \right\} = [E_{wc} + E_n(\mathbf{g})] F^{K}_{n,0}(\mathbf{g}) \\ \sum_{\mathbf{g}'} \left\{ \left\langle \text{WC}; X^{K'}_{\mathbf{g}} \middle| H^{X}_{wc} + V_{ehx} \middle| \text{WC}; X^{K'}_{\mathbf{g}'} \right\rangle F^{K'}_{n,0}(\mathbf{g}') + \left\langle \text{WC}; X^{K'}_{\mathbf{g}} \middle| V_{ehx} \middle| \text{WC}; X^{K}_{\mathbf{g}'} \right\rangle F^{K}_{n,0}(\mathbf{g}') \right\} = [E_{wc} + E_n(\mathbf{g})] F^{K'}_{n,0}(\mathbf{g}) \end{cases} \tag{77}$$

Due to the $e^{-i2\varphi_{\mathbf{g}}}$ phase in Eq. (75), we can simplify Eq. (77) by multiplying a phase $e^{i2\varphi}$ on $X^{K'}_{\mathbf{g}}$ (*i.e.* a $\mathbf{g}$-dependent gauge transformation). Eq. (77) will become:

$$\begin{cases} \sum_{\mathbf{g}'} \left\langle \text{WC}; X^{K}_{\mathbf{g}} \middle| H^{X}_{wc} \middle| \text{WC}; X^{K}_{\mathbf{g}'} \right\rangle F^{K}_{n,0}(\mathbf{g}') + \left( \frac{J}{K} \right) |\mathbf{g}| F^{K}_{n,0}(\mathbf{g}) + \left( \frac{J}{K} \right) |\mathbf{g}| F^{K'}_{n,0}(\mathbf{g}) = [E_{wc} + E^{K}_n(\mathbf{g})] F^{K}_{n,0}(\mathbf{g}) \\ \sum_{\mathbf{g}'} \left\langle \text{WC}; X^{K'}_{\mathbf{g}} \middle| H^{X}_{wc} \middle| \text{WC}; X^{K'}_{\mathbf{g}'} \right\rangle F^{K'}_{n,0}(\mathbf{g}') + \left( \frac{J}{K} \right) |\mathbf{g}| F^{K'}_{n,0}(\mathbf{g}) + \left( \frac{J}{K} \right) |\mathbf{g}| F^{K}_{n,0}(\mathbf{g}) = [E_{wc} + E^{K}_n(\mathbf{g})] F^{K'}_{n,0}(\mathbf{g}) \end{cases} \tag{78}$$

Due to the right-handed (R) chiral symmetry of the KK exciton, it is convenient to define symmetrized K′K′ exciton basis states that transform according to the right-handed chiral symmetry:

$$\left| X^{K'}_s \right\rangle_R = \frac{1}{\sqrt{d_s}} \sum_{\mathbf{g}_s} e^{i2\varphi_{\mathbf{g}}} \left| \text{WC}; X^{K'}_{\mathbf{g}_s} \right\rangle. \tag{79}$$

Here the subscript $R$ denotes right-handed chiral symmetry; the subscript s denotes the star of $\mathbf{g}_s$ that represents the exciton CM wave vector; the summation includes all $\mathbf{g}_s$ in the s-th star; $d_s$ is the number of $\mathbf{g}_s$ in the star. We note the state $\left| X^{K'}_{s=0} \right\rangle_R$ does not exist because it is impossible to construct an angular function $e^{i2\varphi_g}$ for $\mathbf{g} = \mathbf{0}$ that converts the left-handed chiral symmetry to the right-handed symmetry.

Similarly, the symmetrized KK exciton basis states that transform according to the right-handed chiral symmetry ($e^{i\varphi_g}$) are defined as

$$|X_s^K\rangle_R = \frac{1}{\sqrt{d_s}}\sum_{\mathbf{g_s}} |\text{WC}; X_{\mathbf{g_s}}^K\rangle. \tag{80}$$

Therefore, we only need to compute the matrix elements between the symmetrized basis states with right-hand chiral symmetry. It turns out that all matrix elements have the same value:

$$_R\langle X_s^K|V_{ehx}|X_s^{K'}\rangle_R = {_R}\langle X_s^K|V_{ehx}|X_s^K\rangle_R = {_R}\langle X_s^{K'}|V_{ehx}|X_s^{K'}\rangle_R \approx J|\mathbf{g_s}|/K \tag{81}$$

Here, we have neglected the effect of defects that can break the point group symmetry.

We define the right-handed coupled exciton state $|\psi_R\rangle$ as a linear combination of $|X_s^K\rangle_R$ with $s = 0, 1$ and $|X_s^{K'}\rangle_R$ with $s = 1$:

$$|\psi_R\rangle = C_{0R}^K|X_{s=0}^K\rangle_R + C_{1R}^K|X_{s=1}^K\rangle_R + C_{1R}^{K'}|X_{s=1}^{K'}\rangle_R \tag{82}$$

Here $C_{0R}^K$, $C_{1R}^K$, $C_{1R}^{K'}$ are the expansion coefficients.

The matrix elements of $H_{wc}^X$ in the symmetrized basis read

$$_R\langle X_s^K|H_{wc}^X|X_{s'}^K\rangle_R = \left(E_0^X + \frac{\hbar^2 \mathbf{g_0^2}}{2m_X}\right)\delta_{s,s'} + \sqrt{\frac{d_s}{d_{s'}}}\sum_{\{\mathbf{g_{s'}}\}}\langle \text{WC}; X_{\mathbf{g_s}}^K|v|\text{WC}; X_{\mathbf{g_{s'}}}^K\rangle \equiv H_{ss'R}^K \tag{83}$$

$$_R\langle X_s^{K'}|H_{wc}^X|X_{s'}^{K'}\rangle_R = \left(E_0^X + \frac{\hbar^2 \mathbf{g_0^2}}{2m_X}\right)\delta_{s,s'} + \sqrt{\frac{d_s}{d_{s'}}}\sum_{\{\mathbf{g_{s'}}\}}\langle \text{WC}; X_{\mathbf{g_s}}^{K'}|v|\text{WC}; X_{\mathbf{g_{s'}}}^{K'}\rangle e^{+i2\left(\varphi_{\mathbf{g_{s'}}} - \varphi_{\mathbf{g_s}}\right)} \equiv H_{ss'R}^{K'} \tag{84}$$

Here s and s′ denote any two stars with wave vector collections $\{\mathbf{g_s}\}$ and $\{\mathbf{g_{s'}}\}$, respectively.

We note that the energy $E_0^X$ of the primary exciton at zero CM momentum varies with the WC particle density $n_{wc}$ due to the screening effect. Therefore, $H_{ss'R}^K$ and $H_{ss'R}^{K'}$ are functions of $n_{wc}$. To account for the screening effect of the ambient carriers, we use a phenomenological density-dependent expression of the exciton energy

$$E_0^X(n_{wc}) = E_g - E_B(n_{wc}) = E_g - E_B(0)/[1 + \chi_X(n_{wc})]. \tag{85}$$

Here $E_g$ is the band gap; $E_B(n_{wc})$ is the density-dependent exciton binding energy; The phenomenological susceptibility $\chi_X(n_{wc})$ is given by

$$\chi_X(n_{wc}) = 1000n_{wc}^2(1 - f_c n_{wc}) \tag{86}$$

Here $n_{wc}$ is in units of $nm^{-2}$. $f_c = 43.5$ (on the hole side) and $22.5$ (on the electron side) is a parameter obtained by fitting the measured density-dependent peak energy of the primary exciton in our experiment.

After addressing the density dependence of $E_0^X$, let's consider Eqs. (83-84). For our current problem, the predominant interaction occurs only between excitonic states in the zeroth star ($s = 0$, including only $\mathbf{g} = 0$) and the first star ($s = 1$, including six $\mathbf{g_1}$ vectors) and also between states in the first star. Therefore, we can limit $s$ and $s'$ to the zeroth and first star in Eqs. (83-84) and simplify the Hamiltonian into a 3×3 matrix. Eq. (78) for the exciton with the right-handed chiral symmetry, on the basis of $\left\{ |X_{s=0}^K\rangle_R; |X_{s=1}^K\rangle_R; |X_{s=1}^{K'}\rangle_R \right\}$, becomes:

$$\begin{pmatrix} H_{00R}^K & H_{01R}^K & 0 \\ H_{10R}^K & H_{11R}^K + \frac{Jg_1}{K} & \frac{Jg_1}{K} \\ 0 & \frac{Jg_1}{K} & H_{11R}^{K'} + \frac{Jg_1}{K} \end{pmatrix} \begin{pmatrix} C_{0R}^K \\ C_{1R}^K \\ C_{1R}^{K'} \end{pmatrix} = (E_{wc} + E_R) \begin{pmatrix} C_{0R}^K \\ C_{1R}^K \\ C_{1R}^{K'} \end{pmatrix} \qquad (87)$$

Here $\left( C_{0R}^K, C_{1R}^K, C_{1R}^{K'} \right)$ and $E_R$ are the eigen vector and eigen energy of the exciton that includes both the Umklapp scattering effect and the intervalley mixing effect. The solution for the exciton with left-handed (L) chiral symmetry can be similarly obtained by changing the "$R$" subscript into "$L$" and swap the K and K′ superscripts in Eqs. (78 – 84, 87).

### 6.2. Energy dispersion of Umklapp states

The Umklapp coupling terms $H_{01R}^K$ and $H_{10R}^K$ are much smaller than the kinetic energy of the exciton CM motion. By comparing the results with and without these two terms, we find that the eigen energy difference is tiny (<10 μeV). Therefore, we can neglect these two terms in the calculation of the eigen energies (but they must be included in the calculation of eigen vectors to give the Umklapp scattering effect). By neglecting the $H_{01R}^K$ and $H_{10R}^K$ terms, the energies of two Umklapp excitons (labeled by "$u1$" and "$u2$") have the analytic expressions (S5)

$$E_{u1}(g_1) = E_0^X(n_{wc}) + \frac{\hbar^2 g_1^2}{2m_X} \qquad (88)$$

$$E_{u2}(g_1) = E_0^X(n_{wc}) + \frac{\hbar^2 g_1^2}{2m_X} + \frac{2Jg_1}{K} \qquad (89)$$

Here $E_0^X(n_{wc})$ is the energy of the primary exciton. The $E_{u1}(g_1)$ and $E_{u2}(g_1)$ energies correspond to the $A_{hu1}^0$ and $A_{hu2}^0$ lines on the hole side in our experimental map in Fig. 2 (on the electron side, our experiment only resolves the $A_{eu2}^0$ line). The Umklapp state with $E_{u1}$ ($E_{u2}$) corresponds to the anti-symmetric (symmetric) combination of the KK and K'K' excitons $\left[i.e.\ \left(|X_{g_1}^K\rangle \mp |X_{g_1}^{K'}\rangle\right)/\sqrt{2}\right]$, and the energy separation between them is $2Jg_1/K$, which increases linearly with the Umklapp wave vector $g_1$. By using the relationship $g_1 = \frac{4\pi}{\sqrt{3}a_{wc}} = 2\pi\sqrt{\frac{2}{\sqrt{3}}n_{wc}}$ between $g_1$ and the WC particle density ($n_{wc}$), we can write Eqs. (88,89) as:

$$E_{u1}(n_{wc}) = E_0^X(n_{wc}) + \frac{\hbar^2 n_{wc}}{\sqrt{3}m_X} \tag{90}$$

$$E_{u2}(n_{wc}) = E_0^X(n_{wc}) + \frac{\hbar^2 n_{wc}}{\sqrt{3}m_X} + 4\pi\frac{J}{K}\sqrt{\frac{2}{\sqrt{3}}n_{wc}} \tag{91}$$

We have numerically solved Eq. (87) at varying WC particle density. Fig. S12 displays the density-dependent energy $E_0^X(n_{wc})$ of the primary exciton ($A^0$) as well as energies $E_{u1}(n_{wc})$, $E_{u2}(n_{wc})$ of Umklapp excitons ($A_{u1}^0$, $A_{u2}^0$), plotted relative to $E_0^X(0)$, under zero magnetic field. In this calculation, the energies of the $A^0$ and $A_{u1}$ states don't depend on the intervalley coupling strength $J$, but the $A_{u2}$ energy does. Fig. S12 displays two $A_{u2}$ lines for $J = 160$ and $180\ meV$, which correspond to the best fits for the experimental electron and hole Umklapp lines ($A_{eu2}^0$, $A_{hu2}^0$) in Fig. 2, respectively. At $n_{wc} = 5.6 \times 10^{11}cm^{-2}$ (corresponding to $r_s \approx 16$), these two $J$ values yield a calculated $A_{u1}^0 - A^0$ energy separation of ~8.15 meV, consistent with the measured $A_{hu1}^0 - A^0$ gap of 9.2 ± 5.9 meV. At $n_{wc} = 2.2 \times 10^{11}cm^{-2}$ (corresponding to $r_s \approx 25$), they yield a calculated $A_{u2}^0 - A^0$ separation of ~10.4 meV, consistent with the measured $A_{hu2}^0 - A^0$ gap of 7.2 ± 4.1 meV and the measured $A_{eu2}^0 - A^0$ gap of 8.1 ± 4.2 meV. Fig. S12 also displays a third $A_{u2}^0$ line for $J = 300\ meV$, which was used in Ref. (*S5*). At this larger $J$ value, the calculated $A_{u2}^0 - A^0$ energy separation increases to ~19 meV (near $n_{wc} = 3 \times 10^{11}cm^{-2}$), which does not match our experimental observation.

The $A_{u1}^0 - A^0$ and $A_{u2}^0 - A^0$ energy separations exhibit different density dependence (Fig. S12). The $A_{u1}^0 - A^0$ separation increases linearly with the WC particle density ($n_{wc}$) because $g_1^2 \propto n_{wc}$ in its dispersion $E_{u1}(g_1) = \hbar^2 g_1^2/2m_X$ [Eq. (88)]. But the $A_{u2}^0 - A^0$ separation exhibits an initial sublinear density dependence because the $g_1$ term dominates at low density in its dispersion $E_{u2}(g_1) = \hbar^2 g_1^2/2m_X + 2Jg_1/K$ [Eq. (89)]. As the density increases, however, the $g_1^2$ term becomes more important, causing the $A_{u2}^0 - A^0$ separation to gradually adopt a linear density dependence. This aligns with our experimental observation (conducted in relatively high density), where both the $A_{u1}^0 - A^0$ and $A_{u2}^0 - A^0$ energy separations exhibit approximately linear density dependence (Fig. 2D).

After solving Eq. (87) for the expansion coefficients $C_{1R,u1}^{K}$ and $C_{1R,u2}^{K}$, the Umklapp scattering probability is given by

$$\begin{cases} P_{u1}(n_{wc}) = \left|C_{1R,u1}^{K}\right|^2 \\ P_{u2}(n_{wc}) = \left|C_{1R,u2}^{K}\right|^2 \end{cases} \quad (92)$$

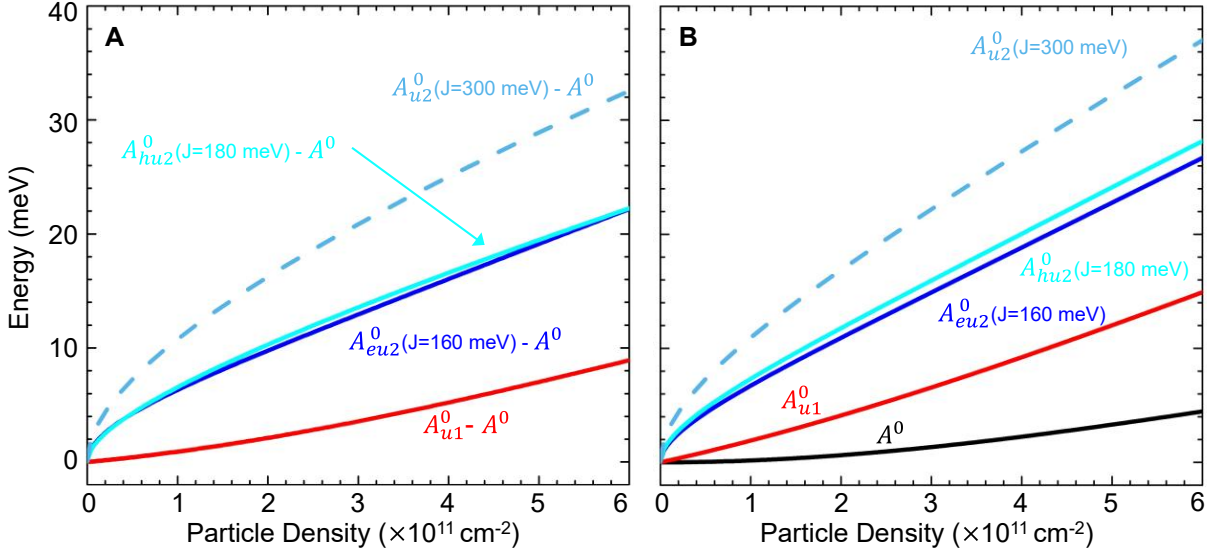

**Figure S12**. (A) Calculated energy separation between the Umklapp excitons ($A_{u1}^{0}$, $A_{u2}^{0}$) and the primary exciton ($A^{0}$) at varying WC particle density in monolayer WSe$_2$ at zero magnetic field. (B) Similar plot of energies relative to the $A^{0}$ energy at zero density, including the energy shift due to screening effect. In both plots, the $A^{0}$ and $A_{u1}^{0}$ energies don't depend on the intervalley coupling strength $J$, but the $A_{u2}$ energy does. We show three $A_{u2}^{0}$ lines for $J = 160$, 180, and 300 $meV$.

### 6.3. Effect of valley disorder in the Wigner crystal

Under zero magnetic field, the charges in a Wigner crystal are evenly distributed in K and K' valleys, since they are degenerate. Because the nearest-neighbor exchange interaction between two charges in the Wigner crystal is very weak, we expect the spatial distribution of charges in the two valleys to be random. So, the Wigner crystal considered here has a charge order (i.e. one charge in each WC cell) but valley disorder. One may wonder if such a valley disorder will ruin the observation of the Umklapp peaks. Our analysis below shows that the Umklapp peaks still exist despite the valley disorder.

In the four cases considered in Section 6.1, the exciton-WC scattering is described by Eqs. (41), (42), (43) and (48). These four equations share the same form:

$$\left\langle \text{WC}; X^{K'}_{\mathbf{k}_1+\mathbf{g}} \middle| v^X_{wc} \middle| \text{WC}; X^{K'}_{\mathbf{k}_1+\mathbf{g}'} \right\rangle = \frac{1}{A} \sum_{\mathbf{k},\mathbf{k}',\mathbf{q},\mathbf{q}',j} \tilde{\varphi}_0(\mathbf{k}) \tilde{\varphi}_0(\mathbf{k}') \tilde{\psi}_{wc}(\mathbf{q}) \tilde{\psi}_{wc}(\mathbf{q}') e^{i\Delta \mathbf{g} \cdot \mathbf{R}_j} \, U_i(\Delta \mathbf{g}) \delta_{\Delta \mathbf{q}, -\Delta \mathbf{g}} \quad (93)$$

Here $U_i(\Delta \mathbf{g})$ denotes the kernel of interaction for Cases $i = 1 - 4$. For Cases $3 - 4$, we need to change all K' in Eq. (93) into K because these two cases deal with K-valley exciton.

In Eq. (93), $\mathbf{R}_j$ denotes the center of the j-th WC site, and the only site-dependent term is the phase factor $e^{i\Delta \mathbf{g} \cdot \mathbf{R}_j}$. The Umklapp scattering occurs when $\Delta \mathbf{g}$ matches a reciprocal lattice vector of the Wigner crystal, which is determined by the charge distribution only, regardless of the valley disorder. When $\Delta \mathbf{g}$ equals a WC reciprocal lattice vector, $e^{i\Delta \mathbf{g} \cdot \mathbf{R}_j} = 1$ and all the site $\mathbf{R}_j$ have equal contribution. The summation over j in Eq. (93) becomes a simple multiplication of $N_{\text{wc}}/2$, since only one half of the WC particle resides in the relevant valley of each of the case. Then Eq. (93) becomes:

$$\left\langle \text{WC}; X^{K'}_{\mathbf{k}_1+\mathbf{g}} \middle| v^X_{wc} \middle| \text{WC}; X^{K'}_{\mathbf{k}_1+\mathbf{g}'} \right\rangle = \frac{n_{\text{wc}}}{2} \sum_{\mathbf{k},\mathbf{k}',\mathbf{q},\mathbf{q}'} \tilde{\varphi}_0(\mathbf{k}) \tilde{\varphi}_0(\mathbf{k}') \tilde{\psi}_{wc}(\mathbf{q}) \tilde{\psi}_{wc}(\mathbf{q}') \, U_i(\Delta \mathbf{g}) \delta_{\Delta \mathbf{q}, -\Delta \mathbf{g}} \quad (94)$$

Here $n_{\text{wc}}/2$ denotes the one half of the WC particle density. It is interesting to note that the matrix element in Eq. (94) does not depend on the detailed distribution of the valley index on the WC sites. That is, even if the valley indices on the WC sites are random, the result is the same and the Umklapp scattering remains robust even with valley disorder.

At zero magnetic field, both valleys have the same density ($n_{\text{wc}}/2$). To consider the Umklapp scattering, we need to include four other cases related to Cases 1-4 by the time reversal symmetry (that is, switching the valley indices of both the exciton and WC particles in Fig. 3). These four time-reversal cases, denoted as Cases 1'-4', have the same scattering probability as Cases 1-4, respectively. The Umklapp scattering on the electron side is contributed by Cases 1 and 3', and the scattering probability ($P_{eu1}$, $P_{eu2}$) of each Umklapp line equals the sum of the scattering probabilities in Cases 1 and 3. Similarly, the Umklapp scattering on the hole side is contributed by Cases 2 and 4', and the scattering probability ($P_{hu1}$, $P_{hu2}$) of each Umklapp line equals the sum of the scattering probabilities in Cases 2 and 4.

$$\begin{cases} P_{eu1} = P_{u1;\,Case\,1}(n_{wc}/2) + P_{u1;\,Case\,3}(n_{wc}/2) \\ P_{eu2} = P_{u2;\,Case\,1}(n_{wc}/2) + P_{u2;\,Case\,3}(n_{wc}/2) \\ P_{hu1} = P_{u1;\,Case\,2}(n_{wc}/2) + P_{u1;\,Case\,4}(n_{wc}/2) \\ P_{hu1} = P_{u2;\,Case\,2}(n_{wc}/2) + P_{u2;\,Case\,4}(n_{wc}/2) \end{cases} \tag{95}$$

Here $P_{u1}$ and $P_{u2}$ are defined in Eq. (92) for different Cases 1-4; $n_{wc}$ denotes the total WC particle density at zero magnetic field. The factor 1/2 is due to the evenly distributed population in K and K' valleys as shown in Eq. (94).

### 6.4. Simulation of Umklapp lines induced by exciton-WC scattering

Next, we discuss the oscillator strength of the exciton Umklapp lines. After we obtain the eigen states of Eq. (87), we calculate the absorption spectrum of the dominant Umklapp peak for right-handed polarization as:

$$S_R(\omega) \propto \sum_n |\langle \psi_R^n | \mathbf{e}_R \cdot \mathbf{p} | 0 \rangle|^2 \delta(\hbar\omega - E_n) \tag{96}$$

Here the summation includes all three eigen states of $\psi_R^n$ ($n = 1 - 3$) with energy $E_n$ obtained by solving Eq. (87); $\mathbf{e}_R$ is the polarization vector for right-handed light; $|0\rangle$ denotes the system before the creation of exciton. The absorption spectrum for the left-handed polarization can be expressed by changing the "$R$" subscript into "$L$". To address the screening effect of the WC particles on the exciton oscillator strength, we phenomenologically multiply the oscillator strength of the free exciton at zero density a screening function. Namely,

$$O_X(n_{wc}) = O_X(0)/(1 + f_s n_{wc}), \tag{97}$$

where $f_s = 400 nm^2$ is determined by fitting the optical conductivity of the primary excitonic states extracted from the reflectance contrast data.

Figure S13 presents the calculated oscillator strength of the primary and Umklapp excitons as a function of WC particle density $n_{wc}$ at zero magnetic field for both Models 1 and 2. Fig. S14 displays the color map of the calculated optical conductivity of the primary and Umklapp excitons for Model 1. Our calculation includes the intervalley scattering as well as the influence of the WC charge screening on the exciton oscillator strength. Our results show two remarkable features. First, the hole-WC Umklapp peaks are one to two orders of magnitude stronger than the electron-WC Umklapp peaks, because the former involves exciton-WC scattering in Case 4 that includes long-range exchange interaction, dominating the scattering processes in Cases 1-3 with short-range interactions. Second, the second Umklapp line ($A_{eu2}^0$, $A_{hu2}^0$) is about one order of magnitude weaker than the first Umklapp line ($A_{eu1}^0$, $A_{hu1}^0$) on both the electron and hole sides, because it has a much larger energy separation from the principal exciton line.

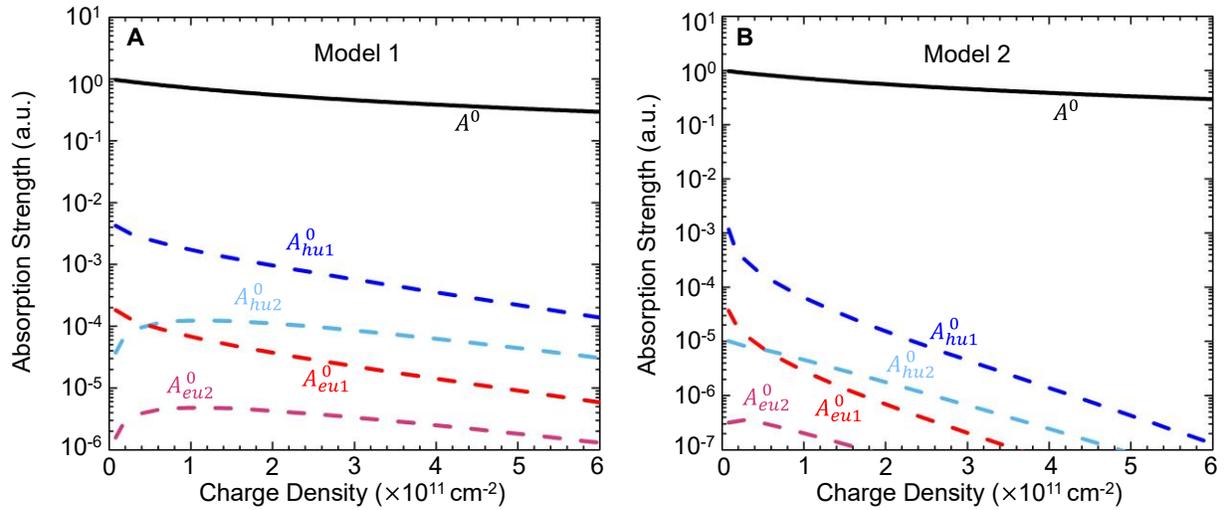

**Figure S13**. Calculated absorption strength of the primary and Umklapp excitons as a function of WC particle density for monolayer WSe₂ at zero magnetic field. (A) Results of Model 1. (B) Resulted of Model 2. $J = 160$ meV is used for both models.

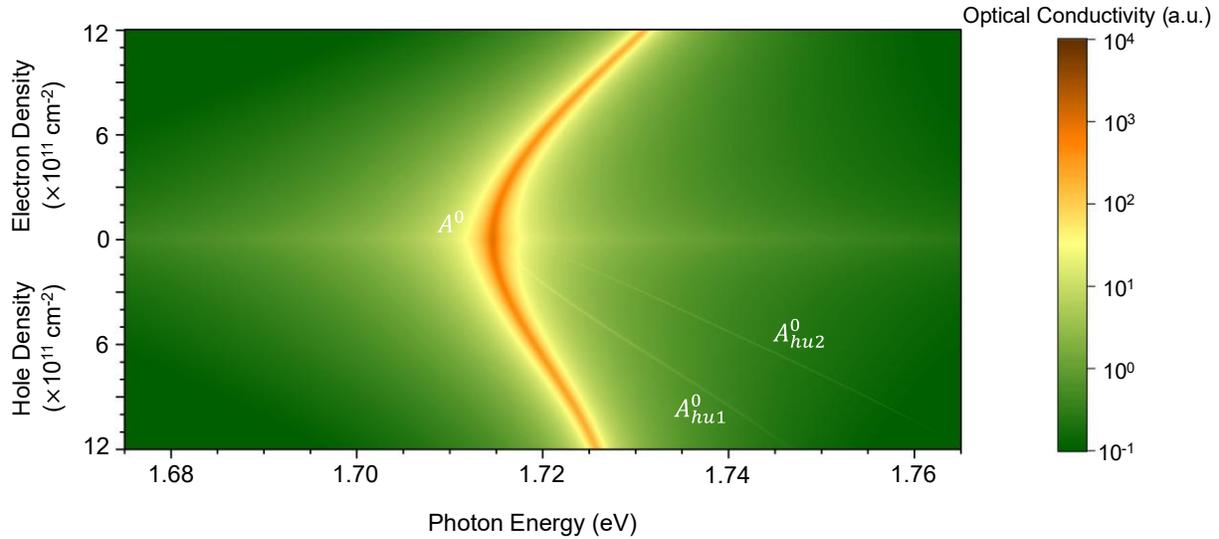

**Figure S14**. Color map of the calculated optical conductivity for the primary and Umklapp excitons as a function of WC particle density for monolayer WSe₂ at zero magnetic field for Model 1. We use $J = 160$ meV on the electron side and 180 meV on the hole side. This figure is the same as Fig. 4A in the main paper.

While our simulation in Figs. S13 and S14 can account for the presence of two Umklapp lines on the hole side, it cannot reproduce the Umklapp lines on the electron side. Our experimental result (Fig. 2C) reveals one exciton Umklapp line and two exciton-polaron Umklapp lines on the electron side. In particular, the electron-side exciton Umklapp line ($A_{eu2}^0$) has comparable strength with the hole-side exciton Umklapp lines ($A_{hu2}^0$) in our experiment. These discrepancies suggest that other mechanisms play an important role in the exciton-WC scattering. In the following section, we will show that the exciton-polaron effect can drastically change the exciton-WC scattering strength and better simulate to our observation.

## 7. Polaron effect on exciton Umklapp scattering with Wigner crystals

### 7.1. Polaron effect without considering intervalley coupling

When an exciton interacts with a Fermi sea, it can form an exciton polaron, which is a complex quasiparticle dressed with a polarization of Fermi sea (*S19-21*). After the Wigner crystal is formed, the Fermi-sea picture is no longer adequate, but the polaron effect is still present because the exciton can polarize the Wigner crystal to form the exciton polaron. In this section, we will discuss how the polaron effect will influence the Umklapp scattering processes.

#### 7.1.1 Exciton-polaron Hamiltonian

We consider the following exciton-polaron Hamiltonian within the framework of effective-mass approximation:

$$H_P = H_X(\mathbf{r}_{12}, \mathbf{R}_X) - \frac{\hbar^2}{2m^*}\nabla_3^2 - v(\mathbf{r}_{13}) + v(\mathbf{r}_{23}) - \frac{\hbar^2}{2m^*}\nabla_4^2 + V_{T4} \equiv H_T - \frac{\hbar^2}{2m^*}\nabla_4^2 + V_{T4} \qquad (98)$$

Here $\mathbf{R}_X$ denotes CM coordinate of the exciton; $\mathbf{r}_{ij}$ denotes the relative coordinate between two particles; The subscripts $i, j = 1, 2, 3, 4$ denote the four interacting particles in our consideration, including the electron and hole (1, 2) in the exciton, the WC particle (3) in the trion, and the polarized WC described by a missing particle (4); $v(\mathbf{r}_{ij})$ denotes the Coulomb interaction between particle $i$ and $j$. $m^*$ denotes the effective mass of the WC particle. In the expression, the first term is the exciton Hamiltonian ($H_X$) as defined in Eq. (5); The first four terms constitute the trion Hamiltonian ($H_T$). The last term

$$V_{T4} = -v(\mathbf{r}_{14}) + v(\mathbf{r}_{24}) + v(\mathbf{r}_{34}) \qquad (99)$$

describes the net Coulomb interaction between the trion (T) and the surrounding polarized charge density (particle 4).

### 7.1.2. Exciton-polaron wavefunction

Next, we consider the exciton-polaron wavefunction associated with the Hamiltonian in Eq. (98). When an exciton is photogenerated with zero momentum, it can bind with a charge at a WC site to form a bound trion. This trion can induce polarization of charges at surrounding sites to form a tetron. The exciton and trion wavefunctions with zero momentum are called exciton and trion internal-motion wavefunctions, respectively. The exciton internal-motion wavefunction $\varphi_0(\mathbf{r}_{12})$ is defined in Eq. (11). The trion internal-motion wavefunction $\varphi_T(\mathbf{r}_{12}, \mathbf{r}_{13})$ can be expressed as a linear combination of products of Slater-type orbitals as (*S21*)

$$\varphi_T(\mathbf{r}_{12}, \mathbf{r}_{13}) = \sum_{n_1, n_2, m} C_{n_1, n_2, m} f_{n_1, m}(\mathbf{r}_{12}) f_{n_2, -m}(\mathbf{r}_{13}). \tag{100}$$

Here $f_{nm}(r_j) = r_j^s e^{im\varphi_j} e^{-\alpha_n r_j}$, where $s = 1$ for $m \neq 0$ and $s = 0$ for $m = 0$. The coefficients $C_{n_1, n_2, m}$ are solved numerically by using the Rayleigh-Ritz variational method within the effective-mass approximation as described in Ref. (*S21, 22*).

For a tetron localized in a WC site, its wavefunction can be approximated as the product of the trion center-of-mass (CM) motion wavefunction and the trion internal-motion wavefunction $\varphi_T(\mathbf{r}_{12}, \mathbf{r}_{13})$, and the additional correlation function of the form $f(\mathbf{r}_{13} - \mathbf{r}_4, \mathbf{r}_{12} - \mathbf{r}_4)$. The CM-motion wavefunction of the localized trion is approximated by a Gaussian function $e^{-bR_T^2}$, where $\mathbf{R}_T = (m_X \mathbf{R}_X + m^* \mathbf{r}_3)/m_T \equiv \gamma_X \mathbf{R}_X + \gamma_3 \mathbf{r}_3$ is the CM coordinate of the trion and $m_T = (m_X + m^*)$ is the total effective mass of the trion. The total wavefunction of the localized tetron is approximately described by

$$\Psi_3(\mathbf{R}_T, \mathbf{r}_{12}, \mathbf{r}_{13}, \mathbf{r}_4) \approx e^{-bR_T^2} F(\mathbf{r}_{12}, \mathbf{r}_{13}, \mathbf{r}_4), \tag{101}$$

where $F(\mathbf{r}_{12}, \mathbf{r}_{13}, \mathbf{r}_4) = \varphi_T(\mathbf{r}_{12}, \mathbf{r}_{13}) f(\mathbf{r}_{13} - \mathbf{r}_4, \mathbf{r}_{12} - \mathbf{r}_4)$. The effect of the correlation function $f(\mathbf{r}_{12} - \mathbf{r}_4, \mathbf{r}_{13} - \mathbf{r}_4)$ on the tetron binding energy will be included empirically and found to be rather small in the density range of interest.

A tetron with CM wave vector $\mathbf{g}_1$ that matches a reciprocal lattice vector of the Wigner crystal can be coupled to light via Umklapp scattering. Such a finite-momentum tetron can travel between WC sites. Its wavefunction is approximated as the product of $F(\mathbf{r}_{12}, \mathbf{r}_{13}, \mathbf{r}_4)$ and the linear combination of the CM wavefunction $e^{i\mathbf{g}_1 \cdot \mathbf{R}_X}$ of a traveling exciton and the CM wavefunction $e^{-bR_T^2}$ of a localized trion at a WC site. Due to the point-group symmetry, it is convenient to define a symmetrized tetron state for the first star as $|\Im_1\rangle = \frac{1}{\sqrt{d_1}}\sum_{\mathbf{g}_1}|\Im_{\mathbf{g}_1}\rangle$, where $\mathbf{g}_1$ denote the CM wave vectors $\mathbf{g}$ of the tetron basis state in the first star and $d_1$ is the number of $\mathbf{g}_1$ in the first star. We write the symmetrized tetron wavefunction in the first star as

$$\langle \mathbf{R}_X; \mathbf{r}_{12}, \mathbf{r}_{13}, \mathbf{r}_4|\Im_1\rangle \approx \left(\frac{c_1}{\sqrt{d_1 A}}\sum_{\mathbf{g}_1} e^{i\mathbf{g}_1\cdot\mathbf{R}_X} + \frac{c_2}{\sqrt{N_{wc}}}\sum_j \sqrt{\frac{2b}{\pi}} e^{-b(\mathbf{R}_T - \mathbf{R}_j)^2}\right) F(\mathbf{r}_{12}, \mathbf{r}_{13}, \mathbf{r}_4) \qquad (102)$$

When $\mathbf{g}$ matches a reciprocal lattice vector of the Wigner crystal, the corresponding tetron state is an eigen state of the polaron Hamiltonian at the center of the WC Brillouin zone; hence, it must be orthogonal to other tetron states with different $\mathbf{g}$ (including $\mathbf{g} = \mathbf{0}$). The orthogonality relation $\langle \Im_0|\Im_1\rangle = 0$ gives:

$$c_1 \sqrt{\frac{2b}{d_1 A_{wc}\pi}} \int d\mathbf{R}_T \, e^{-bR_T^2}\sum_{\mathbf{g}_1} e^{i\mathbf{g}_1\cdot\mathbf{R}_T/\gamma_X} \int d\mathbf{r}_{12}d\mathbf{r}_{13}d\mathbf{r}_4 \, e^{-i\mathbf{g}_1\cdot\mathbf{r}_3\gamma_3/\gamma_X}|F(\mathbf{r}_{12}, \mathbf{r}_{13}, \mathbf{r}_4)|^2 + c_2 = 0$$

$$(103)$$

Since the phase factor $e^{-i\mathbf{g}_1\cdot\mathbf{r}_3\gamma_3/\gamma_X}$ varies smoothly with $\mathbf{r}_3$ in the trion internal motion, we can approximate it as 1 inside the integral. By using the normalization condition $\int d\mathbf{r}_{12}d\mathbf{r}_{13}d\mathbf{r}_4\,|F(\mathbf{r}_{12}, \mathbf{r}_{13}, \mathbf{r}_4)|^2 = 1$ and taking the integration over $\mathbf{R}_T$, Eq. (103) gives

$$c_2 = -c_1\sqrt{\frac{2\pi d_1}{bA_{wc}}}\, e^{-\frac{\mathbf{g}_1^2}{4b\gamma_X^2}} \qquad (104)$$

By using the normalization condition $c_1^2 + c_2^2 = 1$, we obtain

$$c_1 = \left(1 + \frac{2\pi d_1}{bA_{wc}}e^{-\frac{\mathbf{g}_1^2}{2b\gamma_X^2}}\right)^{-1/2} \qquad (105)$$

$$c_2 = -\sqrt{\frac{2\pi d_1}{bA_{wc}}}e^{-\frac{\mathbf{g}_1^2}{4b\gamma_X^2}}\left(1 + \frac{2\pi d_1}{bA_{wc}}e^{-\frac{\mathbf{g}_1^2}{2b\gamma_X^2}}\right)^{-1/2} \qquad (106)$$

Our above model, which describes the Umklapp tetron states, can be summarized by the schematic diagrams in Fig. S15. When a travelling exciton is uncoupled to the WC particle, we can describe the exciton by the wavefunction $\varphi_0(\mathbf{r}_{12})e^{i\mathbf{k}\cdot\mathbf{R}_X}$ and the WC particle by the localized wavefunction $e^{-\beta r_3^2}$, as illustrated in Fig. S15A. When the exciton is coupled to the WC particle, they can form a zero-momentum tetron localized at the WC site. The wavefunction of such a tetron can be approximated by the product of the tetron internal-motion wavefunction $F(\mathbf{r}_{12}, \mathbf{r}_{13}, \mathbf{r}_4)$ and the trion CM wavefunction $e^{-bR_T^2}$, as illustrated in Fig. S15B. When the tetron gains momentum to travel between WC sites, its wavefunction is approximated as the product of $F(\mathbf{r}_{12}, \mathbf{r}_{13}, \mathbf{r}_4)$ and the linear combination of the CM wavefunction $e^{i\mathbf{g_1}\cdot\mathbf{R}_X}$ of a traveling exciton and the CM wavefunction $e^{-bR_T^2}$ of a localized trion at a WC site, as illustrated in Fig. S15C. We found that by using $b = 2\beta$ our theory can predict the exciton-polaron effect on exciton Umklapp lines in reasonable agreement with experiment.

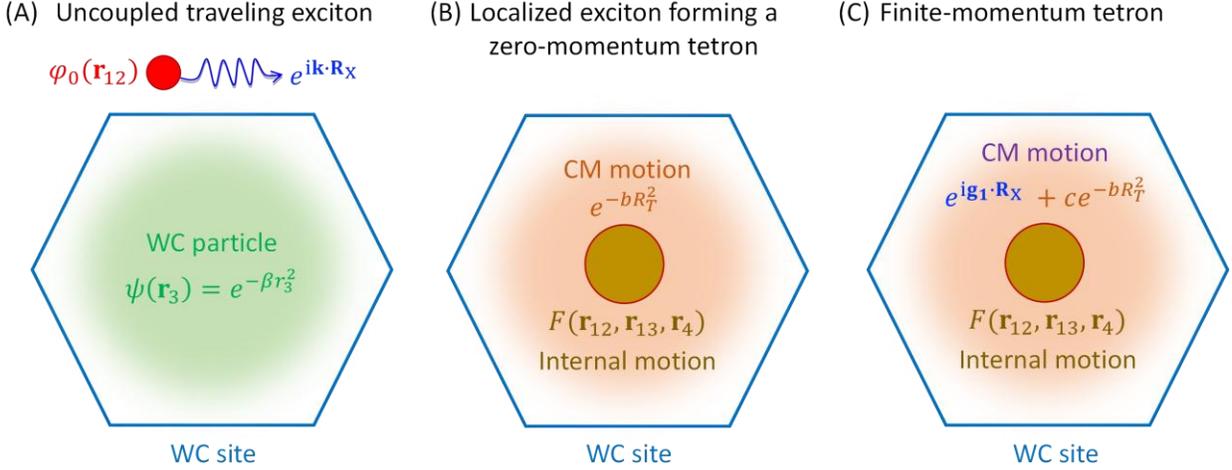

**Figure S15**. Schematic of tetron formation in our model. (A) When a traveling exciton is uncoupled to the WC particle, its wavefunction is a product of the wavefunctions for the internal motion $\varphi_0(\mathbf{r}_{12})$ and center-of-mass (CM) motion $e^{i\mathbf{k}\cdot\mathbf{R}_X}$, represented by the red disk and blue wavy line, respectively. The WC particle has a Gaussian wavefunction $e^{-\beta r_3^2}$, represented by the green disk. (B) After the exciton is localized in the WC site, it forms a tetron with zero CM momentum. The tetron wavefunction is approximated by the product of the internal-motion wavefunction $F(\mathbf{r}_{12}, \mathbf{r}_{13}, \mathbf{r}_4)$ and the trion CM wavefunction $e^{-bR_T^2}$, represented by the darker and lighter brown disks, respectively. (C) When the tetron gains momentum to travel between WC sites, its wavefunction is described as a product of the tetron internal-motion wavefunction $F(\mathbf{r}_{12}, \mathbf{r}_{13}, \mathbf{r}_4)$ and a CM wavefunction with a form of linear combination $e^{i\mathbf{g_1}\cdot\mathbf{R}_X} + ce^{-bR_T^2}$.

### 7.1.3. Exciton-polaron matrix elements

As described in Section 2, the Wigner crystal consists of a periodic array of localized charges, whose density distribution is described by a Gaussian function $\eta(\mathbf{r} - \mathbf{R}_j)$ (Eq. 12). When an exciton is generated near site $j$ of the Wigner crystal, it can bind with the WC particle there to form a trion centered at site $j$ if the WC electron (hole) and the electron (hole) in the exciton are in different valleys (spins). This can occur only in Cases 1-3. As the trion charge density differs from the original charge density $\eta(\mathbf{r} - \mathbf{R}_j)$ at site $j$, it will further polarize the surrounding WC charges to form a polaron. Such a localized polaron state can interact with the exciton state to form two prominent polaron states; the lower branch is commonly referred to as the "trion" or "attractive polaron" while the upper branch is referred to as the "exciton" or "repulsive polaron" in the literature (*S19-21*). Our experiment shows that both the lower and upper polaron branches can exhibit Umklapp scattering features.

For the KK exciton configuration, the polaron states ($P$) with zero momentum can be expressed as a linear combination of the exciton state (X) (first term) and the tetron and higher-order states (denoted by $\mathfrak{I}$) (second term) as follows:

$$|P\rangle = \sum_{\mathbf{g}} C_{\mathbf{g}}^{X} X_{\mathbf{g}}^{\dagger} |WC\rangle + \frac{1}{\sqrt{\bar{N}_{wc}}} \sum_{\mathbf{g}} C_{\mathbf{g}}^{\mathfrak{I}} \sum_{j} e^{i\mathbf{g}\cdot\mathbf{R}_j} T_{j\mathbf{g}}^{\dagger} a_{j,\sigma} |WC\rangle$$

$$= \sum_{s} C_{sR}^{K} |X_{s}^{K}\rangle_R + \sum_{s} C_{sR}^{\mathfrak{I}} |\mathfrak{I}_s\rangle_R$$

$$\approx C_{0R}^{K} |X_{s=0}^{K}\rangle_R + C_{1R}^{K} |X_{s=1}^{K}\rangle_R + C_{0R}^{\mathfrak{I}} |\mathfrak{I}_{s=0}\rangle_R + C_{1R}^{\mathfrak{I}} |\mathfrak{I}_{s=1}\rangle_R. \qquad (107)$$

Here $\bar{N}_{wc} = N_{wc}/2$, since a trion can only be formed at one half of the $N_{wc}$ sites. In the first line of Eq. (107), $|WC\rangle$ denotes the quantum state of the Wigner crystal defined in Eq. (26); $X_{\mathbf{g}}^{\dagger}$ creates an exciton with a reciprocal WC lattice vector $\mathbf{g}$; $C_{\mathbf{g}}^{X}$ is the corresponding expansion coefficient; $T_{j\mathbf{g}}^{\dagger}$ creates a trion (associated with the exciton at $\mathbf{g}$) localized at site $j$; $a_{j,\sigma}$ annihilates a WC particle at site $j$; σ denotes the spin state of the WC particle at site $j$. $C_{\mathbf{g}}^{\mathfrak{I}}$ is the corresponding expansion coefficient for a tetron state at $\mathbf{g}$; Here we have neglected the polarization of the Wigner crystal surrounding the trion at site $j$, since they are far apart.

The second line of Eq. (107) applies when we only consider the right-handed (R) symmetrized exciton basis $|X_{s}^{K}\rangle_R$ as defined in Eq. (80) and the associated symmetrized tetron state $|\mathfrak{I}_s\rangle_R$. Similar to Eq. (80), the symmetrized tetron state is defined as

$$|\mathfrak{I}_s\rangle_R = \frac{1}{\sqrt{d_s}} \sum_{\mathbf{g}_s} |\mathfrak{I}_{\mathbf{g}_s}\rangle_R. \qquad (108)$$

This state is analogous to the trion-FS hole complex as described in Ref. (*S21*) for the case of an exciton coupled to a Fermi sea.

In the third line of Eq. (106), we limit the exciton and tetron basis to the zeroth star ($s = 0$) and first star ($s = 1$) of $\mathbf{g}$ with corresponding expansion coefficients $C_{0R}^{K}$ and $C_{1R}^{K}$ for exciton and $C_{0R}^{\mathfrak{I}}$ and $C_{1R}^{\mathfrak{I}}$ for tetron, because higher stars are found to contribute negligibly to our results.

By using the Hamiltonian in Eq. (98) and the basis in Eq. (106), we can write the Schrödinger equation of the polaron as a set of four coupled equations:

$$\begin{cases} \sum_{s'=0,1} [\,_R\langle X_s^K | H_P | X_{s'}^K \rangle_R \, C_{s'R}^K + \,_R\langle X_s^K | H_P | \mathfrak{I}_{s'} \rangle_R \, C_{s'R}^{\mathfrak{I}}] = E_P C_{sR}^K \\ \sum_{s'=0,1} [\,_R\langle \mathfrak{I}_s | H_P | X_{s'}^K \rangle_R \, C_{s'R}^K + \,_R\langle \mathfrak{I}_s | H_P | \mathfrak{I}_{s'} \rangle_R \, C_{s'R}^{\mathfrak{I}}] = E_P C_{sR}^{\mathfrak{I}} \end{cases} \quad (109)$$

Here we only consider $s, s' = 0, 1$. $E_P$ denotes the eigen energy of the exciton polaron. The matrix elements between two exciton states $_R\langle X_s^K | H_P | X_{s'}^K \rangle_R$ are given by $H_{ss'R}^K$ as defined in Eq. (83). For the convenience of presentation, we define four matrix elements between the exciton and tetron states

$$U_{X\mathfrak{I}}^{ss'} = \,_R\langle X_s^K | H_P | \mathfrak{I}_{s'} \rangle_R = \frac{1}{\sqrt{d_s}} \sum_{\mathbf{g}_s} \langle \text{WC}; X_{\mathbf{g}_s}^K | H_P | \mathfrak{I}_{s'} \rangle_R \quad \text{with s, s}' = 0, 1 \quad (110)$$

Here $d_{s=0} = 1$ and $d_{s=1} = 6$ are the total number of g vectors in the s-th star. All of the g vectors in the same star have the same magnitude due to symmetry. For s, s' = 0, 1, Eq. (110) can be written as:

$$\begin{cases} U_{X\mathfrak{I}}^{00} = \,_R\langle X_0^K | H_P | \mathfrak{I}_0 \rangle_R = \langle \text{WC}; X_{\mathbf{g}=0}^K | H_P | \mathfrak{I}_0 \rangle_R = U_{X\mathfrak{I}}(g = 0) \\ U_{X\mathfrak{I}}^{10} = \,_R\langle X_1^K | H_P | \mathfrak{I}_0 \rangle_R = \frac{1}{\sqrt{6}} \sum_{\mathbf{g}_1} \langle \text{WC}; X_{\mathbf{g}_1}^K | H_P | \mathfrak{I}_0 \rangle_R = \sqrt{6} U_{X\mathfrak{I}}(g_1) \\ U_{X\mathfrak{I}}^{01} = \,_R\langle X_0^K | H_P | \mathfrak{I}_1 \rangle_R = \langle \text{WC}; X_{\mathbf{g}=0}^K | H_P | \mathfrak{I}_1 \rangle_R \\ U_{X\mathfrak{I}}^{11} = \,_R\langle X_1^K | H_P | \mathfrak{I}_1 \rangle_R = \frac{1}{\sqrt{6}} \sum_{\mathbf{g}_1} \langle \text{WC}; X_{\mathbf{g}_1}^K | H_P | \mathfrak{I}_1 \rangle_R \end{cases} \quad (111)$$

Here $U_{X\mathfrak{I}}(g = 0)$ is the matrix element between the exciton with CM vector g = 0 and the tetron state with g = 0; $U_{X\mathfrak{I}}(g_1)$ is the matrix element between the unsymmetrized finite-momentum (CM vector $\mathbf{g}$) exciton state with wavefunction $\varphi_0(\mathbf{r}_{12}) e^{i\mathbf{g}\cdot\mathbf{R}_X}/\sqrt{A}$ and the zero-momentum tetron state with wavefunction $e^{-bR_T^2} F(\mathbf{r}_{12}, \mathbf{r}_{13}, \mathbf{r}_4)$. They can be expressed as:

$$U_{X\mathfrak{I}}(\mathbf{g}) = \langle \text{WC}; X_{\mathbf{g}}^K | H_P | \mathfrak{I}_0 \rangle_R$$

$$= \frac{1}{\sqrt{A}} \sum_j \sqrt{\frac{2b}{\pi}} \int d\mathbf{R}_T \, e^{-bR_T^2} e^{i\mathbf{g}\cdot\mathbf{R}_X/\gamma_X} \int d\mathbf{r}_{12} d\mathbf{r}_3 d\mathbf{r}_4 \, \varphi_0(\mathbf{r}_{12}) H_P \, F(\mathbf{r}_{12}, \mathbf{r}_{13}, \mathbf{r}_4)$$

$$= \frac{1}{\sqrt{2A_{wc}}} \sqrt{\frac{2b}{\pi}} \int d\mathbf{R}_T \, e^{-bR_T^2} e^{i\mathbf{g}\cdot\frac{\mathbf{R}_T}{\gamma_X}} \int d\mathbf{r}_{12} d\mathbf{r}_3 d\mathbf{r}_4 \, e^{i\mathbf{g}\cdot\frac{\mathbf{r}_3\gamma_3}{\gamma_X}} \varphi_0(\mathbf{r}_{12}) H_P \, F(\mathbf{r}_{12}, \mathbf{r}_{13}, \mathbf{r}_4) \quad (112)$$

The function $F(\mathbf{r}_{12}, \mathbf{r}_{13}, \mathbf{r}_4)$ describes the trion internal motion and its correlation with the polarization of surrounding charges in the Wigner crystal (with coordinate labeled by $\mathbf{r}_4$). We note that sum over $j$ in the second line of Eq. (112) only covers half of the $N_{wc}$ sites.

In Eq. (112), when we set $\mathbf{g} = 0$, we can do the integration over $\mathbf{R}_T$ to obtain

$$U_{X\mathfrak{I}}^{00} = U_{X\mathfrak{I}}(0) = \sqrt{\pi/(bA_{wc})} \int d\mathbf{r}_{12} d\mathbf{r}_3 d\mathbf{r}_4 \varphi_0(\mathbf{r}_{12}) H_P F(\mathbf{r}_{12}, \mathbf{r}_3, \mathbf{r}_4) \tag{113}$$

When we set $\mathbf{g} = \mathbf{g}_1$ and approximate the slowly varying phase factor in Eq. (112) as $e^{i\mathbf{g}_1 \cdot \mathbf{r}_3 \gamma_3 / \gamma_X} = 1$, we obtain

$$U_{X\mathfrak{I}}^{10} = \sqrt{6} U_{X\mathfrak{I}}(\mathbf{g}_1) = \sqrt{6} U_{X\mathfrak{I}}(\mathbf{g} = 0) e^{-\frac{\mathbf{g}_1^2}{4b\gamma_X^2}} \tag{114}$$

As the unit-cell area $A_{wc}$ is inversely proportional to the WC particle density $n_{wc}$, $U_{X\mathfrak{I}}(0)$ is linearly proportional to $\sqrt{n_{wc}}$. Therefore, we can write

$$U_{X\mathfrak{I}}(0) = \bar{U}_{X\mathfrak{I}} \sqrt{n_{wc}/2} \tag{115}$$

Here we have a factor 1/2 because only half of the WC sites are available to form a valley tetron when the WC particles occupy two valleys at zero magnetic field. The constant $\bar{U}_{X\mathfrak{I}}$ is treated as an empirical parameter that does not depend on the WC particle density ($n_{wc}$).

Next, we discuss $U_{X\mathfrak{I}}^{01}$ and $U_{X\mathfrak{I}}^{11}$ in Eq. (111). Using the wavefunction of the Umklapp tetron state given in Eq. (102), we obtain

$$U_{X\mathfrak{I}}^{01} = {}_R\langle X_0^K | H_P | \mathfrak{I}_1 \rangle_R \approx c_2 U_{X\mathfrak{I}}(0) \tag{116}$$

$$U_{X\mathfrak{I}}^{11} = {}_R\langle X_1^K | H_P | \mathfrak{I}_1 \rangle_R \approx \left( c_1 \sqrt{\frac{bA_{wc}}{2\pi}} + \sqrt{6} c_2 e^{-\frac{\mathbf{g}_1^2}{4b\gamma_X^2}} \right) U_{X\mathfrak{I}}(0) \tag{117}$$

The diagonal matrix element of the tetron state is

$$_R\langle \mathfrak{I}_s | H_P | \mathfrak{I}_s \rangle_R = E_s^T + \Delta_{T4} \equiv E_s^{\mathfrak{I}} \tag{118}$$

Here $E_s^{\mathfrak{I}}$ denotes the energy of the tetron associated with exciton at the $s$-$th$ star of $\mathbf{g}$; $\Delta_{T4}$ is the energy correction due to the coupling between the trion and the surrounding polarization (particle 4). We expect $\Delta_{T4} < 0$ since the tetron energy ($E_s^{\mathfrak{I}}$) should be lower than the trion energy $E_s^T$.

After we obtain the four matrix elements in Eq. (111), the Schrödinger equation in Eq. (109) can be rewritten as a matrix equation:

$$\begin{pmatrix} E_0^X & 0 & U_{X\mathfrak{I}}^{00} & U_{X\mathfrak{I}}^{01} \\ 0 & E_1^X & U_{X\mathfrak{I}}^{10} & U_{X\mathfrak{I}}^{11} \\ U_{X\mathfrak{I}}^{00} & U_{X\mathfrak{I}}^{10} & E_0^{\mathfrak{I}} & 0 \\ U_{X\mathfrak{I}}^{01} & U_{X\mathfrak{I}}^{11} & 0 & E_1^{\mathfrak{I}} \end{pmatrix} \begin{pmatrix} C_{0R}^K \\ C_{1R}^K \\ C_{0R}^{\mathfrak{I}} \\ C_{1R}^{\mathfrak{I}} \end{pmatrix} = E_P \begin{pmatrix} C_{0R}^K \\ C_{1R}^K \\ C_{0R}^{\mathfrak{I}} \\ C_{1R}^{\mathfrak{I}} \end{pmatrix} \tag{119}$$

We note that $U_{X\mathfrak{Z}}^{00}$, $U_{X\mathfrak{Z}}^{01}$, $U_{X\mathfrak{Z}}^{10}$, $U_{X\mathfrak{Z}}^{11}$ are all related to $U_{X\mathfrak{Z}}(0)$ as shown in Eqs. (114-117). Hence, their submatrix can be written as:

$$\begin{pmatrix} U_{X\mathfrak{Z}}^{00} & U_{X\mathfrak{Z}}^{01} \\ U_{X\mathfrak{Z}}^{10} & U_{X\mathfrak{Z}}^{11} \end{pmatrix} = \begin{pmatrix} 1 & c_2 \\ \sqrt{6}e^{-\frac{g_1^2}{4b\gamma_X^2}} & c_1\sqrt{\frac{bA_{wc}}{2\pi}} + \sqrt{6}c_2 e^{-\frac{g_1^2}{4b\gamma_X^2}} \end{pmatrix} U_{X\mathfrak{Z}}(0) \qquad (120)$$

In Eq. (119), $E_0^X$ is the exciton energy at zero CM momentum as considered in Eq. (28).

$$E_1^X = \hbar^2 |\mathbf{g}_1|^2 / 2m_X + E_0^X \qquad (121)$$

is the exciton energy at CM wave vector $\mathbf{g}_1$. The coupling between bare exciton and the Wigner crystal as given by $H_{01R}^K$ in Eq. (83) will be included separately, since it gives negligible contribution as discussed in Cases 1-3 in Section 4. For Case 4, the $H_{01R}^K$ term becomes significant, but the exciton polaron is not formed in this case because the exciton hole and the WC holes are in the same valley.

## 7.2. Polaron effect with intervalley coupling

Next, we will include the effect of intervalley coupling in our calculation. The tetron states with s = 0, 1 in the K′ valley do not couple to the right-handed states in the K valley because they involve either different Fermi seas or different polarizations; So, there is no intervalley coupling between the K′-valley and K-valley tetron states. In our system, the dominant intervalley coupling effect arises from the coupling between the Umklapp KK and K′K ′exciton states $\left|X_{s=1}^K\right\rangle_R$, $\left|X_{s=1}^{K'}\right\rangle_R$ of the same circular polarization. To this end, we add another basis state $\left|X_{s=1}^{K'}\right\rangle_R$ into the polaron state $|P\rangle$ in Eq. (107). For the convenience of interpreting the physics, we re-arrange the basis in the order of $\left\{ |X_{s=0}^K\rangle_R, |\mathfrak{I}_{s=0}\rangle_R, |\mathfrak{I}_{s=1}\rangle_R, |X_{s=1}^K\rangle_R, |X_{s=1}^{K'}\rangle_R \right\}$. Then, we include the intervalley coupling term $Jg_1/K$ ($S18$) as in Eq. (75,87). The revised matrix equation for the polaron state $|P\rangle$ on the new basis reads as:

$$\begin{pmatrix} E_0^X & U_{X\mathfrak{Z}}^{00} & U_{X\mathfrak{Z}}^{01} & 0 & 0 \\ U_{X\mathfrak{Z}}^{00} & E_0^{\mathfrak{Z}} & 0 & U_{X\mathfrak{Z}}^{10} & 0 \\ U_{X\mathfrak{Z}}^{01} & 0 & E_1^{\mathfrak{Z}} + \frac{Jg_1}{K} & U_{X\mathfrak{Z}}^{11} & 0 \\ 0 & U_{X\mathfrak{Z}}^{10} & U_{X\mathfrak{Z}}^{11} & E_1^X + \frac{Jg_1}{K} & \frac{Jg_1}{K} \\ 0 & 0 & 0 & \frac{Jg_1}{K} & E_1^X + \frac{Jg_1}{K} \end{pmatrix} \begin{pmatrix} C_{0R}^K \\ C_{0R}^{\mathfrak{Z}} \\ C_{1R}^{\mathfrak{Z}} \\ C_{1R}^K \\ C_{1R}^{K'} \end{pmatrix} = E_P \begin{pmatrix} C_{0R}^K \\ C_{0R}^{\mathfrak{Z}} \\ C_{1R}^{\mathfrak{Z}} \\ C_{1R}^K \\ C_{1R}^{K'} \end{pmatrix} \qquad (122)$$

Here $E_P$ denotes the eigenvalue of the exciton polaron $|P\rangle$. Solving this matrix equation will give five eigen states, labeled by index $i$ = 1-5:

$$|P_i\rangle = C^{\mathrm{K}}_{0R,i}|X^{\mathrm{K}}_{\mathsf{s}=0}\rangle_R + C^{\mathfrak{I}}_{0R,i}|\mathfrak{I}_{\mathsf{s}=0}\rangle_R + C^{\mathfrak{I}}_{1R,i}|\mathfrak{I}_{\mathsf{s}=1}\rangle_R + C^{\mathrm{K}}_{1R,i}|X^{\mathrm{K}}_{\mathsf{s}=1}\rangle_R + C^{\mathrm{K}'}_{1R,i}|X^{\mathrm{K}'}_{\mathsf{s}=1}\rangle_R \quad (123)$$

We may interpret Eq. (122) in the following way. The first $2 \times 2$ block, neglecting the Zeeman shift, has the expression of

$$H_{polaron} = \begin{pmatrix} E^X_{\mathbf{0}} & U^{00}_{X\mathfrak{I}} \\ U^{00}_{X\mathfrak{I}} & E^{\mathfrak{I}}_0 \end{pmatrix} \quad (124)$$

This is a simplified matrix representation of the exciton-polaron Hamiltonian in the basis of exciton and tetron $\{|X^{\mathrm{K}}_{\mathsf{s}=0}\rangle_R, |\mathfrak{I}_{\mathbf{0}}\rangle_R\}$ at zero CM momentum. $U^{00}_{X\mathfrak{I}}$ represents the coupling between the exciton and tetron states. The two eigenstates correspond to the primary attractive and repulsive polarons, which are bright in the spectrum.

The next $3 \times 3$ block, neglecting the Zeeman shift, has the expression

$$H_{Umklapp} = \begin{pmatrix} E^{\mathfrak{I}}_1 + \frac{Jg_1}{K} & U^{11}_{X\mathfrak{I}} & 0 \\ U^{11}_{X\mathfrak{I}} & E^X_1 + \frac{Jg_1}{K} & \frac{Jg_1}{K} \\ 0 & \frac{Jg_1}{K} & E^X_1 + \frac{Jg_1}{K} \end{pmatrix} \quad (125)$$

This is a simplified matrix representation of the Hamiltonian of Umklapp exciton polarons in the basis of three Umklapp states $\left\{|\mathfrak{I}_1\rangle_R, |X^{\mathrm{K}}_{\mathsf{s}=1}\rangle_R, |X^{\mathrm{K}'}_{\mathsf{s}=1}\rangle_R\right\}$ at the first star of CM wave vectors. It is important to note that, unlike the two bright polaron states described by Eq. (121), these three Umklapp states here are dark because they have finite CM momentum. These Umklapp states are brightened by coupling to the polaron states through the $U^{01}_{X\mathfrak{I}}$ and $U^{10}_{X\mathfrak{I}}$ terms in Eq. (120).

### 7.3. Simulations with only polaron effect at zero magnetic field

In this section, we will calculate the absorption strength of the Umklapp states. In these calculations, we include only the polaron effect, represented by the $U_{XI}$ terms in Eq. (122), while neglecting all exciton–WC scattering processes, corresponding to the $H^{\mathrm{K}}_{01R}$ and $H^{\mathrm{K}}_{10R}$ terms in Eq. (87). We will show that the polaron effect can brighten the Umklapp states in a different way from the exciton-WC scattering effect. We compute the exciton energy ($E^X_{\mathbf{g}}$) and trion energy ($E_T$) by the variational method described in Ref. (S22). We consider zero magnetic field (i.e. no Zeeman shift and the WC particles are evenly distributed between two valleys). The calculations for the electron and hole sides are different because they involve different number of tetrons. On the hole side, we use an effective mass of $0.44\ m_0$ for the valence band of monolayer WSe$_2$. Our calculation obtains a trion binding energy of 18.5 meV, which matches our experimental value (~19 meV) near zero density.

### 7.3.1. Simulation of the Umklapp lines on the hole side

As the hole side involves only one type of tetron, the calculation can be conducted by using the Hamiltonian in Eq. (122). After solving Eq. (122), we obtain the polaron eigen states $|P_i\rangle$ in Eq. (123). The absorption strength of $|P_i\rangle$ is calculated with the Fermi's golden rule as:

$$S^i \propto |C_{0R,i}^{\text{K}} \langle WC|\hat{\mathbf{e}}_R \cdot \mathbf{p}|X_{s=0}^{\text{K}}\rangle_R|^2 + |C_{0R,i}^{\Im} \langle WC|\hat{\mathbf{e}}_R \cdot \mathbf{p}|\Im_{\mathbf{0}}\rangle_R|^2. \qquad (126)$$

Here $C_{0R,i}^{\text{K}}$ and $C_{0R,i}^{\Im}$, as defined in Eq. (123), are the expansion coefficients of the bright components $\{|X_{s=0}^{\text{K}}\rangle_R, |\Im_{\mathbf{0}}\rangle_R\}$ that contribute to the absorption. $\langle WC|\hat{\mathbf{e}}_R \cdot \mathbf{p}|X_{s=0}^{\text{K}}\rangle_R$ and $\langle WC|\hat{\mathbf{e}}_R \cdot \mathbf{p}|\Im_{\mathbf{0}}\rangle_R$ denote the optical transition matrix element of the bright exciton and tetron, respectively. $\hat{\mathbf{e}}_R$ denotes the polarization vector for the right-handed light. The tetron optical transition matrix is difficult to calculate, since it involves a detailed description of the polarization of Wigner crystal induced by the exciton. Here, we shall treat the ratio

$$\left|\langle WC|\hat{\mathbf{e}}_R \cdot \mathbf{p}|\Im_{\mathbf{0}}\rangle_R / \langle WC|\hat{\mathbf{e}}_R \cdot \mathbf{p}|X_{s=0}^{\text{K}}\rangle_R\right|^2 \equiv \bar{R}_{\Im X} n_{wc} \qquad (127)$$

as an empirical parameter. For light with the left-handed ($L$) polarization, we simply replace the subscript $R$ by $L$, and the Hamiltonian matrix remains unchanged.

The oscillator strength in Eq. (126) does not include any screening effect of the WC particles. In realistic conditions, the injected electrons or holes will screen the Coulomb interaction and reduce the exciton oscillator strength. We describe phenomenologically the density-dependent oscillator strength of the exciton under such screening effect by the expression

$$O_X(n_{wc}) = \begin{cases} \frac{1}{1+f_s n_{wc}} \left|\langle WC|\hat{\mathbf{e}}_R \cdot \mathbf{p}|X_{s=0}^{\text{K}}\rangle_R\right|^2 & \text{for } n_{wc} \leq n_0 \\ O_X(n_0)e^{-\zeta_X (n_{nw}-n_0)} & \text{for } n_{wc} > n_0 \end{cases} \qquad (128)$$

Here $n_0$, $f_s$ and $\zeta_X$ are empirical parameters. We set $n_0 = 6 \times 10^{11}\text{cm}^{-2}$ because the Umklapp signal subsides near this density in our experiment. We adjust the values of $f_s$ and $\zeta_X$ to fit our density-dependent optical intensity of the primary exciton peak ($A^0$) in our experiment. The best-fit $f_s$ and $\zeta_X$ are presented in Table S2. After including the screening effect of the WC particles, the oscillator strength in Eq. (126) becomes

$$S^i \propto |C_{0R,i}^{K}|^2 O_h(n_{wc}) + |C_{0R,i}^{\Im}|^2 \bar{R}_{\Im X} n_{wc} O_h(0) \tag{129}$$

At the absence of magnetic field, the WC particles are distributed evenly between the K and K′ valleys with opposite spins. One half of the WC sites are associated with spin-up band and the other half spin-down bands. In the case of hole Wigner crystal, when we consider a KK exciton in spin-up conduction and valence valleys, it can only form a polaron on a WC site of spin-down valley. The WC site with spin-up valleys won't contribute to the polaron formation.

In order to obtain realistic spectra of exciton polarons, we need to determine a series of fitting parameters, including $\bar{U}_{X\Im}$ and $\bar{R}_{\Im X}$ in Eq. (115, 127). We determine the value of $\bar{U}_{X\Im}$ by fitting the density-dependent exciton-polaron energies at zero magnetic field. In our experiment, the energies of the primary exciton-polaron branches shift with increasing hole density ($n_{wc}$). These energy shifts come from two factors: (1) the energy shift of the underlying exciton and tetron components due to the screening effect; (2) the increasing repulsion between the exciton and tetron with increasing carrier density. We describe Factor (1) phenomenologically by assuming a density-dependent exciton energy $E_0^X(n_{wc}) = E_0^X(0)/[1 + \chi_X(n_{wc})]$ with $\chi_X(n_{wc})$ given in Eq. (86).

The tetron energy is $E_0^{\Im} = E_0^T + \Delta_{T4}$ according to Eq. (118). We describe Factor (2) by using the fitting parameter $\bar{U}_{X\Im}$ to quantify the exciton-tetron repulsion. With the parameters $f_c$ [defined in Eq. (86)] and $\bar{U}_{X\Im}$, we can calculate the energies of the exciton-polaron upper and lower branches. We determine the values of these parameters by fitting the measured energies of the upper and lower polaron lines in our experiment. Furthermore, we determine the parameter $\bar{R}_{\Im X}$ in Eq. (127) by fitting the density-dependent ratio of oscillator strengths between the lower and upper branches of the primary exciton polaron under zero magnetic field. Table S2 presents the best-fit values of $\bar{U}_{X\Im}$, $\bar{R}_{\Im X}$, $f_s$ and $\zeta_X$ as well as $f_c$ defined in Eq. (86) for our experimental results on the hole side.

| $\bar{U}_{X\Im}$ (eV · nm) | $\bar{R}_{\Im X}$ (nm²) | $f_c$ (nm²) | $f_s$ (nm²) | $\zeta_X$ (nm²) |
|---|---|---|---|---|
| 0.025 | 28 | 43.5 | 400 | 0.15 |

**Table S2**. The best-fit values of $\bar{U}_{X\Im}$, $\bar{R}_{\Im X}$, $f_c$, $f_s$, and $\zeta_X$ for the hole side.

The best-fit parameters in Table S2 allow us to artificially follow the observed energy shift and reduction of oscillator strength of the primary upper and lower exciton-polaron branches on the hole side. Eq. (129) allows us to calculate the oscillator strength $S^i$. By combining these methods, we can simulate the full density-dependent absorption map on the hole side, including both the primary exciton-polaron states and the Umklapp states.

Figure S16 presents the calculated density-dependent energies and absorption strength of the two primary exciton polaron states ($A^0$, $A^+$) and the three Umklapp states ($A_{U1}^0$, $A_{U2}^0$, $A_U^+$) on the hole side. We note, in the limit of zero carrier density, the $A_{U1}^0$ and $A_{U2}^0$ correspond to the antisymmetric and symmetric combinations $(|X_{s=1}^K\rangle_R \mp |X_{s=1}^{K'}\rangle_R)/\sqrt{2}$ of the Umklapp excitons in the K and K' valleys, respectively; $A_{U1}^0$ and $A_{U2}^0$ have an energy separation of $2Jg_1/K$ due to intervalley coupling.

Figure S17 displays the simulation map of the optical conductivity on the hole side, plotted in a log scale to reveal the weak Umklapp lines. We can visualize two Umklapp lines associated with the upper exciton polaron branch, arising from the intervalley coupling effect. However, these two Umklapp lines are too weak to account for our experimental observation, indicating that the polaron effect alone is insufficient to account for the Umklapp lines on the hole side. On the other hand, our simulation shows an Umklapp line associated with the lower polaron branch on the hole side. But this Umklapp line is not resolved in our experiment (Fig. 2), presumably due to the strong background signal in that spectral region.

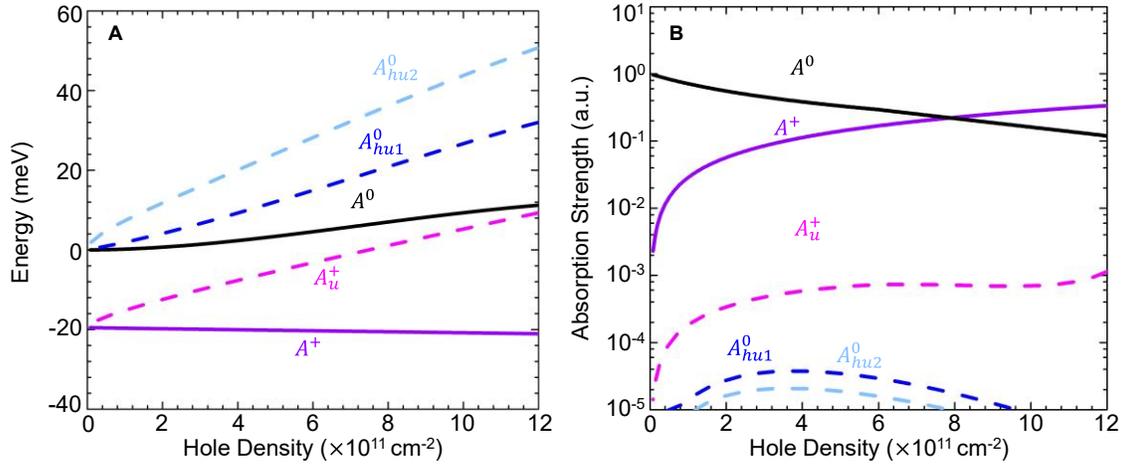

**Figure S16**. (A) Calculated energy and (B) absorption strength of the primary exciton polaron states ($A^0$, $A^+$; solid lines) and the Umklapp states ($A_{hu1}^0$, $A_{hu2}^0$, $A_u^+$; dashed lines) as a function of Wigner-crystal hole density at zero magnetic field. We use $J = 180$ meV for the intervalley coupling strength. The calculation includes the polaron effect but no exciton-WC scattering effect.

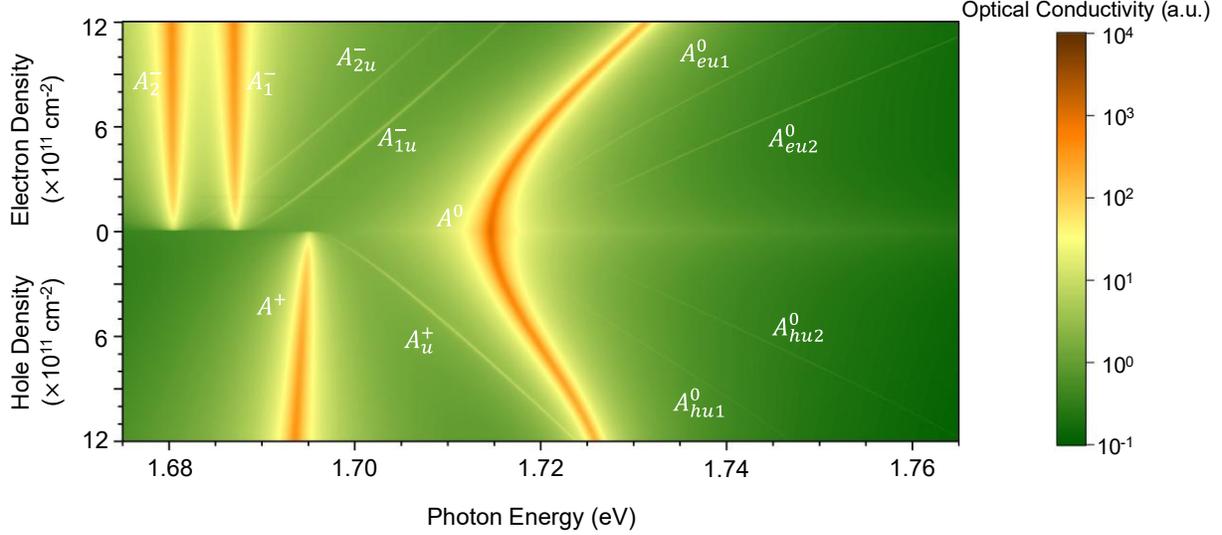

**Figure S17.** Simulation of the optical conductivity including only the polaron effect but no exciton-WC scattering effect. This figure is the same as Fig. 4B in the main paper.

### 7.3.2. Simulation of the Umklapp lines on the electron side

For the electron Wigner crystal, we need to consider two lower branches of exciton polarons ($A_1^-$, $A_2^-$), which are observed in our experiment. Since the KK bright exciton includes an electron in the upper K conduction valley (spin up), it can form the $A_1^-$ trion (commonly called the spin-triplet trion) with a WC electron residing the lower K′ conduction valley (spin up), or form the $A_2^-$ trion (commonly called the spin-singlet trion) with a WC electron residing the lower K conduction valley (spin down) (*S23*). The formation of $A_1^-$ trion is hardly affected by the Pauli exclusion principle because the two spin-up electrons reside in opposite valleys. Using the long-range direct Coulomb interactions and the short-range exchange interactions described in Section 4, we can calculate the $A_2^-$ trion state via the variational method as described in Ref. (*S22*). We adopt an effective mass of 0.46 $m_0$ for the lowest conduction band of monolayer WSe$_2$, which takes into account the non-parabolicity effect. Our calculation obtains a trion binding energy of 21 (26) meV for the $A_1^-$ ($A_2^-$) trion. These values are significantly lower than the experimentally observed binding energies of ~27 (34) meV for the $A_1^-$ ($A_2^-$) state. It was pointed out in Ref. (*S23*), 6-body states can partcipate in the exciton-polaron ground state that provides another mechanism to increase the polaron binding energy, which can be included in the term as given in Eq. (118) . Here, we simply adopt the empirical values as their binding energies, drawn from the experimental observation.

Similar to Eq. (123) on the hole side, the polaron states ($P_0$) with zero momentum and right-hand polarization in the presence of an electron Wigner crystal are given approximately by

$$|P\rangle \approx C_{0R}^{K}|X_{s=0}^{K}\rangle_R + C_{0R}^{\mathfrak{I}_1}|\mathfrak{I}_{s=0}^1\rangle_R + C_{0R}^{\mathfrak{I}_2}|\mathfrak{I}_{s=0}^2\rangle_R + C_{1R}^{\mathfrak{I}_1}|\mathfrak{I}_{s=1}^1\rangle_R + C_{1R}^{\mathfrak{I}_2}|\mathfrak{I}_{s=1}^2\rangle_R + C_{1R}^{K}|X_{s=1}^K\rangle_R + C_{1R}^{K'}|X_{s=1}^{K'}\rangle_R \quad (130)$$

Here $|\mathfrak{I}_s^1\rangle_R$ ($|\mathfrak{I}_s^2\rangle_R$) denotes the tetron state associated with the $A_1^-$ ($A_2^-$) trion at the $s$-$th$ star of the CM vectors. Afterward, we include the effect of intervalley coupling by adding the basis state $|X_{s=1}^{K'}\rangle_R$. By using the expanded basis $\left\{|X_{s=0}^K\rangle_R, |\mathfrak{I}_{s=0}^1\rangle_R, |\mathfrak{I}_{s=0}^2\rangle_R, |\mathfrak{I}_{s=1}^1\rangle_R, |\mathfrak{I}_{s=1}^2\rangle_R, |X_{s=1}^K\rangle_R, |X_{s=1}^{K'}\rangle_R\right\}$, the matrix equation for the coupled states is expressed as:

$$\begin{pmatrix} E_0^X & U_{X\mathfrak{I}_1}^{00} & U_{X\mathfrak{I}_2}^{00} & U_{X\mathfrak{I}_1}^{01} & U_{X\mathfrak{I}_2}^{01} & 0 & 0 \\ U_{X\mathfrak{I}_1}^{00} & E_0^{\mathfrak{I}_1} & 0 & 0 & 0 & U_{X\mathfrak{I}_1}^{10} & 0 \\ U_{X\mathfrak{I}_2}^{00} & 0 & E_0^{\mathfrak{I}_2} & 0 & 0 & U_{X\mathfrak{I}_2}^{10} & 0 \\ U_{X\mathfrak{I}_1}^{01} & 0 & 0 & E_1^{\mathfrak{I}_1} + \frac{Jg_1}{K} & 0 & U_{X\mathfrak{I}_1}^{11} & 0 \\ U_{X\mathfrak{I}_2}^{01} & 0 & 0 & 0 & E_1^{\mathfrak{I}_2} + \frac{Jg_1}{K} & U_{X\mathfrak{I}_2}^{11} & 0 \\ 0 & U_{X\mathfrak{I}_1}^{10} & U_{X\mathfrak{I}_2}^{10} & U_{X\mathfrak{I}_1}^{11} & U_{X\mathfrak{I}_2}^{11} & E_1^X + \frac{Jg_1}{K} & \frac{Jg_1}{K} \\ 0 & 0 & 0 & 0 & 0 & \frac{Jg_1}{K} & E_1^X + \frac{Jg_1}{K} \end{pmatrix} \begin{pmatrix} C_{0R}^K \\ C_{0R}^{\mathfrak{I}_1} \\ C_{0R}^{\mathfrak{I}_2} \\ C_{1R}^{\mathfrak{I}_1} \\ C_{1R}^{\mathfrak{I}_2} \\ C_{1R}^K \\ C_{1R}^{K'} \end{pmatrix} = E_P \begin{pmatrix} C_{0R}^K \\ C_{0R}^{\mathfrak{I}_1} \\ C_{0R}^{\mathfrak{I}_2} \\ C_{1R}^{\mathfrak{I}_1} \\ C_{1R}^{\mathfrak{I}_2} \\ C_{1R}^K \\ C_{1R}^{K'} \end{pmatrix} \quad (131)$$

Here the meanings of the symbols are essentially the same as those in Eq. (122) for the hole side, except that we include two tetron states $\mathfrak{I}_1$ and $\mathfrak{I}_2$. Solving Eq. (131) gives seven eigen polaron states $|P_i\rangle$ labeled by the index $i$ = 1-7:

$$|P_i\rangle = C_{0R,i}^{K}|X_{s=0}^{K}\rangle_R + C_{0R,i}^{\mathfrak{I}_1}|\mathfrak{I}_{s=0}^1\rangle_R + C_{0R,i}^{\mathfrak{I}_2}|\mathfrak{I}_{s=0}^2\rangle_R + C_{1R,i}^{\mathfrak{I}_1}|\mathfrak{I}_{s=1}^1\rangle_R + C_{1R,i}^{\mathfrak{I}_2}|\mathfrak{I}_{s=1}^2\rangle_R + C_{1R,i}^{K}|X_{s=1}^K\rangle_R + C_{1R,i}^{K'}|X_{s=1}^{K'}\rangle_R \quad (132)$$

Similar to Eq. (129), the absorption strength of $|P_i\rangle$, including the screening effect from the WC particles, is

$$S^i \propto |C_{0R,i}^K|^2 O_e(n_{wc}) + |C_{0R,i}^{\mathfrak{I}_1}|^2 \bar{R}_{\mathfrak{I}_1 X} n_{wc} O_e(0) + |C_{0R,i}^{\mathfrak{I}_2}|^2 \bar{R}_{\mathfrak{I}_2 X} n_{wc} O_e(0) \quad (133)$$

Here $O_e(n_{wc})$ are defined in Eq. (128); $\bar{R}_{\mathfrak{I}_1 X}$ and $\bar{R}_{\mathfrak{I}_2 X}$ are fitting parameters defined similarly as in Eq. (127) for the $\mathfrak{I}_1$ and $\mathfrak{I}_2$ states, respectively.

Similar to the case of the hole side, we need to determine a series of fitting parameters, including $\bar{U}_{X\mathfrak{I}_1}$, $\bar{U}_{X\mathfrak{I}_2}$, $\bar{R}_{\mathfrak{I}_1 X}$, $\bar{R}_{\mathfrak{I}_2 X}$, and $f_c$ as defined in Eq. (111, 127, 128, 86), in order to match the primary exciton polaron features on the electron side in our experiment. The parameters $f_s$ and $\zeta_X$, associated with the density dependence of the exciton oscillator strength, have the same values as those on the hole side. The parameters $\bar{U}_{X\mathfrak{I}_1}$ and $\bar{U}_{X\mathfrak{I}_2}$ are determined by fitting the measured energies of the $A_1^-$, $A_2^-$, $A^0$ lines. $\bar{R}_{\mathfrak{I}_1 X}$ and $\bar{R}_{\mathfrak{I}_2 X}$ are determined by fitting the measured density-dependent ratio of oscillator strengths between the $A_1^-$, $A_2^-$ lines and the $A^0$ line. Table S3 presents the best-fit values of these parameters for our experimental results on the electron side.

| $\overline{U}_{X\mathfrak{I}_1} = \overline{U}_{X\mathfrak{I}_2}$ (eV · nm) | $\overline{R}_{\mathfrak{I}_1 X} = \overline{R}_{\mathfrak{I}_2 X}$ (nm² ) | $f_c$ (nm²) | $f_s$ (nm²) | $\zeta_X$ (nm²) |
|---|---|---|---|---|
| 0.04 | 30 | 22.5 | 400 | 0.15 |

**Table S3**. The best-fit values of $\overline{U}_{X\mathfrak{I}_1}$, $\overline{U}_{X\mathfrak{I}_2}$, $\overline{R}_{\mathfrak{I}_1 X}$, $\overline{R}_{\mathfrak{I}_2 X}$, $f_c$, $f_s$, and $\zeta_X$ for the electron side.

Figure S18 presents the calculated density-dependent energies and absorption strength of the three primary exciton polaron states ($A^0$, $A_1^-$, $A_2^-$) and the four Umklapp states ($A_{eU1}^0$, $A_{eU2}^0$, $A_{1U}^-$, $A_{2U}^-$) on the electron side. We also calculate the optical conductivity and show the simulation map on the electron side in Fig. S17. Unlike the previous simulation with exciton-WC scattering but no polaron effect (Fig. S14), which shows no significant Umklapp lines on the electron side, our simulation with the polaron effect (Fig. S17) reveals four Umklapp lines on the electron side. Our experimental results exhibit three Umklapp lines ($A_{2u}^-$, $A_{1u}^-$, $A_{eu2}^0$) while the $A_{eu1}^0$ line is unresolved, presumably because it resides on the strong tail of the primary exciton (Fig. 2C). These features cannot be reproduced by the simulation with only exciton-WC scattering (Fig. S14), but they can be captured by the simulation with polaron effect (Fig. S17). The comparison provides strong evidence that the polaron effect is essential to induce the Umklapp lines on the electron side.

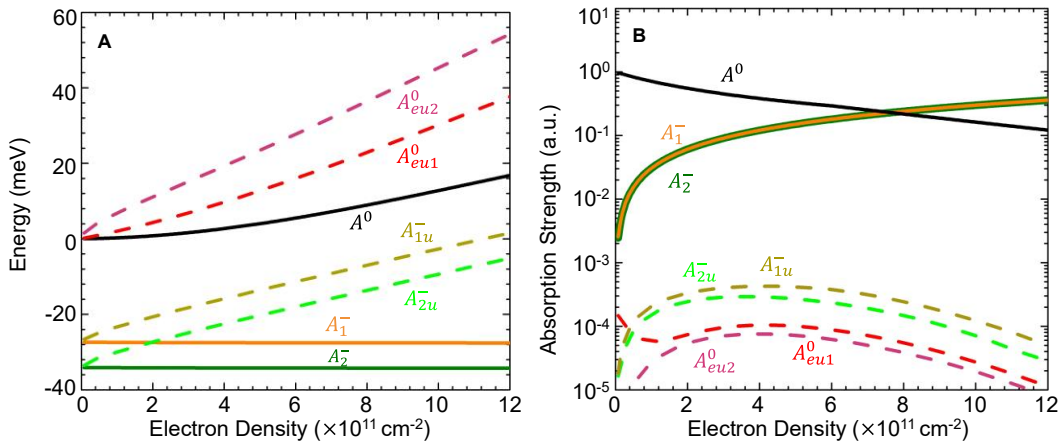

**Figure S18**. (A) Energy and (B) Absorption strength of exciton polaron states of wave vector $\mathbf{k} = 0$ (red curves) and $\mathbf{k} = \mathbf{g_1}$ (green curves) as functions of Wigner-crystal particle density on the electron side at zero magnetic field. We use $J = 160$ meV for the intervalley coupling strength. The calculation includes the polaron effect but no exciton-WC scattering effect.

## 8. Simulations with incoherent combination of exciton-WC scattering and polaron effect

### 8.1. Framework of Simulation

The mechanism to brighten the dark Umklapp states is the main theme of this paper. Our previous discussions consider two brightening mechanisms: (1) the exciton-WC scattering effect discussed in Section 6; (2) the exciton-polaron effect discussed in Section 7. The exciton-WC scattering is contributed by the $H_{01R}^K$ and $H_{01R}^K$ terms in Eq. (83). It induces exciton Umklapp lines, but no polaron Umklapp lines as shown in Fig. S14. Moreover, the exciton Umklapp lines are prominent only on the hole side due to strong exciton-WC scattering in Case 4; they are weak on the electron side because of the weak exciton-WC scattering in Cases 1-2. In contrast, the exciton-polaron effect can brighten both the exciton and polaron Umklapp lines, as shown in Fig. S17. Such brightening is effective on the electron side, where two exciton polarons ($A_1^-$, $A_2^-$) exist. But it becomes less effective on the hole side because of two reasons: (i) only one exciton polaron ($A^+$) exists on the hole side; (ii) No polaron effect exists for Case 4 because the Pauli exclusion principle forbids the polaron formation when the exciton hole and the WC holes occupy the same valley, leaving only Case 3 to contribute to the polaron brightening effect.

When we compare the simulation including only the exciton-WC scattering (Fig. S14) and the simulation including only the polaron effect (Fig. S17), we can see that the two brightening mechanisms contrast and complement each other, that is, the exciton-WC scattering is strong in the region where the polaron effect is weak whereas the polaron effect is strong in regions where the exciton-WC scattering is weak. Therefore, we must consider both effects to adequately account for all the major features in our experiment (Fig. 2C).

In an ideal sample with no disorder, the the exciton-WC scattering effect and polaron effect can add up coherently to produce quantum interference effects. But such a coherent model is not applicable for realistic samples, where defects eliminate the quantum interference between the two effects. Therefore, here we only consider the incoherent combination of the exciton-WC scattering effect and polaron effect, neglecting the quantum interference between them. It turns out that the coherent and incoherent combination give similar results, because the two brightening mechanisms contrast and complement each other. The exciton-WC scattering is strong in the region where the polaron effect is weak, whereas the polaron effect is strong in regions where the exciton-WC scattering is weak. As a result, the quantum interference between them is weak and can be neglected.

Our incoherent model utilizes the first perturbation theory. First, we solve the Hamiltonians in Eqs. (122,131) by setting the exciton-WC coupling terms to $H_{01R}^K = H_{10R}^K = 0$ and obtain the eigen states $|P_i\rangle$ including only the exciton-polaron effect. The absorption strength of the polaron eigen states $|P_i\rangle$ are calculated as $S^i$ in Eq. (129) for the hole side and Eq. (133) for the electron side. Afterward, we treat $H_{01R}^K$, $H_{10R}^K$ as perturbation terms to obtain the new eigen states $|\bar{P}_i\rangle$:

$$|\bar{P}_i\rangle = |P_i\rangle + \sum_{j \neq i} |P_j\rangle \frac{\langle P_j|\Delta H|P_i\rangle}{E_i - E_j} \tag{134}$$

Here $\Delta H$ is a perturbation matrix that contains $H_{01R}^K$ and $H_{10R}^K$ as the only non-zero matrix elements whereas all other matrix elements are zero. We note that the influence of $\Delta H$ on the eigen energies is tiny and can be neglected. The new eigen states $|\bar{P}_i\rangle$ have oscillator strength $\bar{S}^i \propto S^i + \Delta S^i$, where $\Delta S^i$ is the correction term with the expression:

$$\Delta S^i = |C_{1R,i}^K|^2 O_{e,h}(n_{wc}) \left| \frac{\langle WC; X_0^K|H_{01R}^K|WC; X_{g_1}^K\rangle}{\Delta E_i(g_1)} \right|^2 + cross\ terms \tag{135}$$

Here the first term is a direct term that includes $|C_{1R,i}^K|^2$, where $C_{1R,i}^K$ is the expansion coefficient of $|P_i\rangle$. It is present here because the $H_{01R}^K$, $H_{10R}^K$ terms couple the $|X_{s=0}^K\rangle_R$ and $|X_{s=1}^K\rangle_R$ basis states together. All other terms in Eq. (135) are cross terms that represent the interference between the exciton-WC scattering effect and exciton-polaron effect. These cross terms can be removed in realistic samples, where defects randomize carrier phases. Therefore, we only keep the first term in Eq. (135) to account for an incoherent combination of the exciton-WC scattering effect and exciton-polaron effect. We note that in the limit of zero exciton-polaron effect, $C_{1R,i}^K = 1$ and Eq. (135) is reduced to $P(\mathbf{g})$ in Eq. (70) multiplied by oscillator strength $O_{e,h}(n_{wc})$ of the primary exciton.

### 8.2. Simulation Results

Figure S20 presents the simulation map of the absorption spectra in our perturbative model, which eliminates quantum interference between the exciton-WC scattering and the polaron effects. The results successfully capture all the major features in our experiment (Fig. 2C). We also conducted simulations using a fully coherent model that incorporates both exciton-WC scattering and polaron effect within the Hamiltonian, and the results were found to be similar. This is because the two brightening mechanisms contrast and complement each other. The exciton-WC scattering is strong in the region where the polaron effect is weak, whereas the polaron effect is strong in regions where the exciton-WC scattering is weak. As a result, the quantum interference between them is weak and shows negligible effect on the simulation maps.

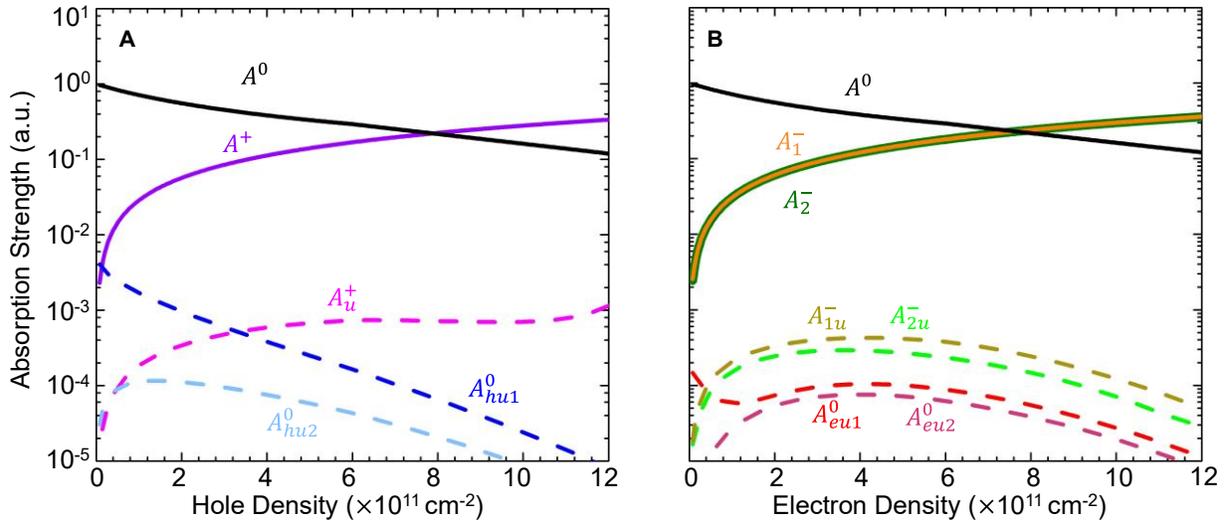

**Figure S19**. (A) Calculated absorption strength of the primary exciton polaron states ($A^0$, $A^+$; solid lines) and the Umklapp states ($A^0_{hu1}$, $A^0_{hu2}$, $A^+_u$; dashed lines) as a function of Wigner-crystal hole density at zero magnetic field. (B) Similar plot for the electron Wigner crystal. We $J = 180$ meV for the hole side and $J = 160$ meV for the electron side. The calculations consider the incoherent combination of the exciton-WC coupling and the polaron effect.

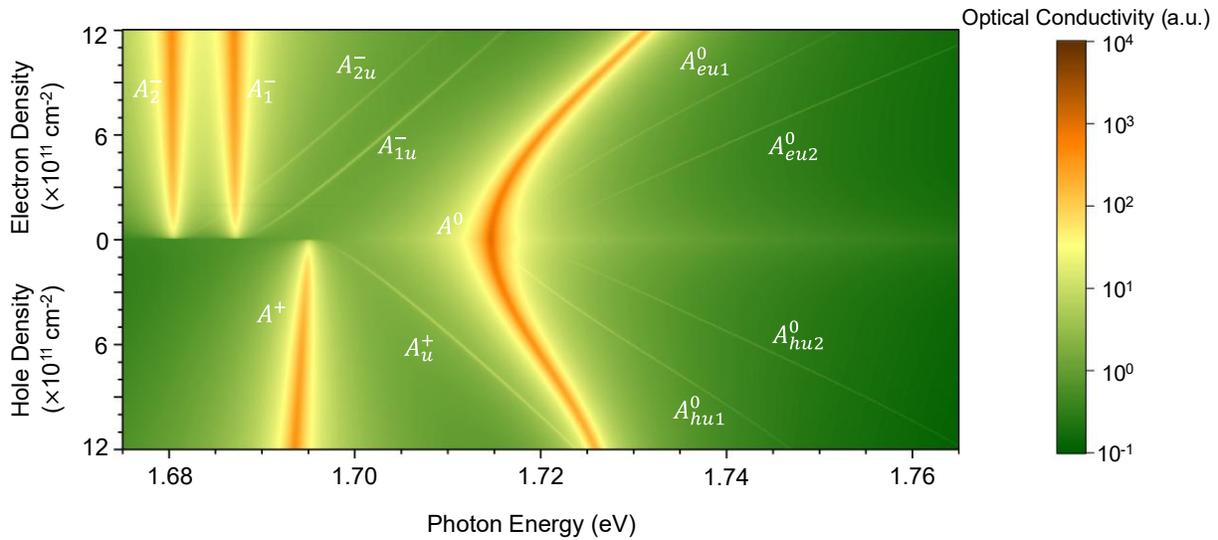

**Figure S20**. Simulation of the optical conductivity considering the incoherent combination of the exciton-WC scattering and polaron effects. This figure is the same as Fig. 4C in the main paper.

By examining the results of our calculations, we can gain important insight into the contribution of different effects and states to brightening the Umklapp states. Here we will analyze the results of Section 8.3 of the incoherent model that better describes realistic samples. Fig. S21 displays the contribution of the exciton-WC scattering (X-WC lines) and of the exciton-polaron effect (XP lines) at varying WC particle density. To further analyze the details of the exciton-polaron effect, we also present the contribution of oscillator strength from different primary states in each Umklapp state (shaded curves).

Figure S21 reveals a few remarkable features. First, the oscillator strength of all the polaron Umklapp lines ($A_{2u}^-$, $A_{1u}^-$, $A_u^+$) is contributed dominantly by the exciton-polaron effect, whereas the exciton-WC scattering effect is tiny (Fig. S21A, B, E). Interestingly, although the exciton-polaron effect plays an essential role, the oscillator strength comes predominantly from the primary exciton ($A^0$), not the tetron states ($A^+$, $A_1^-$, $A_2^-$). We can perceive such behavior by examining the Hamiltonian in Eq. (122) and (131), where the coupling term between the $|\mathfrak{I}_{s=0}\rangle_R$ and $|\mathfrak{I}_{s=1}\rangle_R$ is zero because these two states are defined as the eigen states of the tetron system in the WC lattice. However, the coupling term $U_{X3}^{01}$ between the $|\mathfrak{I}_{s=1}\rangle_R$ and the primary exciton $|X_{s=0}^K\rangle_R$ is finite. Therefore, the oscillator strength of the polaron Umklapp states comes from the primary exciton, while its energies are determined by the zone-folding effect of the polaron dispersion in the presence of Wigner crystal (Fig. 1C).

Second, the oscillator strength of exciton Umklapp lines ($A_{hu1}^0$, $A_{hu2}^0$) on the hole side (Fig. S21F, G) is contributed dominantly by exciton-WC scattering due to the presence of long-range exchange interaction in Case 4.

Third, the oscillator strength of exciton Umklapp lines ($A_{eu1}^0$, $A_{eu2}^0$) on the electron side (Fig. S21C, D) are contributed comparably by both the exciton-WC scattering effect and the polaron effect at low electron density, but the polaron effect becomes dominant at high electron density. The lower-energy $A_{eu1}^0$ always has stronger exciton-WC scattering than the higher-energy $A_{eu2}^0$ because the exciton-WC scattering probability depends on the inverse square of energy (relative to the primary exciton). Therefore, the polaron effect plays a more important role in $A_{eu2}^0$ than $A_{eu1}^0$. As the density increases, the exciton-WC scattering in both $A_{eu1}^0$ and $A_{eu2}^0$ drop and the polaron effect becomes the dominant contribution in both lines. As $A_{eu1}^0$ and $A_{eu2}^0$ have comparable polaron effect, they become similarly bright at high density.

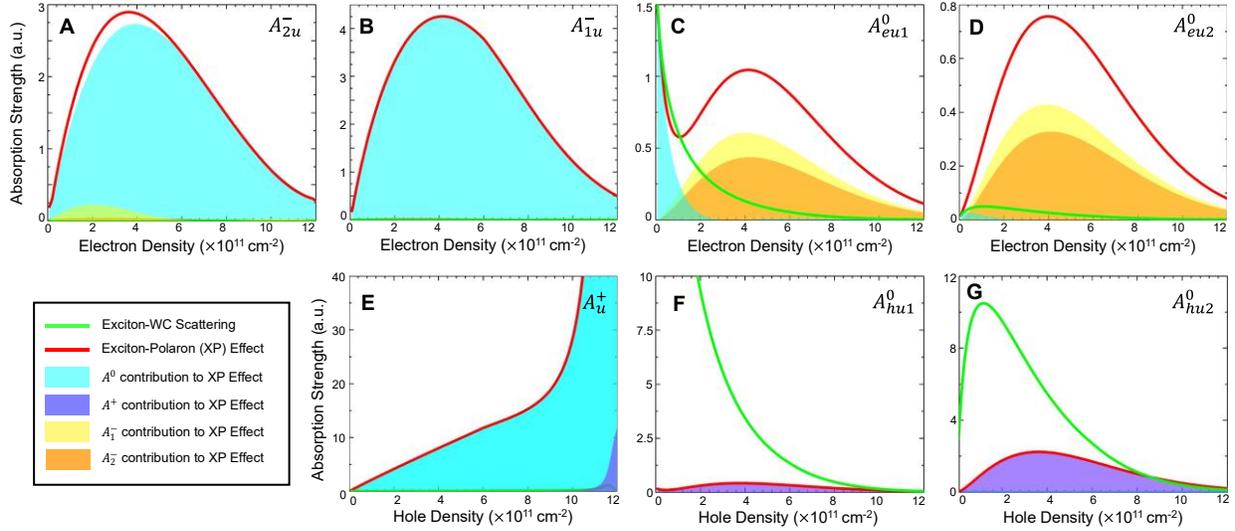

**Figure S21**. (A-D) Calculated density-dependent absorption strength of the $A_{2u}^-$, $A_{1u}^-$, $A_{eu1}^0$, and $A_{eu2}^0$ Umklapp lines on the electron side, arising from two distinct brightening mechanisms. The green line represents absorption strength originating from the exciton-WC scattering, while the red line indicates the absorption strength due to the exciton-polaron effect. The shaded areas correspond to the absorption strength transferred from four primary excitonic states: $A^0$ (aqua), $A^+$ (purple), $A_1^-$ (yellow), and $A_2^-$ (orange), via the exciton-polaron effect. (E-G) Similar plots for the $A_u^+$, $A_{hu1}^0$, and $A_{hu2}^0$ Umklapp lines on the hole side, with the same color coding and distinctions between mechanisms as described for the electron side.

### *8.3. Simulation Results with Realistic Broadening*

In the following section we discuss the process used to convert our purely theoretically calculated data into simulated maps which are used to directly compare to experimental data. Density-dependent dispersion and linewidth of principle peaks are determined by the optical conductivity data extracted from experimental data (detailed in Section I.3). Theoretically predicted Umklapp scattering peaks are then calculated using the extracted peak parameters. The experimental major peaks, calculated Umklapp peaks, and experimentally extracted Lorentzian background (detailed in Section I.3) are all added together to form the simulated real optical conductivity map. KK-relations are used to calculate simulated imaginary optical conductivity which, together with real conductivity, can be used to create a simulated reflection contrast map via the transfer matrix technique (*S1, 2*). Two energy derivatives of the reflection contrast are taken to flatten the background and accentuate the weak Umklapp signals close to the principal peaks. The charge neutrality gap is simulated by spline interpolation between the calculated lowest non-zero density lines. The simulated and experimental second derivative maps are compared, and iterative corrections are necessary to optimize aspects of the Umklapp lines that our theory cannot predict (*i.e.* intervalley coupling strength (J) and broadening of individual Umklapp peaks). Though the simulated and experimental optical conductivity and reflection maps look dissimilar, the final iterative steps of this process lead to simulated second derivative maps that look very similar to their experimental counterparts.

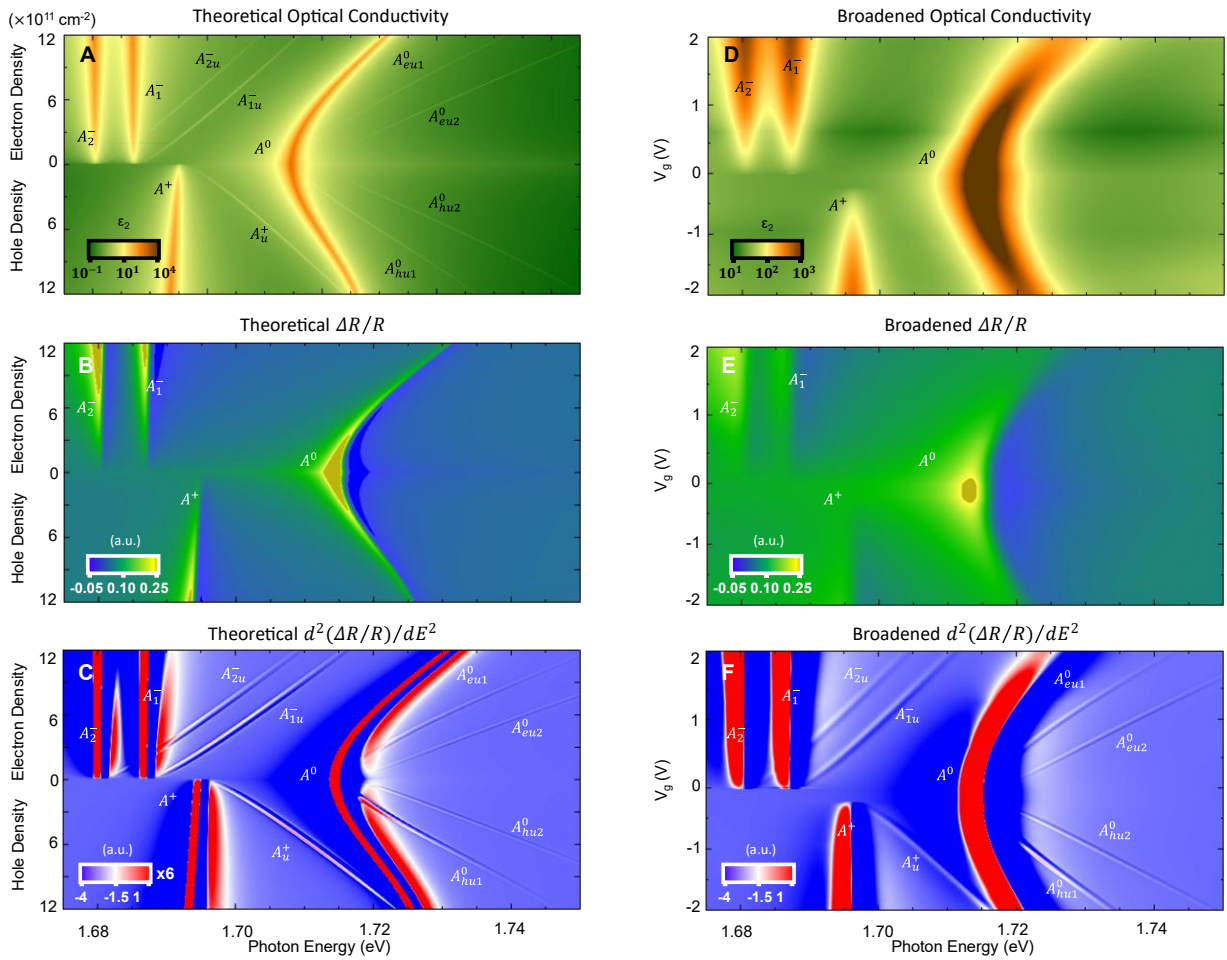

**Figure S22**. (A) Simulated optical conductivity map. (B) Reflectance contrast map calculated from panel A. (C) Second-order energy derivative of panel B. (D) Simulated optical conductivity map including a background contribution (e.g., from defects and higher-energy features) and a density-dependent linewidth, enabling closer comparison with experiment. (E) Reflectance contrast map calculated from panel D. (F) Second-order energy derivative of panel E. Panel F is identical to Fig. 2F in the main text.

## 9. Simulation of Umklapp lines at finite magnetic field

Our calculations in Sections 1-8 consider no magnetic field. In this section, we will present our calculations at finite magnetic field. In all of our calculations Section 9, we neglect the effect of Landau quantization and only consider the valley Zeeman shift and the redistribution of the carriers between the two valleys under magnetic field. At zero magnetic field, the WC particles are evenly distributed between the energy-degenerate K and K' valleys. However, at high magnetic field, the two valleys become non-degenerate and all the WC particles reside in one valley. In the following sections, we will calculate the Umklapp lines with exciton-WC scattering and exciton-polaron effect under magnetic field.

We compare our calculated results with measurements from a different monolayer WSe₂ device under a magnetic field. Because the screening environment differs, the screening effect on the primary exciton is modified. Accordingly, the screening parameter $f_c$ used in Eq. (86) to describe the density-dependent shift of the exciton principal line is adjusted as follows: the best-fit values of $f_c$ are 10 nm² for Cases 1 and 2, 12 nm² for Case 3, and 60 nm² for Case 4.

### 9.1. Calculation with only exciton-WC scattering

We first calculate the Umklapp lines including only the exciton-WC scattering (without any exciton-polaron effect). We have carried out such calculations for zero magnetic field in Section 6, and here we will revise the calculations for a finite magnetic field. A magnetic field (B) can induce Zeeman energy splitting of $\Delta_B = g_B \mu_B B$ between the KK and K'K' excitons, where $g_B$ is the g-factor of the exciton in monolayer WSe₂ and $\mu_B$ is the Bohr magneton. The splitting energy reaches $\Delta_B \approx 4.3$ meV at $B = 17\ T$. After including the Zeeman shifts of $-\Delta_B/2$ for the K valley and $+\Delta_B/2$ for the K' valley, the matrix equation in Eq. (87) becomes:

$$\begin{pmatrix} H_{00R}^{K} - \frac{\Delta_B}{2} & H_{01R}^{K} & 0 \\ H_{10R}^{K} & H_{11R}^{K} - \frac{\Delta_B}{2} + \frac{Jg_1}{K} & \frac{Jg_1}{K} \\ 0 & \frac{Jg_1}{K} & H_{11R}^{K'} + \frac{\Delta_B}{2} + \frac{Jg_1}{K} \end{pmatrix} \begin{pmatrix} C_{0R}^{K} \\ C_{1R}^{K} \\ C_{1R}^{K'} \end{pmatrix} = (E_{wc} + E_R) \begin{pmatrix} C_{0R}^{K} \\ C_{1R}^{K} \\ C_{1R}^{K'} \end{pmatrix} \qquad (136)$$

Here $\left( C_{0R}^{K}, C_{1R}^{K}, C_{1R}^{K'} \right)$ and $E_R$ are the eigen vector and eigen energy of the exciton that includes both the Umklapp scattering effect and the intervalley mixing effect on the basis of $\left\{ |X_{s=0}^{K}\rangle_R; |X_{s=1}^{K}\rangle_R; |X_{s=1}^{K'}\rangle_R \right\}$. The Umklapp coupling terms $H_{01R}^{K}$ and $H_{10R}^{K}$ have negligible influence on the eigen energy, though they must be included in the calculation of eigen vectors to give the Umklapp scattering effect. By neglecting the $H_{01R}^{K}$ and $H_{10R}^{K}$ terms, the energies of the primary exciton (labeled by "0") and two Umklapp excitons (labeled by "$u1$" and "$u2$") have the analytic expressions (*S5*)

Primary exciton energy:    $E_0 = E_0^X(n_{wc}) - \frac{\Delta_B}{2}$    (137)

Umklapp exciton energies: $\begin{cases} E_{u1}(g_1) = E_0^X(n_{wc}) + \frac{\hbar^2 g_1^2}{2m_X} + \frac{Jg_1}{K} - \sqrt{\left(\frac{Jg_1}{K}\right)^2 + \left(\frac{\Delta_B}{2}\right)^2} \\ E_{u2}(g_1) = E_0^X(n_{wc}) + \frac{\hbar^2 g_1^2}{2m_X} + \frac{Jg_1}{K} + \sqrt{\left(\frac{Jg_1}{K}\right)^2 + \left(\frac{\Delta_B}{2}\right)^2} \end{cases}$    (138)

By subtracting Eq. (137) from Eq. (138), we obtain the energy separation between the primary and Umklapp excitons as

$\begin{cases} \Delta E_{u1}(g_1) = \frac{\hbar^2 g_1^2}{2m_X} + \frac{Jg_1}{K} - \sqrt{\left(\frac{Jg_1}{K}\right)^2 + \left(\frac{\Delta_B}{2}\right)^2} + \frac{\Delta_B}{2} \\ \Delta E_{u2}(g_1) = \frac{\hbar^2 g_1^2}{2m_X} + \frac{Jg_1}{K} + \sqrt{\left(\frac{Jg_1}{K}\right)^2 + \left(\frac{\Delta_B}{2}\right)^2} + \frac{\Delta_B}{2} \end{cases}$    (139)

Figure S23A-D (the top row) display the schematics of Cases 1-4 under a finite magnetic field. Fig. S23E-H (the second row) display the corresponding calculated energies of the primary and Umklapp states as functions of WC particle density in Cases 1-4, including both exciton-WC scattering terms ($H_{01R}^K$ and $H_{10R}^K$) and the exciton-polaron effect (which will be discussed in the next section). The energy dispersions are dominantly determined by Eq. (136), while the modification from the exciton-WC scattering and exciton-polaron effect is two to three orders of magnitude smaller and cannot be noticed by visual inspection of the plots. Neglecting the tiny influence of exciton-WC scattering and exciton-polaron effect, Cases 1-2 have the same dispersion, Cases 3-4 also have the same dispersion, and the energy difference between Cases 1-2 and Cases 3-4 are due to the opposite sign of $\Delta_B$ in Eq. (137).

After we solve the eigen states in Eq. (136), we can use the first term in Eq. (135) to calculate the oscillator strength of the eigen states with only exciton-WC scattering. The results for Cases 1-4 are shown in Fig. S23I-L (the third row). Similar to our results in Section 6, the exciton-WC scattering only gives negligible oscillator strength in Cases 1-3 but much stronger oscillator strength in Case 4 due to the existence of long-range exchange interaction in this case.

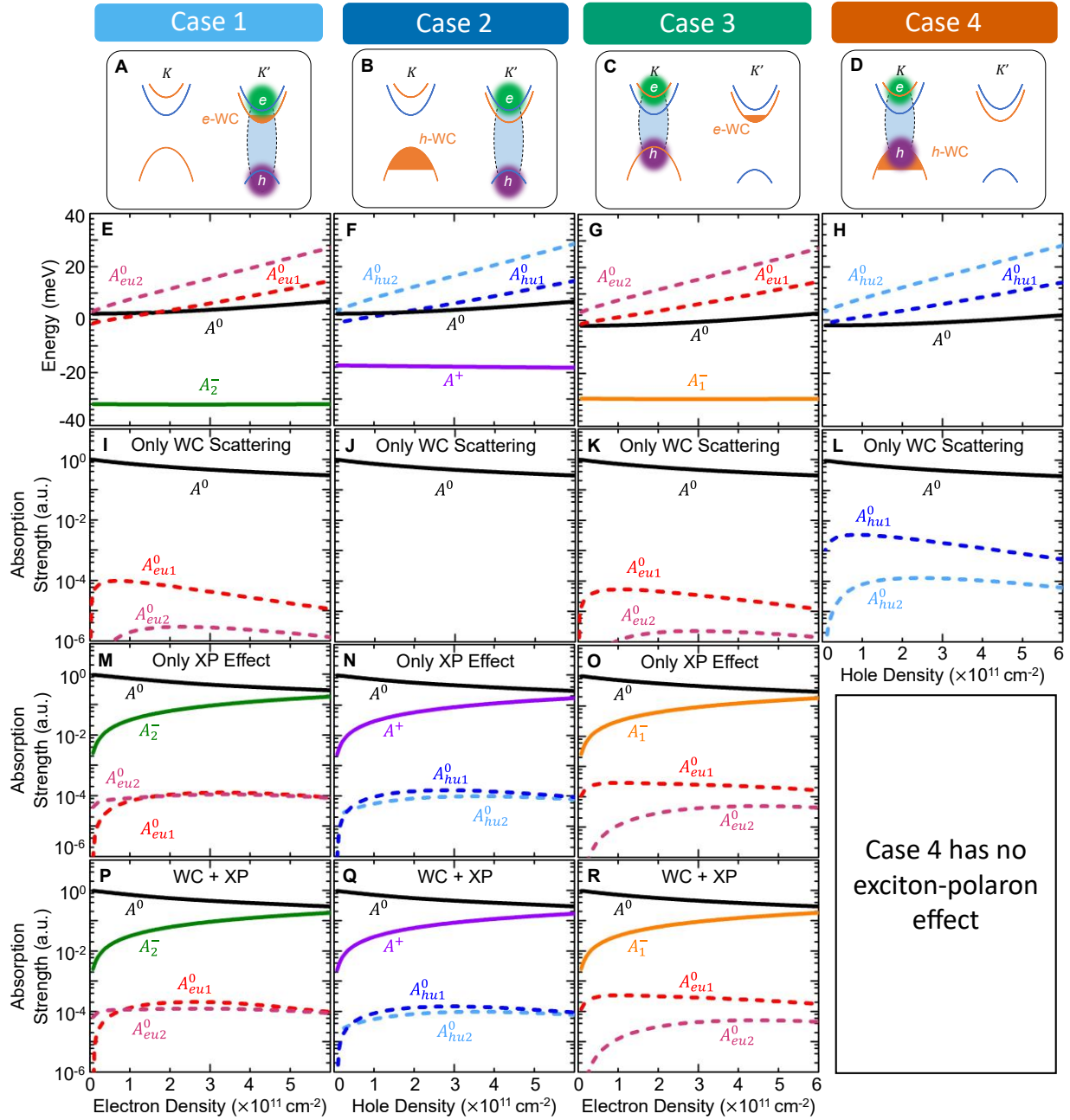

**Figure S23.** (A-D, top row) Schematic of Cases 1-4 under a vertical magnetic field. (E-H, second row) Calculated energy of different excitonic states as functions of WC particle density. (I-L, third row) Calculated density-dependent oscillator strength of different excitonic states with only exciton-WC scattering. (M-O, fourth row) Calculated oscillator strength with only exciton-polaron effect. (P-R, fifth row) Calculated oscillator strength with incoherent combination of exciton-WC scattering and exciton-polaron effect. The plots with polaron effect are vacant for Cases 4 because there is no polaron effect in this case. All calculations are done with B = 17 T magnetic field.

### 9.2. Calculation with only exciton-polaron effect

Next, we calculate the Umklapp lines including only the exciton-polaron effect (no exciton-WC scattering). When a strong magnetic field lifts the valley degeneracy, the injected electrons (holes) reside exclusively in the K' (K) valley, giving rise to Cases 1 – 4 as illustrated in Fig. 3 of the main paper. Cases 1 – 3 allow the formation of exciton polaron, whereas Case 4 does not. The polaron effect can brighten the Umklapp lines in Cases 1 – 3, whereas there is no polaron effect in Case 4. Here we will calculate the oscillator strength of the Umklapp lines including the polaron effect in Cases 1 – 3.

We first do the calculation for Case 1. Case 1 features the intravalley trion ($A_2^-$) with opposite electron spins, whereas the intervalley trion ($A_1^-$) with parallel electron spins is not formed (Fig. 3). Therefore, we can neglect the $\mathfrak{I}_1$ states in our calculation. The corresponding Schrödinger equation can be adopted from Eq. (131) with the following modifications:

(i) We exchange symbols K and K' and change the handedness "$R$" to "$L$" because Case 1 involves K' exciton with left-handed optical polarization. These changes do not affect the exciton-tetron coupling terms.

(ii) When we define $U_{X\mathfrak{I}}$ in Eq. (115) for zero magnetic field, there is a $\sqrt{1/2}$ factor within because only half of the WC sites are available to form a valley tetron when the WC particles occupy two valleys at zero magnetic field. Under strong magnetic field, however, tetrons can be formed in all WC sites since electrons reside exclusively in one valley. To account for this difference, we multiply all the exciton-tetron coupling terms $U_{X\mathfrak{I}_2}$ by a factor of $\sqrt{2}$.

(iii) We add the Zeeman energy $-\Delta_B/2$ for the K valley and $+\Delta_B/2$ for the K' valley in the diagonal matrix elements.

(iv) We remove the $|\mathfrak{I}_{s=0}^1\rangle_R$ and $|\mathfrak{I}_{s=1}^1\rangle_R$ states from the basis in Eq. (131) because $\mathfrak{I}_1$ is not involved in Case 1.

(v) We also remove the $|\mathfrak{I}_{s=1}^2\rangle_R$ state from the basis, because our model cannot describe its coupling properties under strong magnetic field. A prior study shows that, in the presence of a strong magnetic field, the internal motion and the CM motion of a trion can still be decoupled when the CM momentum is small, but they are strongly coupled when the CM momentum is large (*S19*). Our simple model, which assume decoupled internal and CM motion of the trion, is still applicable to describe the $|\mathfrak{I}_{s=0}^2\rangle_R$ with zero CM momentum, but no longer applicable to describe the $|\mathfrak{I}_{s=1}^2\rangle_R$ state with large CM momentum, under strong magnetic field. For simplicity, we remove the $|\mathfrak{I}_{s=1}^2\rangle_R$ state from our model. The removal of the $|\mathfrak{I}_{s=1}^1\rangle_R$ state will eliminate the exciton-polaron Umklapp lines, but have negligible effect on the exciton Umklapp line. As our experiment hardly observe any exciton-polaron Umklapp lines under strong magnetic field, a model without the $|\mathfrak{I}_{s=1}^2\rangle_R$ state is sufficient to account for our experimental findings at strong magnetic field.

With the above changes, the basis becomes $\left\{|X_{s=0}^{K'}\rangle_L, |\mathfrak{I}_{s=0}^2\rangle_L, |X_{s=1}^{K'}\rangle_L, |X_{s=1}^{K}\rangle_L\right\}$ and Eq. (136) becomes

$$
\begin{pmatrix}
E_0^X + \frac{\Delta_B}{2} & \sqrt{2}U_{X\mathfrak{I}_2}^{00} & 0 & 0 \\
\sqrt{2}U_{X\mathfrak{I}_2}^{00} & E_0^{\mathfrak{I}_2} + \frac{\Delta_B}{2} & \sqrt{2}U_{X\mathfrak{I}_2}^{10} & 0 \\
0 & \sqrt{2}U_{X\mathfrak{I}_2}^{10} & E_1^X + \frac{Jg_1}{K} + \frac{\Delta_B}{2} & \frac{Jg_1}{K} \\
0 & 0 & \frac{Jg_1}{K} & E_1^X + \frac{Jg_1}{K} - \frac{\Delta_B}{2}
\end{pmatrix}
\begin{pmatrix}
C_{0L}^{K'} \\
C_{0L}^{\mathfrak{I}_2} \\
C_{1L}^{K'} \\
C_{1L}^{K}
\end{pmatrix}
= E_P
\begin{pmatrix}
C_{0L}^{K'} \\
C_{0L}^{\mathfrak{I}_2} \\
C_{1L}^{K'} \\
C_{1L}^{K}
\end{pmatrix}
\qquad (140)
$$

After setting up the equation for Case 1, we turn to set up the question for Case 2, which features the intervalley trion ($A^+$) with opposite hole spins (Fig. 3). Similar to our treatment in Case 1, we adopt Eq. (117) by multiplying all the exciton-tetron coupling terms $U_{X\mathfrak{I}}$ by a factor of $\sqrt{2}$, exchanging the symbols K and K', changing the handedness "$R$" to "$L$", adding the Zeeman energy terms, and removing the $|\mathfrak{I}_{s=1}\rangle_R$ state from the basis. Eq. (117) becomes

$$
\begin{pmatrix}
E_0^X + \frac{\Delta_B}{2} & \sqrt{2}U_{X\mathfrak{I}}^{00} & 0 & 0 \\
\sqrt{2}U_{X\mathfrak{I}}^{00} & E_0^{\mathfrak{I}} + \frac{\Delta_B}{2} & \sqrt{2}U_{X\mathfrak{I}}^{10} & 0 \\
0 & \sqrt{2}U_{X\mathfrak{I}}^{10} & E_1^X + \frac{Jg_1}{K} + \frac{\Delta_B}{2} & \frac{Jg_1}{K} \\
0 & 0 & \frac{Jg_1}{K} & E_1^X + \frac{Jg_1}{K} - \frac{\Delta_B}{2}
\end{pmatrix}
\begin{pmatrix}
C_{0L}^{K'} \\
C_{0L}^{\mathfrak{I}} \\
C_{1L}^{K'} \\
C_{1L}^{K}
\end{pmatrix}
= E_P
\begin{pmatrix}
C_{0L}^{K'} \\
C_{0L}^{\mathfrak{I}} \\
C_{1L}^{K'} \\
C_{1L}^{K}
\end{pmatrix}
\qquad (141)
$$

Case 3 features the intervalley trion ($A_1^-$) with parallel electron spins, whereas intravalley trion ($A_1^-$) with opposite electron spins is not formed. In this case, we can simply use Eq. (140) but replace $\mathfrak{I}_2$ by $\mathfrak{I}_1$, exchange the symbols K and K', change the handedness "$L$" to "$R$", flip the sign of the Zeeman energies, and remove the $|\mathfrak{I}_{s=1}^1\rangle_R$ state. Eq. (140) becomes

$$
\begin{pmatrix}
E_0^X - \frac{\Delta_B}{2} & \sqrt{2}U_{X\mathfrak{I}_1}^{00} & 0 & 0 \\
\sqrt{2}U_{X\mathfrak{I}_1}^{00} & E_0^{\mathfrak{I}_1} - \frac{\Delta_B}{2} & \sqrt{2}U_{X\mathfrak{I}_1}^{10} & 0 \\
0 & \sqrt{2}U_{X\mathfrak{I}_1}^{10} & E_1^X + \frac{Jg_1}{K} - \frac{\Delta_B}{2} & \frac{Jg_1}{K} \\
0 & 0 & \frac{Jg_1}{K} & E_1^X + \frac{Jg_1}{K} + \frac{\Delta_B}{2}
\end{pmatrix}
\begin{pmatrix}
C_{0R}^{K} \\
C_{0R}^{\mathfrak{I}_1} \\
C_{1R}^{K} \\
C_{1R}^{K'}
\end{pmatrix}
= E_P
\begin{pmatrix}
C_{0R}^{K} \\
C_{0R}^{\mathfrak{I}_1} \\
C_{1R}^{K} \\
C_{1R}^{K'}
\end{pmatrix}
\qquad (142)
$$

Case 4 remains unchanged because this case has no polaron effect. After solving Eqs. (136-138) for Cases 1-3 at B = 17 T, we obtain their eigen vectors and energies. Afterward, we adapt Eq. (126) to calculate their oscillator strengths.

### 9.3. Calculation with incoherent combination of exciton-WC scattering and polaron effect

After we calculate the eigen states with only polaron effect, we incorporate the effect of exciton-WC scattering incoherently in a first-order perturbation theory, similar to the method used in Section 8 for zero magnetic field. Fig. S23 shows the energy and oscillator strength of the primary and Umklapp lines for Cases 1-4 under B = 17 T vertical magnetic field, calculated by incoherent combination of exciton-WC scattering and polaron effect.

Let's first examine our calculated results for Cases 1 and 2, where we detect the K′-valley optical response with left-handed light. Fig. S23E, F shows a finite Zeeman energy separation (~4 meV) between the higher and lower Umklapp lines at B = 17 T. When the charge density approaches zero, the energy of the higher line ($A_{eu2}^0$, $A_{hu2}^0$) converges to that of the primary line ($A^0$), while the energy of the lower line ($A_{eu1}^0$, $A_{hu1}^0$) falls below $A^0$. This behavior arises because the magnetic field induces a Zeeman energy separation ($\Delta_B$) between the K′K′ and KK excitons, which exceeds (or is comparable to) the intervalley coupling strength ($Jg_1/K$) at low (high) density, leading to a significant reduction in intervalley mixing. Consequently, the higher (lower) Umklapp line is primarily contributed by the K′K′ (KK) exciton at the CM wave vector $\mathbf{g_1}$, in contrast to the comparable contributions from both valleys at zero field. Our experiment detects only one Umklapp line for Cases 1-2 at B = 17 T (Fig. 3A). Since its extrapolated energy matches that of the primary exciton at zero density, this observed line must correspond to the higher Umklapp state ($A_{eu2}^0$, $A_{hu2}^0$).

In both Cases 1-2, the oscillator strength of the Umklapp lines is primarily determined by the polaron effect. The two cases exhibit comparable brightness due to similar polaron binding energy and exciton-tetron coupling strength, consistent with our experimental results in Fig. 3A. Our calculations, considering only polaron effect, shows comparable strength between the higher Umklapp line ($A_{eu2}^0$, $A_{hu2}^0$) and the lower line ($A_{eu1}^0$, $A_{hu1}^0$) in our experimental density range (Fig. S23M, N). Their strength remains comparable even when exciton-WC scattering is included (Fig. S23P, Q). Since the higher Umklapp line is further away from the primary exciton line, we expect it to be more visible. This further supports that the observe Umklapp lines in Fig. 3A correspond to the higher Umklapp states ($A_{eu2}^0$, $A_{hu2}^0$).

The situation is different in Case 3, where we detect the K-valley optical response with right-handed light. As the higher (lower) Umklapp line is primarily contributed by the K′K′ (KK) exciton under a strong magnetic field, when the charge density approaches zero, the lower Umklapp line will approach the primary line ($A^0$) while the higher line will lie above $A^0$, as shown in Fig. S23G.

Our experiment detects only one Umklapp line for Case 3 at B = 17 T (Fig. 3B). Since its extrapolated energy matches that of the primary exciton at zero density, this observed line must correspond to the lower Umklapp state ($A_{eu1}^0$).

In Case 3, the tetron state is associated with the KK exciton. The lower Umklapp state ($A_{eu1}^0$), associated with the KK exciton, can couple to the tetron state and gain appreciable absorption strength, but the higher Umklapp state ($A_{eu2}^0$), associated with the K'K' exciton, cannot couple directly to the tetron state except through intervalley mixing. At zero electron density, $A_{eu2}^0$ is dark because of the lack of intervalley mixing. At increasing density, $A_{eu2}^0$ becomes brighter due to increasing intervalley mixing, but it is still considerably dimmer than $A_{eu1}^0$. This can explain why the higher line ($A_{eu2}^0$) is not detected in our experiment for Case 3 (Fig. 3B).

In Case 4, we also detect the K-valley optical response with right-handed light. Fig. S23H displays the density-dependent energy of the higher Umklapp line ($A_{hu2}^0$) and lower Umklapp line ($A_{hu1}^0$) at B = 17 T for Case 4. When the hole density approaches zero, the lower Umklapp line will approach the primary line ($A^0$) while the higher line will lie above $A^0$. Our experiment detects only one Umklapp line for Cases 4 at B = 17 T (Fig. 3B). Since its extrapolated energy matches that of the primary exciton at zero density, this observed line must correspond to the lower Umklapp state ($A_{eu1}^0$).

Unlike Cases 1-3 where the polaron effect dominates over the exciton-WC scattering effect, Case 4 exhibits no polaron effect but strong exciton-WC scattering. Fig. S23L displays the density-dependent oscillator strength of $A_{hu1}^0$ and $A_{hu2}^0$. The lower line ($A_{hu1}^0$) is about an order of magnitude brighter than the higher line ($A_{hu2}^0$) in our experimental density range. This further supports that the observed Umklapp line for Case 4 in our experiment (Fig. 3B) correspond to the lower Umklapp state ($A_{eu1}^0$). Furthermore, our calculations show that the $A_{hu1}^0$ line in Case 4 is about an order of magnitude brighter than all the strongest Umklapp lines in Cases 1-3, consistent with our experimental observation (Fig. 3A-B).

To sum up, when the exciton-polaron effect is included, the absorption strength of exciton Umklapp lines for Cases 1-3 in the high-density regime ($> 3 \times 10^{11} cm^{-2}$) is enhanced by about one order of magnitude. This makes the $A_{eu2}^0$, $A_{hu2}^0$, and $A_{eu1}^0$ lines detectable in Cases 1, 2, and 3, respectively, while the $A_{hu1}$ line in Case 4 is brighter (by about one order of magnitude) even without the exciton-polaron effect.

### *9.4. Simulation maps*

Figure S24 displays the simulation maps of the excitonic optical conductivity of monolayer WSe₂ under B = 17 T vertical magnetic field. Figure S25 displays the second-derivative reflectance contrast maps calculated from the conductivity maps in Figure S24.

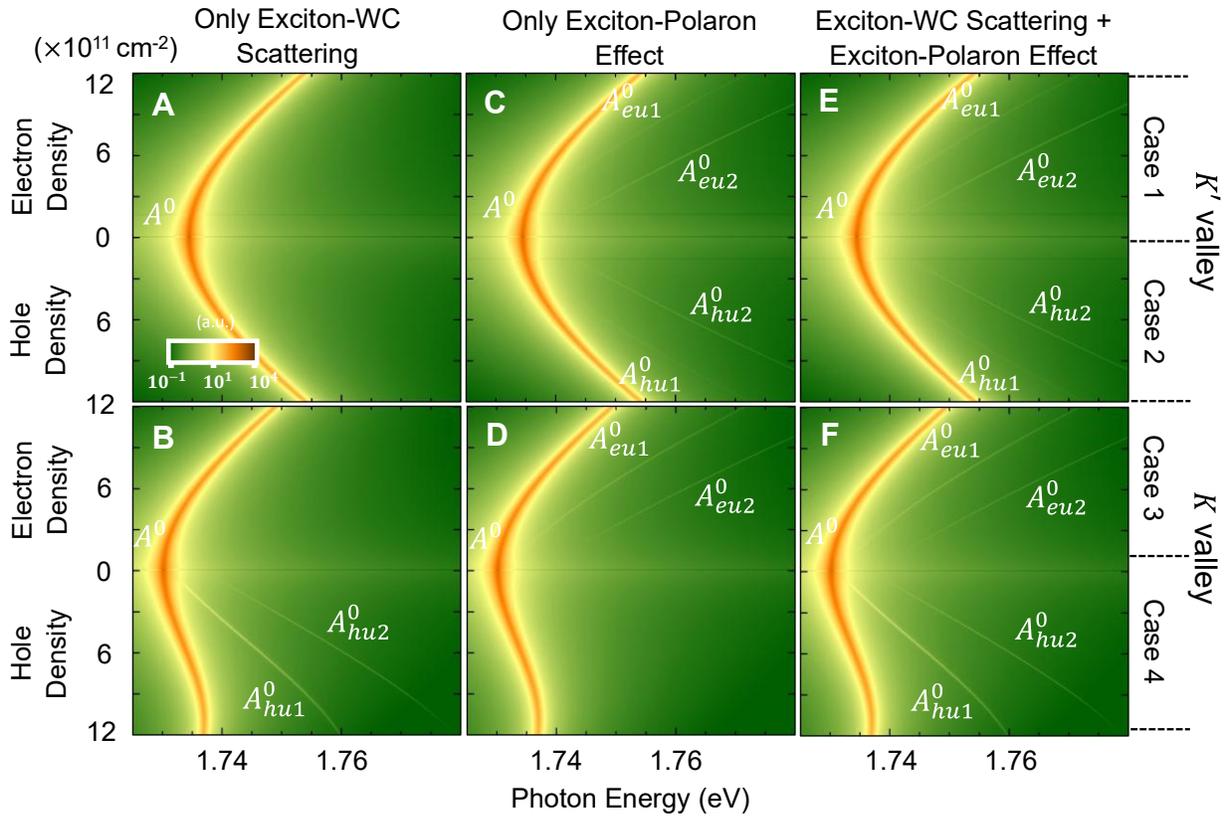

**Figure S24**. Simulation maps of the optical conductivity of the primary and Umklapp lines, plotted in a log scale, for monolayer WSe₂ under B = 17 T vertical magnetic field. The top (bottom) row corresponds to K'-valley (K-valley) optical transition. (A-B) Simulation with only exciton-WC scattering. (C-D) Simulation with only exciton-polaron effect. (E-F) Simulation with incoherent combination of exciton-WC scattering and exciton-polaron effect. All plots here are in log scale.

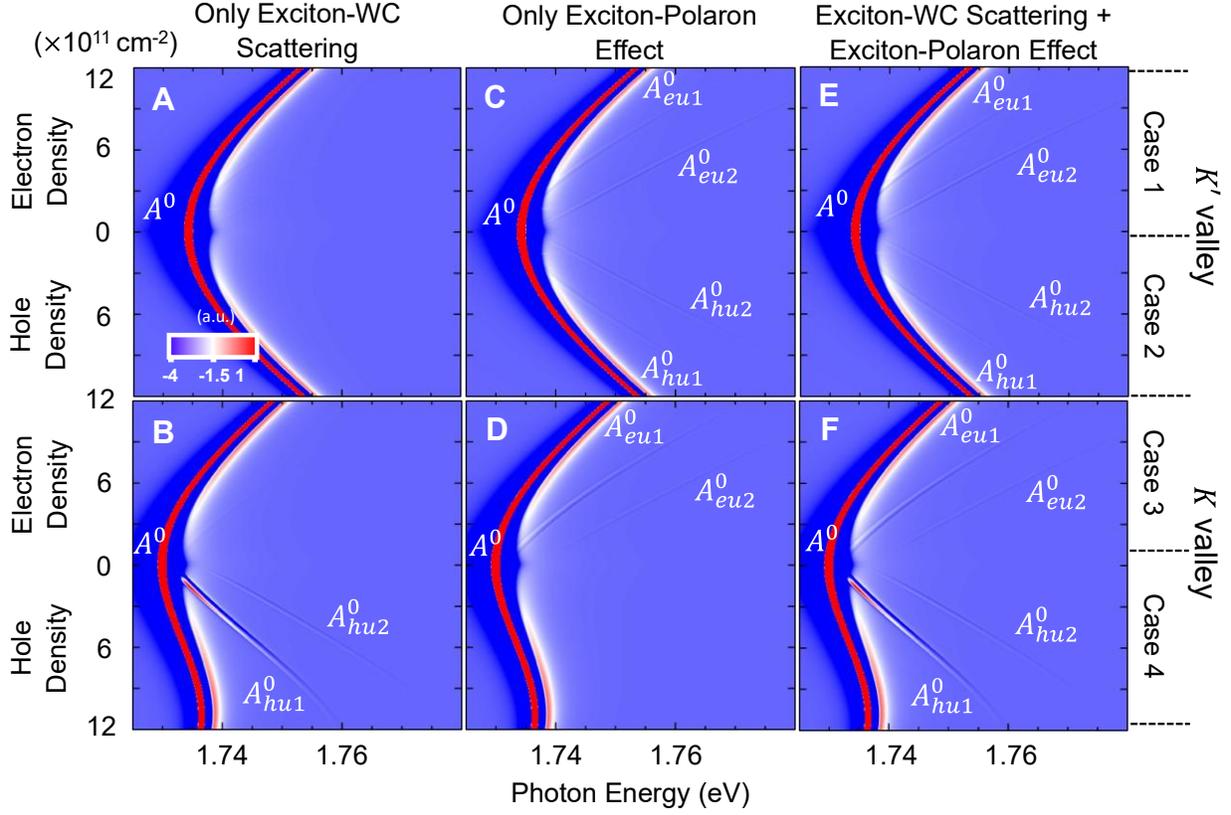

**Figure S25**. Color maps of the second-order energy derivative of reflectance contrast calculated by using the corresponding conductivity maps in Fig. S24. All plots here are in linear scale.

## 10. Summary of our theoretical model

In our theory, we adopt an empirical exciton-polaron model that describes the coupling and relative oscillator strength between the localized tetron and exciton in BN-encapsulated monolayer WSe$_2$ with electron or hole Wigner crystal (WC). Our model predicts the density-dependent spectra of the exciton and exciton-polaron Umklapp states at zero magnetic field, as well as the exciton Umklapp states under a strong magnetic field. Our results align adequately with the experimental observations. Below we summary the key assumptions, empirical parameters, results and limitations of our model.

### 10.1. Assumptions, approximations, and theoretical framework

Our model uses the following assumptions and approximations:

1) In Section 1, The exciton wavefunction is determined using the Rayleigh-Ritz variational method within the effective-mass approximation, as detailed in (*S21*).

2) In Section 2, We assume that the wavefunctions of the WC particles at each WC site can be described by Gaussian functions with a parameter $\beta$ on the exponent.

3) In Section 2, For $r_s \geq 31$, we assume the density-dependent $\beta$ is determined by a variational method based on the Hartree-Fock approximation, neglecting any inter-site overlap. For $r_s < 31$, we assume $\beta$ decreases exponentially with density, with a decay parameter $\zeta$ determined by fitting the observed Umklapp line strength.

4) In Section 3, we approximate the inter-particle interaction using a Keldysh potential (*S7*), supplemented by addition potential terms (listed in Table S1) to describe the short-range inter-particle interaction.

5) In Section 7, our model adopts the following physical picture: a photogenerated exciton can bind with a WC particle to form a localized trion. This trion, combined with the vacancy at the WC site, constitutes a tetron whose ground state has zero momentum. The trion's internal wavefunction is determined using the Rayleigh-Ritz variational method within the effective-mass approximation, as described in Ref. (*S22*).

6) In Section 7.1.2, we approximate the wavefunction of the trion CM motion by $e^{-bR_T^2}$ with $b = 2\beta$, as illustrated in Fig. S15B.

7) As illustrated in Fig. S15B, for a tetron localized in a WC site, we approximate its wavefunction by the product of the tetron internal-motion wavefunction $F(\mathbf{r}_{12}, \mathbf{r}_{13}, \mathbf{r}_4)$ and the trion CM wavefunction $e^{-bR_T^2}$.

8) As illustrated in Fig. S15C, we use a tetron state at finite momentum to describe an exciton hopping between WC sites while simultaneously interacting with WC particles. The energy of this itinerant tetron is approximated as the sum of the kinetic energy of the exciton and the energy of a zero-momentum tetron, assuming the tetron binding energy remains constant with varying momentum.

9) As illustrated in Fig. S15C, when the tetron gains momentum to travel between WC sites, we approximate its wavefunction as the product of $F(\mathbf{r}_{12}, \mathbf{r}_{13}, \mathbf{r}_4)$ and the linear combination of the CM wavefunction $e^{i\mathbf{g}_1 \cdot \mathbf{R}_X}$ of a traveling exciton and the CM wavefunction $e^{-bR_T^2}$ of a localized trion at a WC site.

Furthermore, our observation of optical signatures of Wigner crystals in the regime where the average inter-particle distance $r_s$ falls below the conventional limit of $r_s \geq 31$ suggests that defects and the Keldysh potential may stabilize the Wigner crystals (Section 2).

## 10.2. Empirical parameters in our model

Our model employs a set of empirical parameters to characterize the WC particle density distribution, exciton-tetron coupling, as well as the energies and oscillator strengths of excitonic states. These parameters are summarized below.

(1) In page 13, we introduce a parameter $\zeta = 100 \ nm^2$ to account for the limitations of modeling the WC particle distribution using Gaussian functions $\eta(r)$ in Eq. (12). Our simplified model neglects the overlap between WC particle wavefunctions at neighboring sites, which leads to an underestimate of the Gaussian width in the high-density regime ($r_s < 31$). To address this shortcoming, we modify the standard deviation $s = \sqrt{1/(2\beta)}$ of the Gaussian function by multiplying it with an exponential factor $exp(\zeta/a_{wc}^2)$.

(2) In Eqs. (86) and (95), we use the parameters $f_c$ ($f_s$) to describe the density-dependent screening effect caused by WC particles on the energy (absorption strength) of the primary exciton.

(3) In Eq. (101), we use the parameter $b = 2\beta$ to describe the CM motion of the trion, as illustrated in Fig. S10B.

(4) In Eq. (115), we use the parameter $\overline{U}_{X\mathfrak{Z}}$ to describe the exciton-tetron coupling strength for the $A^+$ polaron, and similarly $\overline{U}_{X\mathfrak{Z}_1} = \overline{U}_{X\mathfrak{Z}_2}$ for the $A_1^-$ and $A_2^-$ polarons.

(5) In Eq. (127), we use the parameter $\overline{R}_{\mathfrak{Z}X}$ to describe the absorption strength ratio between the $A^+$ polaron and the $A^0$ exciton, and similarly $\overline{R}_{\mathfrak{Z}_1X} = \overline{R}_{\mathfrak{Z}_2X}$ for the strength ratio between the $A_1^-$, $A_2^-$ polarons and the $A^0$ exciton.

(6) In Eq. (128), we introduce the parameter $\zeta_X$ to describe the rapid decrease in the absorption strength of the primary exciton at densities exceeding $n_0 = 6 \times 10^{11} cm^{-2}$.

(7) In Eq. (132), we use the experimental values $E_0^{\mathfrak{Z}_1} = -27.4$ meV and $E_0^{\mathfrak{Z}_2} = -34.1$ meV for the energies of the $A_1^-$ and $A_2^-$ polarons relative to the exciton energy at near zero density. We also use parameters $\overline{U}_{X\mathfrak{Z}_1} = \overline{U}_{X\mathfrak{Z}_2}$ to describe the exciton-tetron coupling strength for the $A_1^-$ and $A_2^-$ polarons.

All the parameters mentioned above fall within a physically reasonable range. Except for $\zeta$ and $b$ in Points (1, 3), these parameters are used to fit the primary excitonic features, which are not the central focus of our study. Instead, they establish the background for our simulations and are unrelated to the Umklapp lines. For modelling the Umklapp lines, we employ only two parameters: $\zeta$, to capture the exciton-WC scattering effect in the high-density regime, and $b$, to account for the polaron effect.

### 10.3. Key results of our model

Our model successfully reproduces the observed energy dispersion and relative absorption strengths of the Umklapp lines for both excitons and exciton polarons across various cases at zero and finite magnetic fields. The key results are summarized below.

(1) Our model confirms the existence of quasi-linearly dispersed Umklapp lines for excitons ($A_{eu2}^0$, $A_{hu2}^0$) and exciton polarons ($A_{u1}^-$, $A_{u2}^-$, $A_u^+$), in addition to the quadratically dispersed Umklapp lines of the exciton ($A_{eu1}^0$, $A_{hu1}^0$) as illustrated in Fig. 1 in the main paper.

(2) Our simulation of differential reflectance contrast (Fig. 2F) captures all the observed Umklapp lines from our experiment, including the $A_{eu2}^0$, $A_{u1}^-$, $A_{u2}^-$, $A_{hu1}^0$, and $A_{hu2}^0$ lines (Fig. 2C). Additionally, it predicts the $A_{eu1}^0$ and $A_u^+$ Umklapp lines, which remain unresolved in our experiment, highlighting potential avenues for further research and improvements.

(3) Our simulation of differential reflectance contrast under a strong magnetic field (Fig. 3C-D) captures all the observed Umklapp lines from our experiment, including the $A_{eu1}^0$ $A_{eu2}^0$, $A_{hu1}^0$, and $A_{hu2}^0$ lines (Fig. 3A-B). The simulation allows us to identify the observed Umklapp lines as $A_{eu2}^0$ and $A_{hu2}^0$ in the K-valley response, and as $A_{eu1}^0$ and $A_{hu1}^0$ in K'-valley response.

(4) Our model identifies two distinct mechanisms that brighten the Umklapp lines: the exciton-WC scattering and the exciton-polaron effect.

(5) Our model identifies four distinct cases of band configurations, each contributing differently to the Umklapp lines, as illustrated by the band diagrams for Cases 1-4 in Fig. 3, Fig. S6, and Fig. S23.

(6) The four cases and two brightening mechanisms provide complementary contributions to the Umklapp lines. Cases 1-3 exhibit a strong exciton-polaron effect but minimal exciton-WC scattering, whereas Case 4 lacks the polaron effect but demonstrates strong exciton-WC scattering. This strong exciton-WC scattering in Case 4 is driven by the long-range exchange interaction arising from the overlap of the photogenerated hole and the WC hole in the same valley, as illustrated by the diagrams of Case 4 in Fig. 3.

(7) Separate simulations (Fig. 4) resolve the contributions of the two brightening mechanisms – the exciton-WC scattering and the exciton-polaron effect – to each Umklapp line through separate simulations.

### 10.4. Limitations of our model

Due to the complex nature of our Wigner-crystal system with excitonic states, constructing a fully realistic model from first principles is challenging. Our model is subject to several limitations, as outlined below:

(1) *Simplified Wavefunction Approximation*: The wavefunctions of the Wigner crystal are approximated as a 2D array of Gaussian functions, which is an oversimplification that cannot fully capture the intricacies of the many-body system's wavefunction.

(2) *Neglect of Defects*: Our model does not account for the effects of defects, which can alter the system's wavefunction and play a crucial role in stabilizing the Wigner crystal at high charge densities. To address the behavior of Wigner crystals in the high-density regime ($r_s < 31$), which is beyond the scope of the simplified approach in Point (1), we introduce a density-dependent empirical parameter $\zeta$ to adjust the Gaussian spreading in this regime.

(3) *Binding Energy Discrepancies*: Our model does not accurately reproduce the binding energies of the $A_1^-$ and $A_2^-$ polaron states on the electron side. Instead, we simply adopt the experimentally determined binding energies at low density.

(4) *Empirical exciton-tetron coupling strength*: Our model does not compute the coupling strength between exciton and tetron states in the Wigner crystal from first principles. Instead, the strength is represented by an empirical parameter ($\overline{U}_{X\mathfrak{Z}}$ for the hole side; $\overline{U}_{X\mathfrak{Z}_1} = \overline{U}_{X\mathfrak{Z}_2}$ for the electron side), determined by fitting the measured density-dependent energy separation between the exciton and polaron.

### 10.5. Prospects for improving the theoretical model

Future research could explore the following improvements to our theoretical model.

(1) The WC wavefunction could be calculated more rigorously by employing density-functional theory that incorporates realistic band structure effects.

(2) The influence of defects could be explicitly included in the calculation of WC wavefunction to provide a more accurate representation.

(3) The wavefunctions of trions and polarons could be refined by incorporating key effects from the realistic ban structure, moving beyond the simplified effective mass approximation in our current model.

(4) On the electron side, the polaron wavefunction could be calculated by using a six-particle configuration approach (*S23*), replacing the four-particle configuration used in our current model.

(5) The absolute absorption strength of the Umklapp lines could be explicitly determined, eliminating the need for the empirical parameter $b$ in Eq. (101) of our current model. An effective approach would involve using a more complete basis to describe the CM wavefunction of the tetron, rather than using the simple trial wavefunction, as shown in Eq. (102) of our current model.

## Supplementary References